\pdfoutput=1

\documentclass[11pt,twoside,a4paper,cmspaper,final,collab]{cms-tdr}
\def\svnVersion{370072}\def\cmsCernNoTag{CERN-EP/2016-067}\def\cmsCernDate{\today}\def\cmsMessage{Published in the Journal of High Energy Physics as \href{http://dx.doi.org/10.1007/JHEP10(2016)006}{\doi{10.1007/JHEP10(2016)006.}}}

\begin{document}\cmsNoteHeader{SUS-15-003}

\hyphenation{had-ron-i-za-tion}
\hyphenation{cal-or-i-me-ter}
\hyphenation{de-vices}
\RCS$Revision: 370053 $
\RCS$HeadURL: svn+ssh://svn.cern.ch/reps/tdr2/papers/SUS-15-003/trunk/SUS-15-003.tex $
\RCS$Id: SUS-15-003.tex 370053 2016-10-06 13:55:12Z fgolf $
\newlength\cmsTabSkip\setlength\cmsTabSkip{1.5ex}
\cmsNoteHeader{SUS-15-003}
\title{Search for new physics with the \mttwo variable in all-jets final states produced in
pp collisions at $\sqrt{s}$ = 13\TeV}

\date{\today}
\newcommand{\Lint}{2.3\fbinv\xspace}
\newcommand{\mttwo}{\ensuremath{M_{\mathrm{T2}}}\xspace}
\newcommand{\Met}{\ensuremath{E_{\mathrm{T}}^{\text{miss}}}\xspace}
\newcommand{\Mht}{\ensuremath{H_{\mathrm{T}}^{\text{miss}}}\xspace}
\newcommand{\vMht}{\ensuremath{\vec{H}_{\mathrm{T}}^{\mathrm{miss}}}\xspace}
\newcommand{\dpmin}{\ensuremath{\Delta\phi_{\text{min}}}\xspace}
\newcommand{\njets}{\ensuremath{N_{\mathrm{j}}}\xspace}
\newcommand{\nbtags}{\ensuremath{N_{\PQb}}\xspace}
\newcommand{\MT}{\ensuremath{M_{\mathrm{T}}}\xspace}
\newcommand{\wjets}{\ensuremath{\PW\text{+jets}}\xspace}
\newcommand{\zjets}{\ensuremath{\Z\text{+jets}}\xspace}
\newcommand{\tw}{\ensuremath{\PQt\PW}\xspace}
\newcommand{\ttjets}{\ensuremath{\ttbar\text{+jets}}\xspace}
\newcommand{\ttw}{\ensuremath{\ttbar\PW}\xspace}
\newcommand{\ttz}{\ensuremath{\ttbar\Z}\xspace}
\newcommand{\tth}{\ensuremath{\ttbar\PH}\xspace}
\newcommand{\znunu}{\ensuremath{\Z\to\cPgn\cPagn}\xspace}
\newcommand{\zll}{\ensuremath{\Z\to\ell^+\ell^-}\xspace}
\newcommand{\W}{\PW\xspace}
\providecommand{\NA}{\text{---}\xspace}
\providecommand{\MATNLO}{{\MADGRAPH{}\_a\textsc{mc@nlo}}\xspace}
\abstract{
A search for new physics is performed using events that contain one or more jets, no isolated leptons,
and a large transverse momentum imbalance, as measured through the \mttwo variable,
which is an extension of the
transverse mass in events with two invisible particles. The results are based on a sample of
proton-proton collisions collected at a center-of-mass energy of 13\TeV with the CMS detector
at the LHC, and that corresponds to an integrated luminosity of 2.3\fbinv.
The observed event yields in the data are consistent with predictions
for the standard model backgrounds.
The results are interpreted using simplified models
of supersymmetry and are expressed in terms of limits on the masses of potential
new colored particles. Assuming that the lightest neutralino is stable and has a mass less than
about 500\GeV, gluino masses up to 1550--1750\GeV are excluded at 95\% confidence level,
depending on the gluino decay mechanism. For the scenario of direct production of
squark-antisquark pairs, top squarks with masses up to 800\GeV are excluded, assuming
a 100\% branching fraction for the decay to a top quark and neutralino. Similarly, bottom
squark masses are excluded up to 880\GeV, and masses of light-flavor squarks are excluded up to
600--1260\GeV, depending on the degree of degeneracy of the squark masses.
}
\hypersetup{%
pdfauthor={CMS Collaboration},%
pdftitle={Search for new physics with the MT2 variable in all-jets final states produced in pp collisions at sqrt(s) = 13 TeV},%
pdfsubject={CMS},%
pdfkeywords={CMS, physics}}
\maketitle
\section{Introduction}
\label{sec:intro}
Searches for new physics based on final states with jets and large transverse momentum imbalance are sensitive
to broad classes of new physics models, including supersymmetry (SUSY)~\cite{Ramond:1971gb,Golfand:1971iw,Neveu:1971rx,
Volkov:1972jx,Wess:1973kz,Wess:1974tw,Fayet:1974pd,Nilles:1983ge}.
Such searches were previously conducted by both the CMS~\cite{AlphaT8tev,RA2b_8tev,RA2_8tev,Razor8tev,MT2at8TeV}
and ATLAS~\cite{Atlas3rdGen,Atlas8tevSummary} collaborations, using data from 8 TeV proton-proton (pp) collisions.
They placed lower limits on the masses of pair-produced colored particles near the TeV scale for a broad
range of production and decay scenarios and provided some of the
most stringent constraints on the production of supersymmetric
particles. These searches are particularly interesting at this time as
they are among the first to benefit from the increase in the CERN LHC
center-of-mass energy from 8 to 13 TeV, as shown in two recent analyses
of these final states by ATLAS and CMS~\cite{atlasFullHad13TeV,RA2b13TeV}.
As a consequence of the increase in parton luminosity at 13\TeV, the cross section for the pair production of particles
with the color quantum numbers of a gluon increases by more than a factor of 30 for a particle of mass 1.5 TeV.

In this paper we present results of a search for new physics in events with jets and significant transverse momentum imbalance,
as characterized by the ``stransverse mass" \mttwo, a kinematic variable that was first proposed for use in SUSY
searches in Refs.~\cite{MT2variable,MT2variable2} and used in several Run 1 searches~\cite{MT2at7TeV,MT2at8TeV}.
The search is performed using a data sample corresponding to an integrated luminosity of \Lint of pp collisions collected
at a center-of-mass energy of 13 TeV with the CMS detector at the LHC.

In this analysis we select events with at least one jet and veto events with an identified, isolated lepton.
Signal regions are defined by the number of jets, the number of jets identified as a product of b quark
fragmentation (b-tagged jets), the scalar sum of jet transverse momenta (\HT), and \mttwo.
The observed event yields in these regions are compared with
the background expectation from standard model (SM) processes and the predicted contributions from simplified
supersymmetric models of gluino and squark pair production~\cite{bib-sms-1,bib-sms-2,bib-sms-3,bib-sms-4,Chatrchyan:2013sza}.
\section{The CMS detector}
\label{sec:detector}
The central feature of the CMS apparatus is a superconducting solenoid, 13\unit{m} in length and 6\unit{m} in diameter, which
provides an axial magnetic field of 3.8~T. Within the field volume are several particle detection systems.
Charged-particle trajectories are measured with silicon pixel and strip trackers, covering $0 \leq \phi < 2\pi$ in
azimuth and $\abs{\eta}< 2.5$ in pseudorapidity, where $\eta \equiv -\ln [\tan (\theta/2)]$ and $\theta$ is the
polar angle of the trajectory of the particle with respect to the beam direction.
The transverse momentum, the component of the momentum $p$ in the plane orthogonal to the beam, is defined in
terms of the polar angle as $\pt = p \sin \theta$.
A lead-tungstate crystal electromagnetic calorimeter and a brass and scintillator hadron calorimeter surround
the tracking volume, providing energy  measurements of electrons, photons, and hadronic jets
in the range $\abs{\eta} < 3.0$.
Muons are identified and measured within $\abs{\eta} <  2.4$ by gas-ionization detectors embedded in the steel
flux-return yoke of the solenoid.
Forward calorimeters on each side of the interaction point encompass $3.0 <\abs{\eta} <5.0$.
The detector is nearly hermetic, allowing momentum imbalance measurements in the plane transverse to the beam direction.
A two-tier trigger system selects pp collision events of interest for use in physics analyses.
A more detailed description of the CMS detector is available in Ref.~\cite{JINST}.
\section{Simulated event samples}
\label{sec:evtsel:mc}
Monte Carlo (MC) simulations are used in the estimate of some of the SM backgrounds,
as well as to calculate the selection efficiency for various new physics scenarios.
The main background and control samples (\wjets, \zjets, \ttjets, $\gamma\mathrm{+jets}$, and QCD multijet events), as well as signal samples of gluino and squark pair production, are generated with the \MADGRAPH~5 generator~\cite{madgraph5} interfaced with \PYTHIA~8.2~\cite{pythia8} for fragmentation and parton showering.
Signal processes are generated at leading order with up to two extra partons present in the event.
Other background samples are generated with \MATNLO~2.2 \cite{Alwall:2014hca} ($s$ channel single top, \ttw, \ttz, \tth) and with \POWHEG~v2~\cite{Alioli:2009je, Re:2010bp} ($t$ channel single top, \tw), both interfaced with \PYTHIA 8.2 \cite{pythia8}.

Next-to-leading order (NLO) and next-to-NLO cross sections~\cite{Gavin:2010az, Gavin:2012sy, Czakon:2011xx, Alwall:2014hca, Alioli:2009je, Re:2010bp} are used to normalize the simulated background samples, while NLO plus next-to-leading-logarithm (NLL) calculations~\cite{Borschensky:2014cia} are used for the signal samples.
The NNPDF3.0LO and NNPDF3.0NLO \cite{Ball:2014uwa} parton distribution functions (PDF) are used, respectively, with \MADGRAPH, and with {\POWHEG v2} and \MATNLO.
Standard model processes are simulated using a \textsc{geant4} based model~\cite{geant4} of the CMS detector,
while the simulation of new physics signals is performed using the CMS fast simulation package~\cite{fastsim}.
All simulated events include the effects of pileup, \ie multiple pp collisions within the
same or neighboring bunch crossings, and are processed with the same chain of reconstruction
programs as used for collision data.
\section{Event reconstruction}
\label{sec:evtreco}
Event reconstruction is based on the particle-flow (PF) algorithm~\cite{pflow,pflow2}, which combines information from the tracker, calorimeter, and muon systems to reconstruct and identify PF candidates, \ie charged and neutral hadrons, photons, muons, and electrons.
We select events with at least one reconstructed vertex that is within 24\cm\,(2\cm) of the center of the detector in the direction along (perpendicular to) the beam axis.
In the presence of pileup, usually more than one such vertex is reconstructed. We designate as the primary vertex (PV) the one for which the summed $\pt^{2}$ of the associated charged PF candidates is the largest.

Charged PF candidates associated with the PV and neutral particle candidates are clustered into jets using the
anti-\kt algorithm~\cite{antiKt} with a distance parameter of 0.4.
The jet energy is calibrated using a set of corrections similar to those developed for the 8\TeV data~\cite{JetCalibration8TeV}:
an offset correction accounting for neutral energy arising from
pileup interactions in the area of the reconstructed jet;
a relative correction that makes the jet energy response, \ie the ratio of
the reconstructed to the original jet energy, uniform in \pt and $\eta$;
an absolute correction that restores the average jet energy response to unity;
and a residual correction, applied to account for remaining differences between data and simulation.

Jets originating from b quarks are identified by the combined secondary vertex algorithm~\cite{cmsBTV}.
We use a working point with a tagging efficiency of approximately 65\% for jets originating from b quarks with momenta typical of top quark pair events.
For jets with transverse momentum above approximately 200\GeV, the tagging efficiency decreases roughly linearly, reaching an efficiency of about 45\% at 600\GeV.
The probability to misidentify jets arising from c quarks as b jets is about 12\%, while
the corresponding probability for light-flavor quarks or gluons is about 1.5\%.

The transverse hadronic energy, \HT, is defined as the scalar sum of the magnitudes of the jet transverse momenta,
while the missing transverse hadronic momentum, \Mht, is defined as the negative vector sum of the
transverse momenta of the same jets.
Except for a few cases described later, the construction of higher-level variables and the event categorization are based
on jets with $\pt > 30\GeV$, $\abs{\eta} < 2.5$, and passing loose requirements on the jet composition
designed to reject rare spurious signals arising from noise and failures in the event reconstruction~\cite{cmsJetId}.
The transverse momentum imbalance (\ptvecmiss), whose magnitude is referred to as \ETmiss, is defined as the negative of the
vector sum of the transverse momenta of all reconstructed charged and neutral PF candidates.

Electron candidates are reconstructed as clusters of energy deposits in the electromagnetic calorimeter, matched to tracks in the silicon tracker~\cite{cmsEGM}.  We identify electrons having $\pt > 10\GeV$ by loose requirements on the shape of these energy deposits, on the ratio of energy in associated hadron and electromagnetic calorimeter cells ($H/E$), on the geometric matching between the energy deposits and the associated track, and on the consistency between the energy reconstructed from calorimeter deposits and the momentum measured in the tracker.
In addition, we require that the associated track be consistent with originating from the PV.
The PF algorithm applies a looser set of requirements to identify ``PF electrons'' with even smaller transverse momenta.  We use
it to extend the range of identified electrons down to $\pt > 5\GeV$.

Muon candidates are reconstructed by combining tracks found in the muon system with corresponding tracks in the
silicon detectors.
Candidates are required to be classified as either \textit{Global Muons} or \textit{Tracker Muons}, according to the definitions given in Ref.~\cite{cmsMuon}, when they have $\pt > 10\GeV$.
The associated silicon detector track is required to be consistent with originating from the PV.
The PF algorithm applies looser requirements to identify ``PF muons'' with even smaller transverse momenta. We use it to extend the range of identified muons down to $\pt > 5\GeV$.

The isolation of electrons and muons is defined as the scalar sum of the transverse momenta of all neutral
and charged PF candidates within a cone $\Delta R=\sqrt{(\Delta\eta)^2+(\Delta\phi)^2}$ along the lepton direction. The variable
is corrected for the effects of pileup using an effective area correction~\cite{CMS-PAS-JME-14-001},
and the size of the cone is dependent on the lepton \pt according to:

\begin{align}
  \Delta R &=\begin{cases}
    0.2, & \pt \leq 50\GeV, \\
    \dfrac{10\GeV}{\pt}, & 50 < \pt \le 200\GeV,\\
    0.05, & \pt > 200\GeV.
  \end{cases}\label{eq:miniIso}
\end{align}
The relative lepton isolation is the lepton isolation divided by the lepton \pt.

When selecting PF electrons and muons, as well as isolated PF charged hadrons, a track--only isolation
computed in a larger cone is used. Relative track isolation is calculated using all charged
PF candidates within a cone $\Delta R<0.3$ and longitudinal impact parameter $\abs{\Delta z} < 0.1\unit{cm}$ relative to the PV.

The efficiency for selecting prompt electrons, i.e., electrons from decays of electroweak bosons or SUSY particles, increases
from 65--70\% at a \pt of 10\GeV to 80--90\% at 50\GeV,
and plateaus at 85--95\% above 100\GeV,
where the smaller values are from signal samples with high jet multiplicity and the larger numbers are from \ttjets events.
For prompt muons, the efficiency increases from 75--90\% at a \pt of 10\GeV to
85--95\% at 50\GeV, and plateaus at 95--99\% above 200\GeV.

Photon candidates, used in the estimation of the \znunu background, are reconstructed from deposits in the
electromagnetic calorimeter and are selected using the shower shape variable ($\sigma_{\eta\eta}$) and the ratio $H/E$~\cite{Khachatryan:2015iwa}.  Additionally, we require that their track isolation in a cone $\Delta R <0.3$ be less than 2.5\GeV.

\section{Event selection}
\label{sec:evtsel}
Before assigning events to different signal regions, the baseline selection described in this
section is implemented. Collision events are selected using triggers with different
requirements on \HT, \ETmiss, and \Mht.
Table~\ref{tab:trig} summarizes the triggers and corresponding offline selections, after which the
triggers are found to be $>$98\% efficient.
As shown in the table, events with $\HT <1000\GeV$ are selected with triggers that impose an \ETmiss requirement.
As a consequence, for the low \HT sample we employ a tighter requirement on the offline value of \ETmiss.
\begin{table}
\topcaption{The three signal triggers and the corresponding offline selections. }
\label{tab:trig}
\centering
\begin{tabular}{ll}
\hline
Online trigger selection [\GeVns{}] & Offline selection [\GeVns{}]\\
\hline
$\HT > 800$                    & $\HT>1000$ \& $\ETmiss>30$ \\
$\HT>350$ \& $\ETmiss>100$  & $\HT>450$ \& $\ETmiss>200$ \\
$\Mht > 90$ \& $\ETmiss>90$ \& noise removal criteria & $\HT>200$ \& $\ETmiss>200$ \\ \hline
\end{tabular}
\end{table}

The events passing the selections of Table~\ref{tab:trig} are further divided according
to the total number of jets (\njets) and the number of jets identified
as originating from b quarks (\nbtags). When determining \nbtags,
we lower the jet \pt threshold from 30 to 20\GeV in order to increase sensitivity to potential signal
scenarios with soft decay products.

\begin{figure}[htb]
  \centering
    \includegraphics[width=0.45\textwidth]{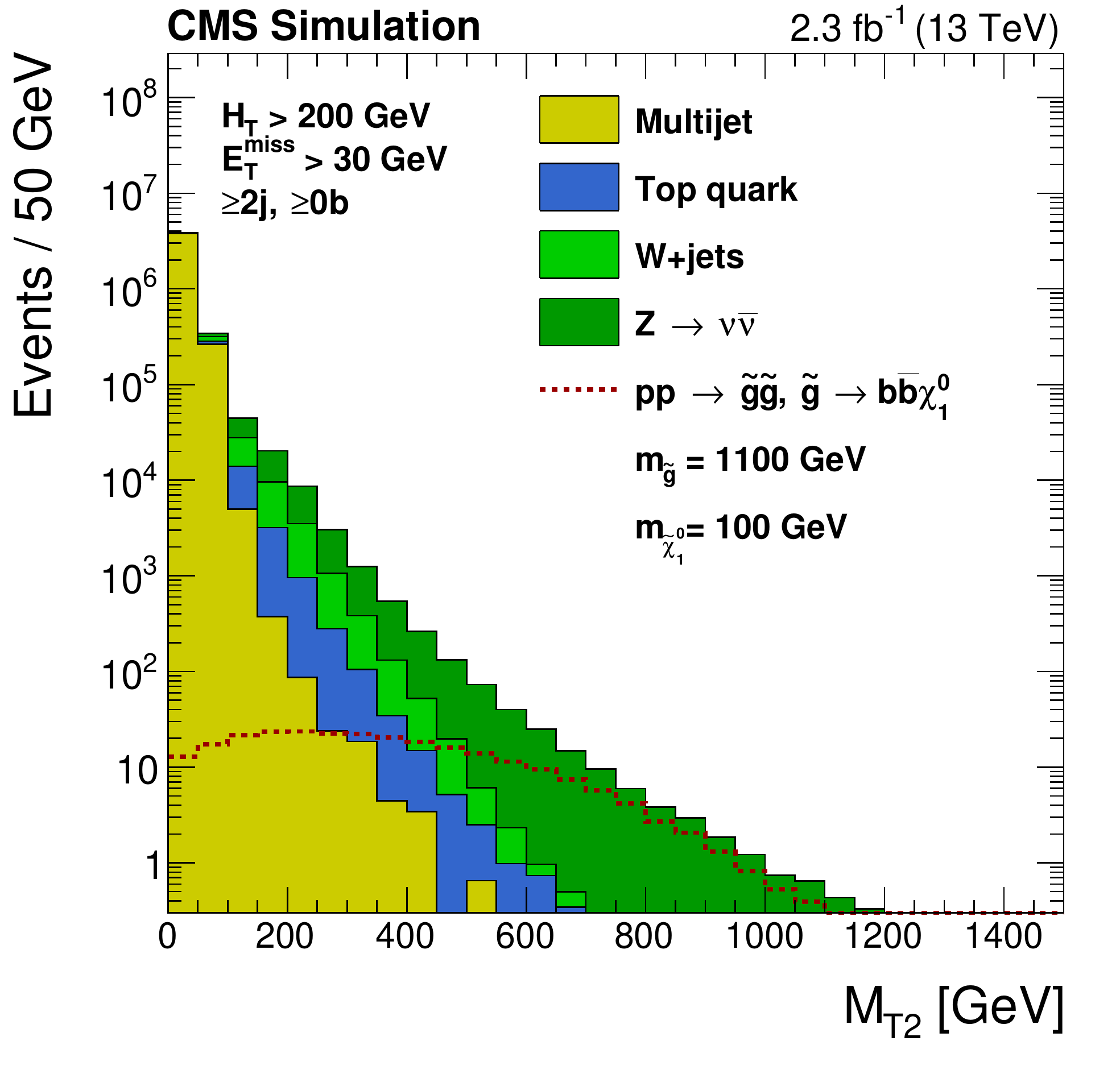}
    \caption{Distribution of the \mttwo variable in simulated background and signal event samples
      after the baseline selection is applied.
      The line shows the expected \mttwo distribution for a signal model of gluino-mediated bottom squark
      production with the masses of gluino and lightest neutralino equal to 1100 and 100\GeV, respectively.
      The simplified signal model is described in Refs.~\cite{bib-sms-1,bib-sms-2,bib-sms-3,bib-sms-4,Chatrchyan:2013sza}
      and in the text.
}
    \label{fig:mt2PR}
\end{figure}

For events with at least two reconstructed jets, we start with the pair having the largest dijet invariant mass
and iteratively cluster all selected jets using a hemisphere algorithm that minimizes the Lund
distance measure~\cite{LundDistRef1,Phythia64} until two stable pseudo-jets are obtained.
The resulting pseudo-jets together with the \ptvecmiss are used to determine the stransverse mass \mttwo~\cite{MT2variable,MT2variable2}.
This kinematic mass variable, which can be considered as a generalization of the
transverse mass variable \MT defined in Ref.~\cite{Arnison:1983rp}, was introduced
as a means to measure the mass of pair-produced particles in situations where both decay to a final state
containing the same type of undetected particle. The variable \mttwo is defined as:

\begin{equation}
\mttwo = \min_{\ptvecmiss{}^{\mathrm{X}(1)} + \ptvecmiss{}^{\mathrm{X}(2)} = \ptvecmiss}
  \left[ \max \left( \MT^{(1)} , \MT^{(2)} \right) \right],
\label{eq.MT2.definition}
\end{equation}
where $\ptvecmiss{}^{\mathrm{X}(i)}$  (with $i=1$,2) are the unknown transverse momenta of the two undetected particles and
$\MT^{(i)}$ the transverse masses obtained by pairing any of the two invisible particles with one of
the two pseudojets.
The minimization is performed over trial momenta of the undetected
particles fulfilling the \ptvecmiss constraint.
Most of the background from QCD multijet events (defined more precisely in Section~\ref{sec:bkgds}) is characterized by
very small values of \mttwo, while a wide class of
new physics models imply large values of stransverse mass. Figure~\ref{fig:mt2PR} shows the \mttwo distributions
expected from simulation for the background processes and one signal model, the gluino-mediated
bottom squark production described in Refs.~\cite{bib-sms-1,bib-sms-2,bib-sms-3,bib-sms-4,Chatrchyan:2013sza} and Section~\ref{sec:results}.
Selections based on the \mttwo variable are a powerful means
to reduce the contribution from multijet events to a subleading component of the total background.
A complete discussion of the \mttwo properties as a discovery variable and details about the exact calculation of the  variable
are given in Refs.~\cite{MT2at7TeV,MT2at8TeV}.

The main selection to suppress the background from multijet production is the requirement $\mttwo >200\GeV$ in events with at least two reconstructed jets.
Even after this requirement, a residual background contribution with larger \mttwo values remains, arising
primarily from events in which the energy of a jet has been severely underestimated.
To further suppress background events resulting from this effect, we require $\dpmin > 0.3$, where \dpmin is
defined as minimum azimuthal angle between the \ptvecmiss vector and up to four highest \pt jets.
For the purpose of the \dpmin calculation only, we consider jets with $\abs{\eta} < 4.7$.
The number and definition of jets entering the \dpmin calculation are chosen to maximize signal to background separation.
In addition, we require that the magnitude of the vector difference in the transverse momentum imbalance determined
using either the selected jets (\vMht) or all PF candidates (\ptvecmiss) satisfy $\abs{\ptvecmiss-\vMht}/\Met< 0.5$.
This requirement protects against large imbalances arising from objects with $\pt< 30\GeV$ or $\abs{\eta} > 2.5$.
Finally, events with possible contributions from beam halo processes or anomalous noise in the calorimeters are rejected using dedicated filters~\cite{cmsJME}.

To reduce the background from SM processes with genuine \ETmiss arising from the decay of a \W boson, we reject events with an identified electron or muon with $\pt> 10\GeV$ and $\abs{\eta} <2.4$.
Only electrons (muons) with a relative isolation less than 0.1\,(0.2) are considered in the veto.
Events are also vetoed if they contain an isolated charged PF candidate (electron, muon or charged hadron) to reject
$\tau$ leptons decaying to leptons or hadrons.
To avoid loss of efficiency in potential signals with large jet multiplicities, events are only vetoed if the transverse mass (\MT) formed by the momentum of the isolated charged PF candidate and \ptvecmiss is less than 100\GeV, consistent with the leptonic decay of a \W boson.
For charged candidates identified as a PF electron or muon, we veto the event if the candidate has $\pt > 5\GeV$ and a relative track isolation of less than 0.2.
For charged candidates identified as a PF hadron, we veto the event if the candidate has $\pt > 10\GeV$ and a relative track isolation of less than 0.1.
\subsection{Signal regions}
\label{sec:evtsel:sr}
Signal regions are defined separately for events with either exactly one jet passing the counting criteria above, or with two or more jets.
Events with $\njets \geq 2$ are categorized based on \HT, \njets, \nbtags as follows:
\begin{itemize}
\item 5 bins in \HT: [200,450], [450, 575], [575, 1000], [1000, 1500], and $>$1500.  These bins, which are expressed in \GeV,
are also referred to as very low \HT, low \HT, medium \HT, high \HT, and extreme \HT regions,
\item 11 bins in \njets and \nbtags: 2--3j \& 0b, 2--3j \& 1b, 2--3j \& 2b, 4--6j \& 0b, 4--6j \& 1b, 4--6j \& 2b, $\geq$7j \& 0b, $\geq$7j \& 1b, $\geq$7j \& 2b, 2--6j \& $\geq$3b, $\geq$7j \& $\geq$3b.
\end{itemize}
Each bin defined by the \HT, \njets, \nbtags requirements above is referred to as a ``topological region''.

Since SUSY events could result in \mttwo distributions harder than the remaining SM backgrounds, we further divide each topological region in bins of \mttwo, expressed in \GeVns, as follows:
\begin{itemize}
\item 3 bins at very-low \HT: [200,300], [300,400], and $>$400,
\item 4 bins at low \HT: [200,300], [300,400], [400,500], and $>$500,
\item 5 bins at medium \HT: [200,300], [300,400], [400,600], [600,800], and $>$800,
\item 5 bins at high \HT: [200,400], [400,600], [600,800], [800, 1000], and $>$1000,
\item 5 bins at extreme \HT: [200,400], [400,600], [600,800], [800,1000], and $>$1000.
\end{itemize}
For events with $\njets = 1$, \ie belonging to the ``monojet'' signal regions, the \mttwo variable
is not defined.
We instead opt for a simpler strategy with signal regions defined by the \pt of the jet and \nbtags:
\begin{itemize}
   \item \nbtags: 0b, $\ge$1b,
   \item 7 bins in jet \pt, indicated in \GeV, which are defined as follows:  [200,250], [250,350], [350,450], [450,575], [575,700], [700,1000], and $>$1000.
\end{itemize}
In order to have more than one event expected in each signal region, the actual \mttwo (or jet \pt) binning is coarser
than indicated above for some of the topological regions. A complete list
of the signal bins is provided in Tables~\ref{tab:mt2bins1}, \ref{tab:mt2bins2}, and \ref{tab:mt2bins3} in Appendix~\ref{app:results}.
In total, we define 172 separate signal regions.
\section{Backgrounds}
\label{sec:bkgds}
There are three sources of SM background to potential new physics signals in a jets plus \ETmiss final state:
\begin{itemize}
\item ``Lost lepton background'':  events with genuine invisible particles, \ie neutrinos, from leptonic \W boson decays where the charged lepton is either out of acceptance, not reconstructed, not identified, or not isolated.
  This background comes from both \wjets and \ttjets events, with a small contribution from single top quark production,
  and is one of the dominant backgrounds in nearly all search regions.
  It is estimated using a one-lepton control sample, obtained by inverting the lepton veto in each topological region.
\item ``\znunu background'':  \zjets events where the \Z boson decays to neutrinos.  This almost irreducible background is
most similar to potential signals.
  It is a major background in nearly all search regions, its importance decreasing for tighter requirements on \nbtags.
  This background is estimated using   $\gamma+$jets and \zll control samples.
\item ``Multijet background'': mostly instrumental background that enters a search region because of either significant mismeasurement of the jet momentum or sources
of anomalous noise in the detector.  There is also a small contribution from events with genuine \ETmiss from neutrinos produced in semi-leptonic decays of charm and bottom quarks.
  To suppress this background we apply the selections described in Section~\ref{sec:evtsel}, after which this type of background is sub-dominant in almost all search regions.
  The background is estimated from a control sample obtained by inverting the \dpmin requirement in each topological region.
\end{itemize}
For all three categories, the event yields in the control regions are translated into background estimates in the signal
regions using ``transfer factors'', either based on simulation or measured in data, which are described in
the next sections.
\subsection{Estimation of the background from leptonic \texorpdfstring{\W}{W} boson decays}
\label{sec:bkgds:ll}
Single-lepton control regions are used to estimate the background arising from leptonic W boson decays
in \wjets and \ttjets processes.
Control region events are selected using the same triggers as for signal regions, and the baseline selections of Section~\ref{sec:evtsel}
are applied with the exception of the lepton veto.
Instead, we require exactly one lepton candidate passing either the PF lepton selection (e or $\mu$ only) or
the lepton selection used in lepton vetoes.
In addition, we require $\MT(\ell,\ptvecmiss) < 100$\GeV to reduce potential contamination from signal.

Selected events are then grouped into the categories described in Section~\ref{sec:evtsel:sr}, binning the single-lepton
control regions in the \HT, \njets, and \nbtags dimensions, but not in \mttwo, to preserve statistical precision.
The binning in \njets and \nbtags is the same as that of the signal regions, except for signal bins with $\njets\geq7$ and $\nbtags\geq1$.
For these signal regions, the background prediction is obtained using a control region with the same \HT selection as the signal and requiring $\njets\geq7$ and $1 \leq \nbtags \leq 2$.
This is motivated by the scarcity of data in control regions with $\njets\geq7$ and $\nbtags\geq2$
as well as potential contamination from signal in bins with $\njets\geq7$ and $\nbtags\geq3$.
For events with $\njets=$ 1, one control region is defined for each bin of jet \pt.

The background yield $N^{\mathrm{SR}}_{1\ell}$ in each signal region SR is obtained from the corresponding
single-lepton yield $N^{\mathrm{CR}}_{1\ell}$ in the control region CR by the application of transfer
factors $R^{0\ell/1\ell}_{\mathrm{MC}}$ and $k_{\mathrm{MC}}$, and
according to the following equation:

\begin{equation}
\label{eq:ll}
  N^{\mathrm{SR}}_{1\ell} \left(\HT,\njets,\nbtags,\mttwo\right) = N^{\mathrm{CR}}_{1\ell} \left(\HT,\njets,\nbtags\right) \, R^{0\ell/1\ell}_{\mathrm{MC}} \left(\HT,\njets,\nbtags\right) \, k_{\mathrm{MC}}\left(\mttwo\right).
\end{equation}
The number of events for which we fail to reconstruct or identify an isolated lepton candidate is obtained via the factor  $R^{0\ell/1\ell}_{\mathrm{MC}} \left(\HT,\njets,\nbtags\right)$, which accounts for lepton acceptance and selection efficiency and the expected contribution from the decay of \W bosons to hadrons through an intermediate $\PGt$ lepton.
The factor $R^{0\ell/1\ell}_{\mathrm{MC}}$ is obtained from simulation and corrected for small measured differences
in lepton efficiency between data and simulation.
The fraction of events in each topological region expected to populate a particular \mttwo bin,
$k_{\mathrm{MC}}\left(\mttwo\right)$, is used to obtain the estimate in each search bin and is also obtained from simulation.

Normalization to data control regions reduces reliance on the MC modeling of most kinematic quantities, except \mttwo.
The uncertainty in $k_{\mathrm{MC}}\left(\mttwo\right)$ is evaluated in simulation by variations of
the important experimental and theoretical parameters.
Reconstruction uncertainties, assessed by varying the tagging efficiency for b quarks, and by evaluating the impact of variations in jet response on the counting of jets and b-tagged jets, \ETmiss, and \mttwo, are typically found to be less than 10\%, but can reach as much as 40\% in some bins.
Renormalization and factorization scales, PDFs~\cite{pdf4lhcRunII}, and the relative composition of \wjets and \ttjets are
varied to assess the dominant theoretical uncertainties, which are found to be as large as 30\%.
Based on these results, for $k_{\mathrm{MC}}\left(\mttwo\right)$ we assign a shape uncertainty that reaches 40\% in the highest bins of \mttwo.

\begin{figure}[htb]
  \centering
    \includegraphics[width=0.45\textwidth]{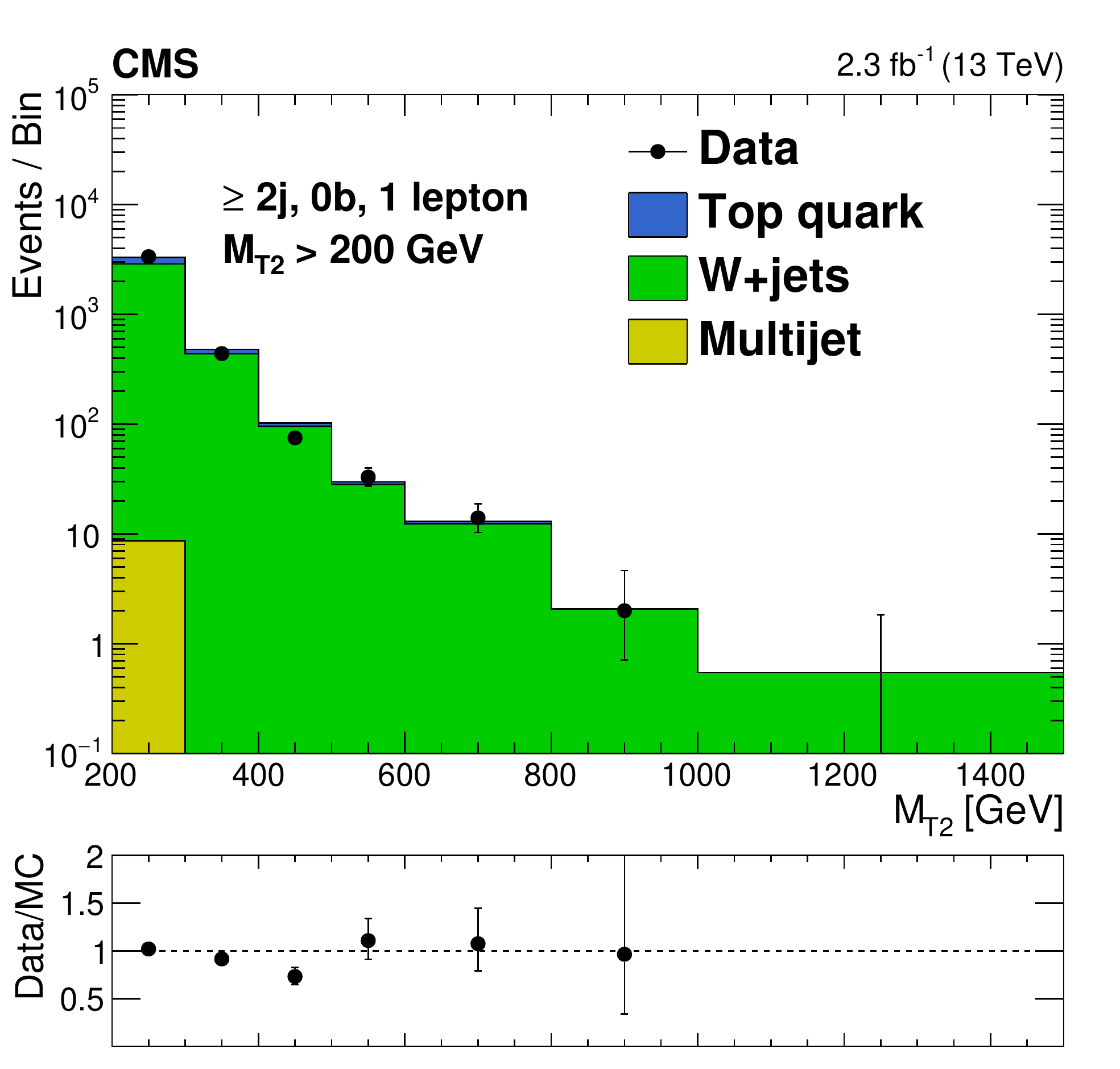}
    \includegraphics[width=0.45\textwidth]{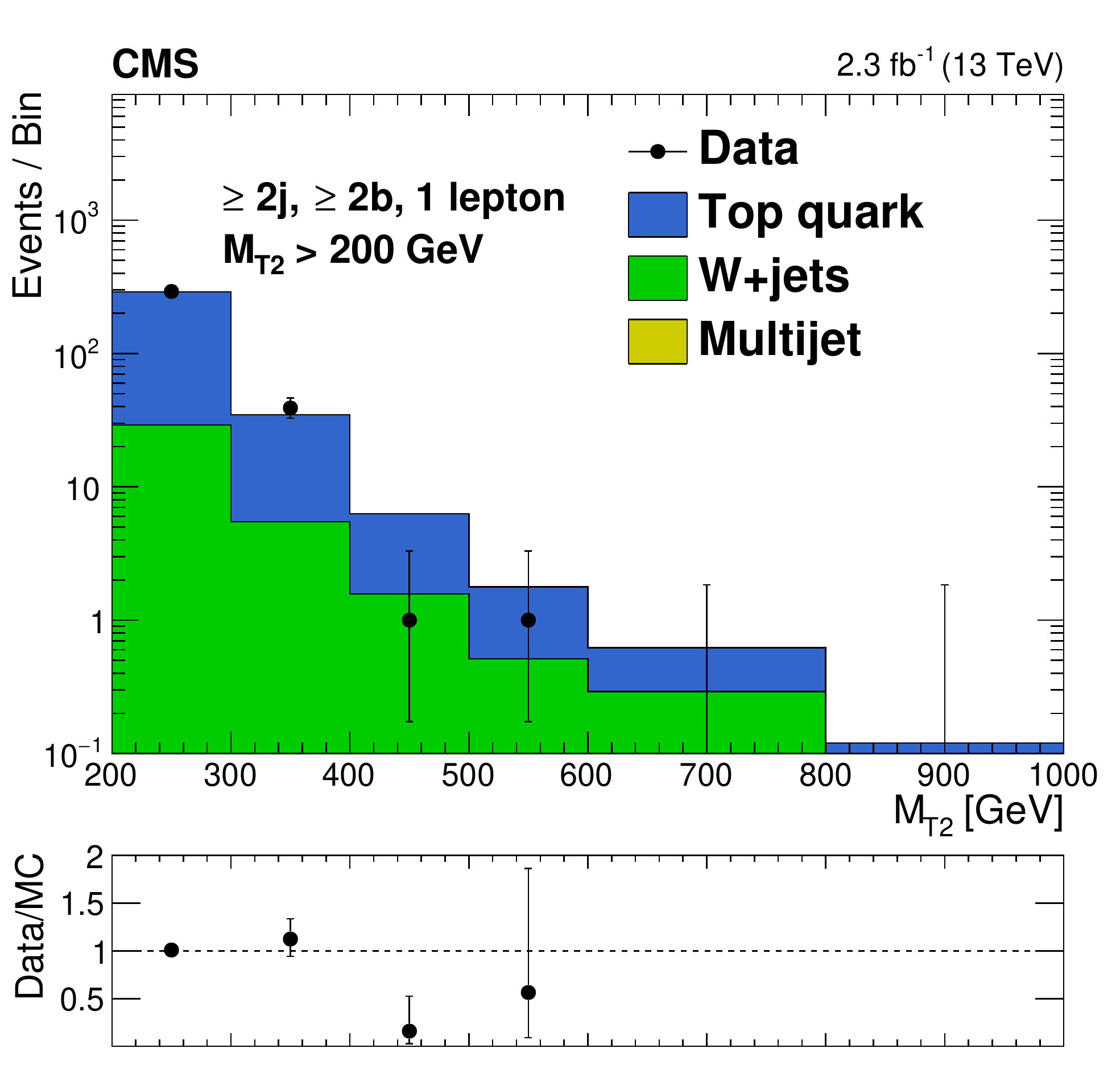}
    \caption{Comparison between simulation and data in the \mttwo observable.
      The left and right plots correspond to control samples enriched in \wjets and \ttjets, respectively.
      The sum of the distributions from simulation is scaled to have the same integral as the corresponding histograms from data.
      The uncertainties shown are statistical only.}
    \label{fig:ll_mt2}
\end{figure}

The MC modeling of the \mttwo distribution is checked in data using control regions enriched in events originating from either \wjets or \ttjets, as shown in the left and right plots of Fig.~\ref{fig:ll_mt2}, respectively.
An additional check is performed by comparing the standard estimate with that obtained by replacing the factor $k_{\mathrm{MC}}\left(\mttwo\right)$ in Eq.~(\ref{eq:ll}), with an extra dimension in the binning of the control region, which becomes $N^{\mathrm{CR}}_{1\ell} \left(\HT,\njets,\nbtags,\mttwo\right)$.  The two estimates agree within the statistical precision permitted by the size of the control regions.

The single-lepton control regions typically have 1--2 times as many events expected as compared to the corresponding signal region.
The statistical uncertainty in this event yield ranges from 1 to 100\%, depending on the region, and is propagated to the final uncertainty in the background estimate.
The transfer factor $R^{0\ell/1\ell}_{\mathrm{MC}}$ depends on the MC modeling of the lepton veto and \MT selection efficiencies.
Leptonic \Z boson decays are used to evaluate the MC modeling of lepton selection efficiencies, and the resulting uncertainty propagated to the background estimate is found to be as large as 7\%.
The \MT selection efficiency is cross-checked using a similar dilepton sample and removing one of the leptons to mimic events where the \W boson decays to a lepton, and an uncertainty of 3\% is assigned by comparing data to simulation.
The uncertainty in the MC modeling of the lepton acceptance, assessed by varying the renormalization and factorization scales and PDF sets,
is found to be as large as 5\%.
Finally, the uncertainty in the b tagging efficiency and the jet energy scale
is typically less than 10\%, although it can be as large as 40\% in some bins.

The effect of signal contributions to the lost-lepton control samples can be non negligible in some parts of
signal parameter space, and is taken into account in the interpretations presented in Section~\ref{sec:results}.
Such a contribution would cause an overestimate of the lost-lepton background in the signal regions.
In order to account for this effect, which is typically small but can become as large as 20\% in
some compressed scenarios, the predicted signal yield in each signal region is
corrected by the amount by which the background would be overestimated.
\subsection{Estimation of the background from \texorpdfstring{$Z(\nu\overline{\nu})+$}{Z->lnu}jets}
\label{sec:bkgds:zinv}
The \znunu background is estimated using a $\gamma+$jets control sample selected using a single-photon trigger.
We select events where the photon has $\pt > 180\GeV$, to mimic the implicit requirement on the \pt
of the \Z boson arising from the baseline selection $\mttwo > 200\GeV$, and $\abs{\eta} < 2.5$.
The full baseline selection requirements are made based on kinematic variables re-calculated after removing the photon from the event, to replicate the \znunu kinematics.

\begin{figure}[htb]
  \centering
    \includegraphics[width=0.45\textwidth]{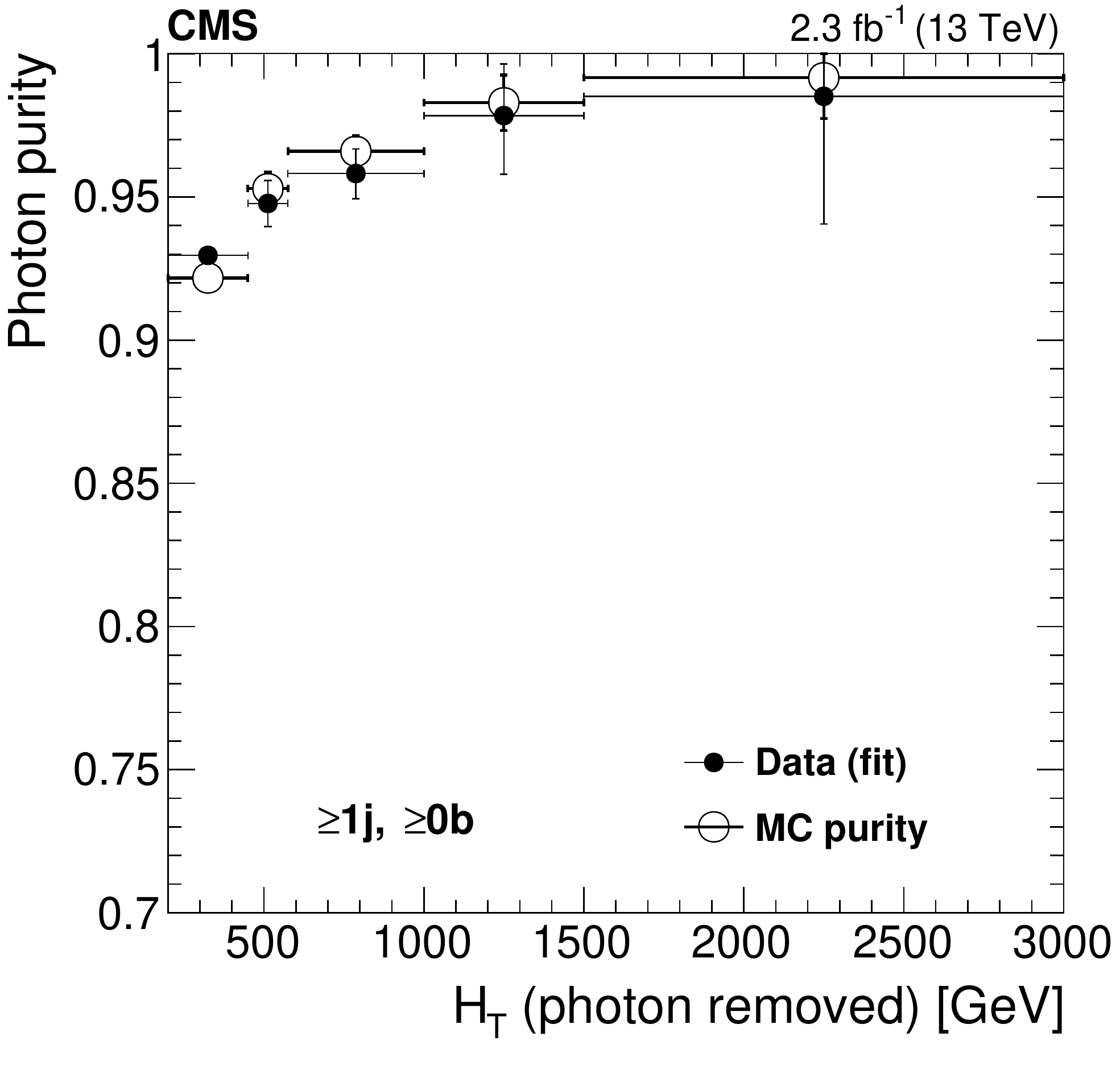}
    \includegraphics[width=0.45\textwidth]{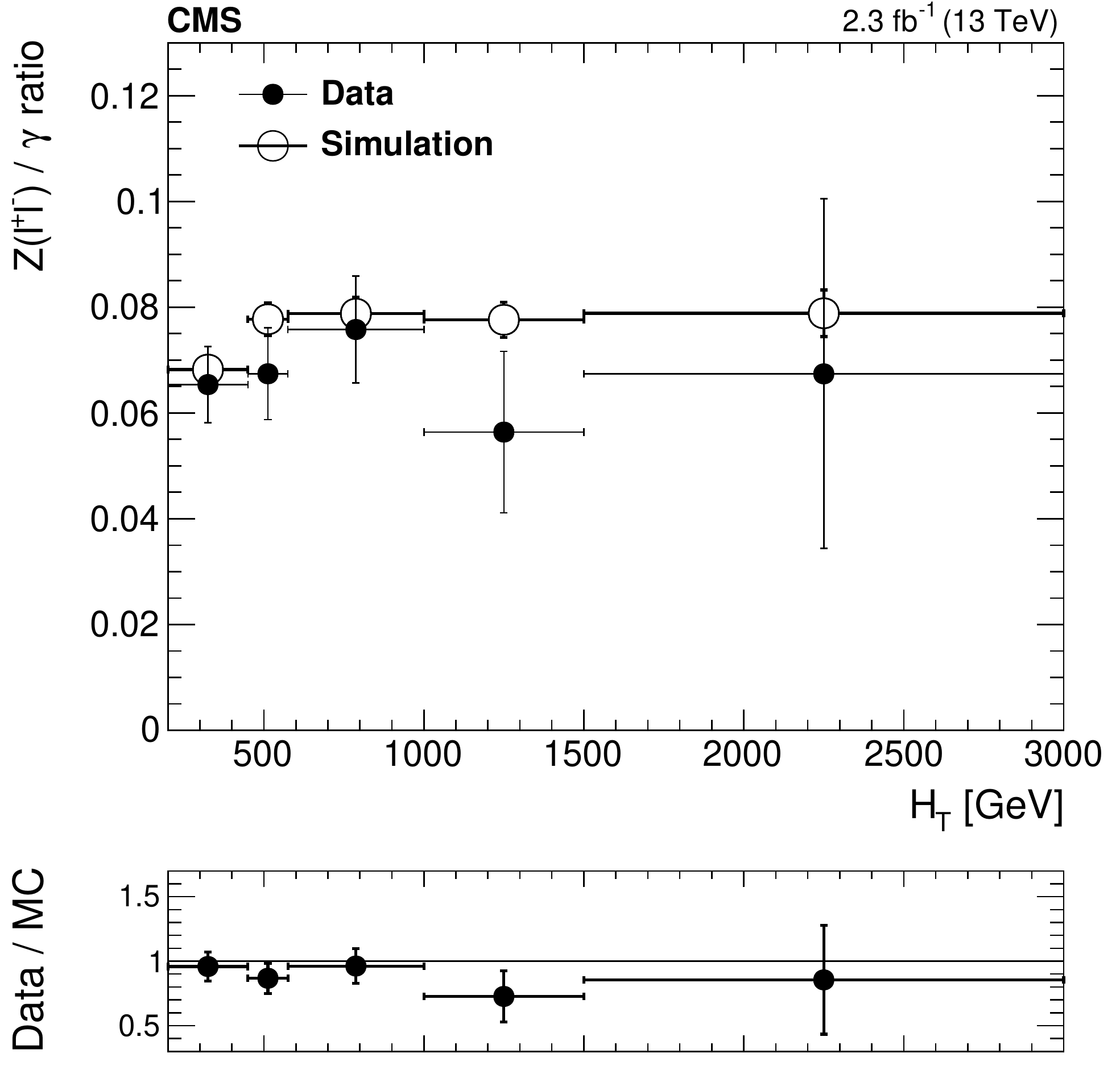}\\
    \caption{The left plot shows the photon purity, $P_{\gamma}$, measured in data for the single-photon
      control sample compared with the values extracted from simulation. The right plots
    show the $\Z/\gamma$ ratio in simulation and data as a function of \HT (upper plot),
    and the corresponding double ratio (lower plot).}
    \label{fig:zinv}
\end{figure}

Adopting a similar strategy as that used for the estimation of the lost-lepton background, selected events are
then grouped into the categories described in Section~\ref{sec:evtsel:sr}, binning the photon
control regions in the \HT, \njets, and \nbtags dimensions, but not in \mttwo, to preserve statistical precision.  For events with $\njets =1$, one control region is defined for each bin of jet \pt.
The background estimate $N^{\mathrm{SR}}_{\Z\to\PGn\PAGn}$ in each signal bin is obtained from the
events yield $N^{\mathrm{CR}}_{\gamma}$ in the control region by the application of transfer
factors according to Eq.~(\ref{eq:zinv}):

\begin{equation}
\label{eq:zinv}
\begin{split}
  N^{\mathrm{SR}}_{\Z\to\PGn\PAGn} & \left(\HT,\njets,\nbtags,\mttwo\right) = \\
 &N^{\mathrm{CR}}_{\gamma} \left(\HT,\njets,\nbtags\right) \, P_{\gamma}\left(\HT,\njets,\nbtags\right) \, f \, R^{\Z/\gamma}_{\mathrm{MC}} \left(\HT,\njets,\nbtags\right) \, k_{\mathrm{MC}}\left(\mttwo\right).
\end{split}
\end{equation}
The prompt-photon purity, $P_{\gamma}$, which accounts for photons arising from meson decays, is measured in data by
performing a template fit of the charged-hadron isolation distribution for each \HT, \njets, and \nbtags region.
The shape of the template for prompt photons is obtained from data by measuring the charged-hadron activity in
cones well-separated from the photon and any jet.
The isolation template for background photons arising from meson decays, which happen normally within hadronic jets, is also obtained from data using photon candidates that fail the $\sigma_{\eta\eta}$ requirement.
A prompt photon purity of 90--100\%, as measured in data, is well reproduced by simulation as seen in the left plot of Fig.~\ref{fig:zinv}.  A separate determination of the prompt photon purity using a tight-to-loose ratio method~\cite{SUS-10-004} obtained from the charged-hadron isolation sideband is found to yield consistent results.

The \znunu background in each bin of \HT, \njets, and \nbtags is obtained from the corresponding photon control region yield via the factor $R^{\Z/\gamma}_{\mathrm{MC}}$,
which accounts for the photon acceptance and selection efficiency and the ratio of cross sections for the production of Z+jets and $\gamma+$jets events.

The ratio $R^{\Z/\gamma}_{\mathrm{MC}}$ is obtained from $\gamma+$jet events simulated with \MADGRAPH with an implicit requirement $\Delta R > 0.4$ between the prompt photon and the nearest parton.
As no such requirement can be made in data, a correction factor $f = 0.92$ is applied to account for the fraction of selected photons passing the $\Delta R$ requirement.
This factor is determined from studies with samples of \MADGRAPH{}+\PYTHIA and \PYTHIA-only multijet events, the latter having no explicit requirement on the separation between the photon and the nearest parton.
The ratio $R^{\Z/\gamma}_{\mathrm{MC}}$ obtained from simulation is validated in data using \zll events.
In this validation, the baseline selection is applied to the \zll sample after removing the reconstructed leptons from the event, to replicate the kinematics of \znunu, and the top-quark background contamination is subtracted.
The upper right plot of Fig.~\ref{fig:zinv} shows the $R^{\Z/\gamma}$ ratios in simulation and in data, while
the double ratio, $R^{\zll/\gamma}_{\mathrm{data}}/R^{\zll/\gamma}_{\mathrm{MC}}$, is shown in the
lower right plot. The values are shown in bins of \HT, after corrections to account for measured differences
between data and simulation in lepton and photon selection efficiencies and in b tagging.
The double ratio shows no significant trend as a function of \HT, and a correction factor of 0.95 is applied to $R^{\Z/\gamma}_{\mathrm{MC}}$ to account for the observed deviation from unity.
Similarly, the double ratio as a function of \njets and \nbtags shows no significant trends and is found to be consistent with unity after the same correction factor is applied.

As in the case of the estimate of the single-lepton background, normalization to data control regions reduces reliance on the MC modeling to a single dimension, \mttwo.
The fraction of events in each topological region expected to populate a particular \mttwo bin, $k_{\mathrm{MC}}\left(\mttwo\right)$, is used to obtain the estimate in each search bin.
The uncertainty in this fraction in each \mttwo bin is evaluated in simulation by variations of the important experimental and theoretical quantities.
Theoretical uncertainties represent the largest contribution, and are assessed by variations of the renormalization and factorization scales and PDF sets.
Smaller contributions from reconstruction uncertainties are determined by varying the b-tagging efficiency
and the mistag rate, and by evaluating the impact of variations in jet energy response on the counting of jets and b-tagged jets, \ETmiss, and \mttwo.  Experimental and theoretical uncertainties in $k_{\mathrm{MC}}\left(\mttwo\right)$ total as much as 30\% at large values of \mttwo. Based on these results,  we assign an uncertainty for $k_{\mathrm{MC}}\left(\mttwo\right)$ that reaches 40\% in the highest bins of \mttwo.
\begin{figure}[htb]
  \centering
    \includegraphics[width=0.45\textwidth]{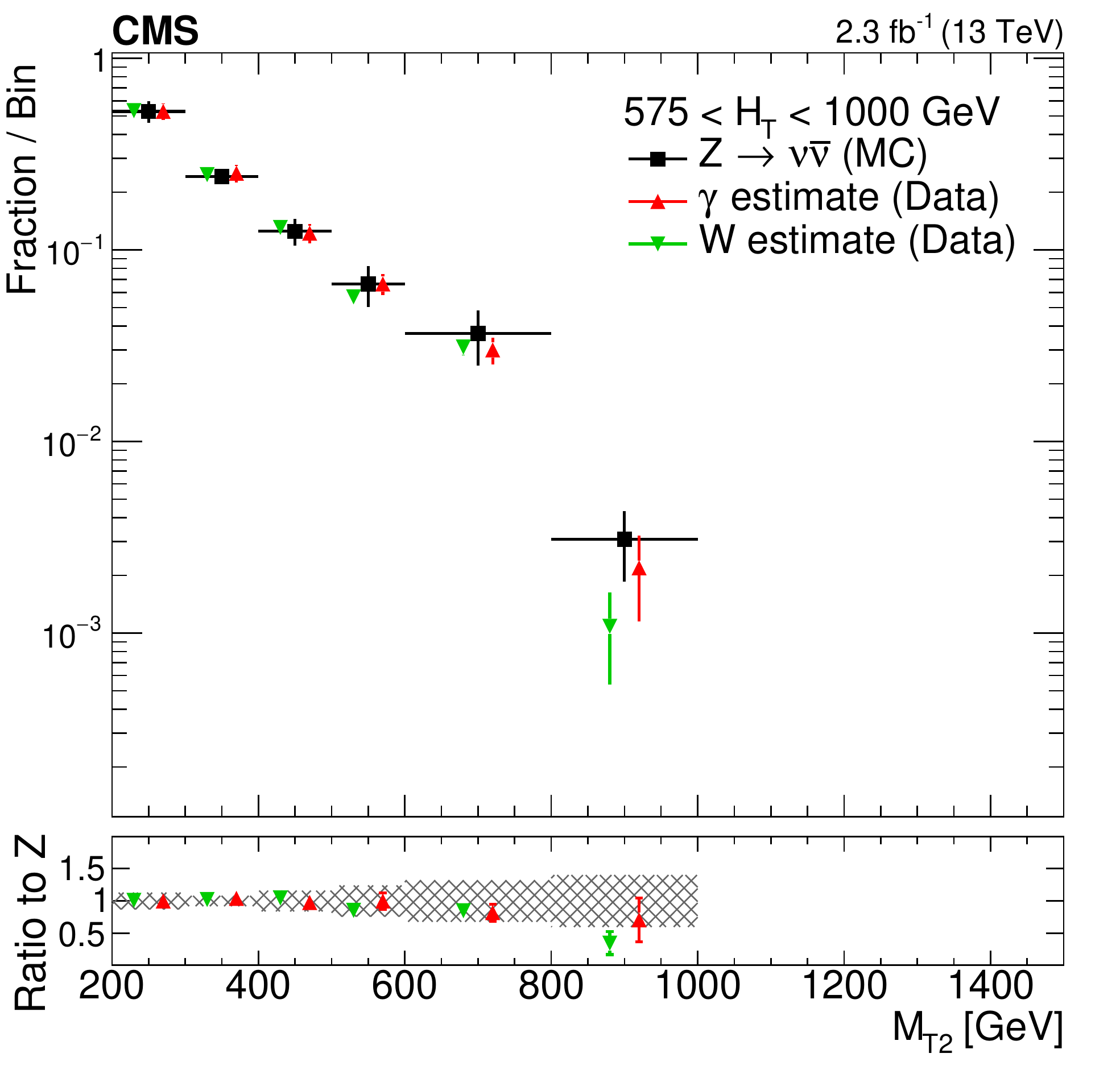}
    \includegraphics[width=0.45\textwidth]{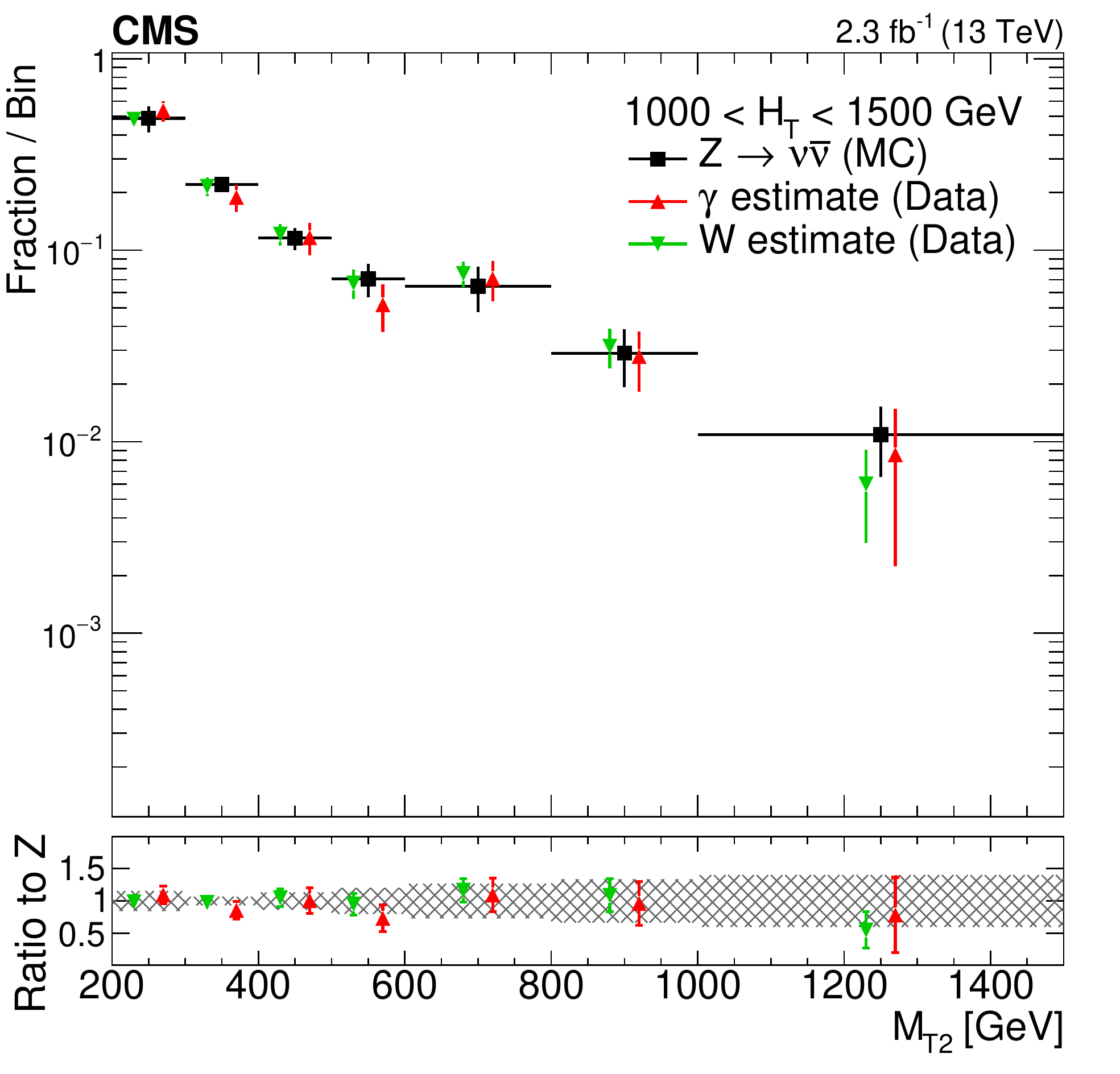}\\
    \caption{The shape of the \mttwo distribution from \znunu simulation compared to shapes extracted from $\gamma$ and $W$
      data control samples in the medium- (left plot) and high-\HT regions (right plot).
      The \mttwo distributions in the data control samples are obtained after removing the reconstructed $\gamma$ or lepton from the event, to replicate the kinematics of \znunu.
      The ratio of the shapes derived from data to the \znunu simulation shape is shown in the lower plots, where the shaded band represents the uncertainty in the MC modeling of the \mttwo variable. Data points are shifted horizontally by $\pm$20 GeV to make the vertical error bars more visible.}
    \label{fig:zinv_mt2}
\end{figure}

The MC modeling of the \mttwo variable is checked in data using highly populated control samples of $\gamma+$jets and $\PW\to\ell\nu$ events.
Figure~\ref{fig:zinv_mt2} shows good agreement between the \mttwo distribution obtained from these samples with that from \znunu simulation in the medium- and high-\HT regions.
In this comparison, the $\gamma+$jets sample is corrected based on $P_{\gamma}$, $f$, and $R^{\Z/\gamma}_{\mathrm{MC}}$, while the \W boson sample is corrected for top quark background contamination and rescaled by a $R^{\Z/\W}_{\mathrm{MC}}$ factor analogous to $R^{\Z/\gamma}_{\mathrm{MC}}$.
Similarly to what is done for the lost-lepton background, an additional check is performed by comparing the standard estimate with that obtained by replacing the factor $k_{\mathrm{MC}}\left(\mttwo\right)$ in Eq.~(\ref{eq:zinv}) with an extra dimension in the binning of the control region, which becomes $N^{\mathrm{CR}}_{\gamma} \left(\HT,\njets,\nbtags,\mttwo\right)$. These two estimates agree within the statistical precision permitted by the size of the control regions.

The single-photon control regions typically have 2--3 times as many events as compared to the corresponding signal regions.
The statistical uncertainty in this yield ranges from 1 to 100\%, depending on the region, and is propagated in the final estimate.
The dominant uncertainty in the MC modeling of $R^{\Z/\gamma}_{\mathrm{MC}}$  comes from the validation of the ratio using \zll events.
One-dimensional projections of the double ratio are constructed---separately in bins of number of jets, number of b-tagged jets, and \HT (Fig.~\ref{fig:zinv}, right)---and an uncertainty in $R^{\Z/\gamma}_{\mathrm{MC}}$ in each bin of \njets, \nbtags, and \HT is determined by adding in quadrature the uncertainty in the ratio $R^{\Z\to\ell\ell/\gamma}$ from the corresponding bins of the one-dimensional projections.
As sufficient data are not available to evaluate the double ratio for regions with $\nbtags \ge 3$, and as no trends are visible in the \nbtags distribution for $\nbtags < 3$, we assign twice the uncertainty obtained in the nearest bin, \ie $\nbtags = 2$.
This uncertainty ranges from 10 to 100\%, depending on the search region.
An additional 11\% uncertainty in the transfer factor, based on the observed offset of the double ratio from unity, is added in quadrature with the above.

The uncertainty in the measurement of the prompt photon purity includes a statistical contribution from yields
in the isolation sideband that is typically 5--10\%, but can reach as much as 100\% for search regions requiring
extreme values of \HT or large \njets.
An additional 5\% uncertainty is derived from variations in purity caused by modifications
of the signal and background templates, and from a ``closure test'' of the method in simulation.
We indicate with closure test a measurement of the ability of the method to predict correctly the true number of
background events when applied to simulated samples.
Finally, an uncertainty of 8\% is assigned to cover differences in the correction
fraction $f$ observed between \MADGRAPH{}+\PYTHIA and \PYTHIA-only simulations.
\subsection{Estimation of the multijet background}
\label{sec:bkgds:qcd}
The multijet background consists predominantly of light-flavor and gluon multijet events.
Though this background is expected to be small after requiring $\mttwo> 200\GeV$, we estimate any residual contribution based on data control samples.
For events with at least two jets, a multijet-enriched control region is obtained in each \HT bin by inverting the \dpmin requirement described in Section~\ref{sec:evtsel}.
For the high- and extreme-\HT bins, control region events are selected using the same trigger as for signal events.
For lower-\HT regions, the online \ETmiss requirement precludes the use of the signal trigger, and the
control sample is instead selected using prescaled \HT triggers with lower thresholds. Prescaled triggers accept
only a fixed fraction of the events that satisfy their selection criteria.
\begin{figure}[htb]
  \centering
    \includegraphics[width=0.45\textwidth]{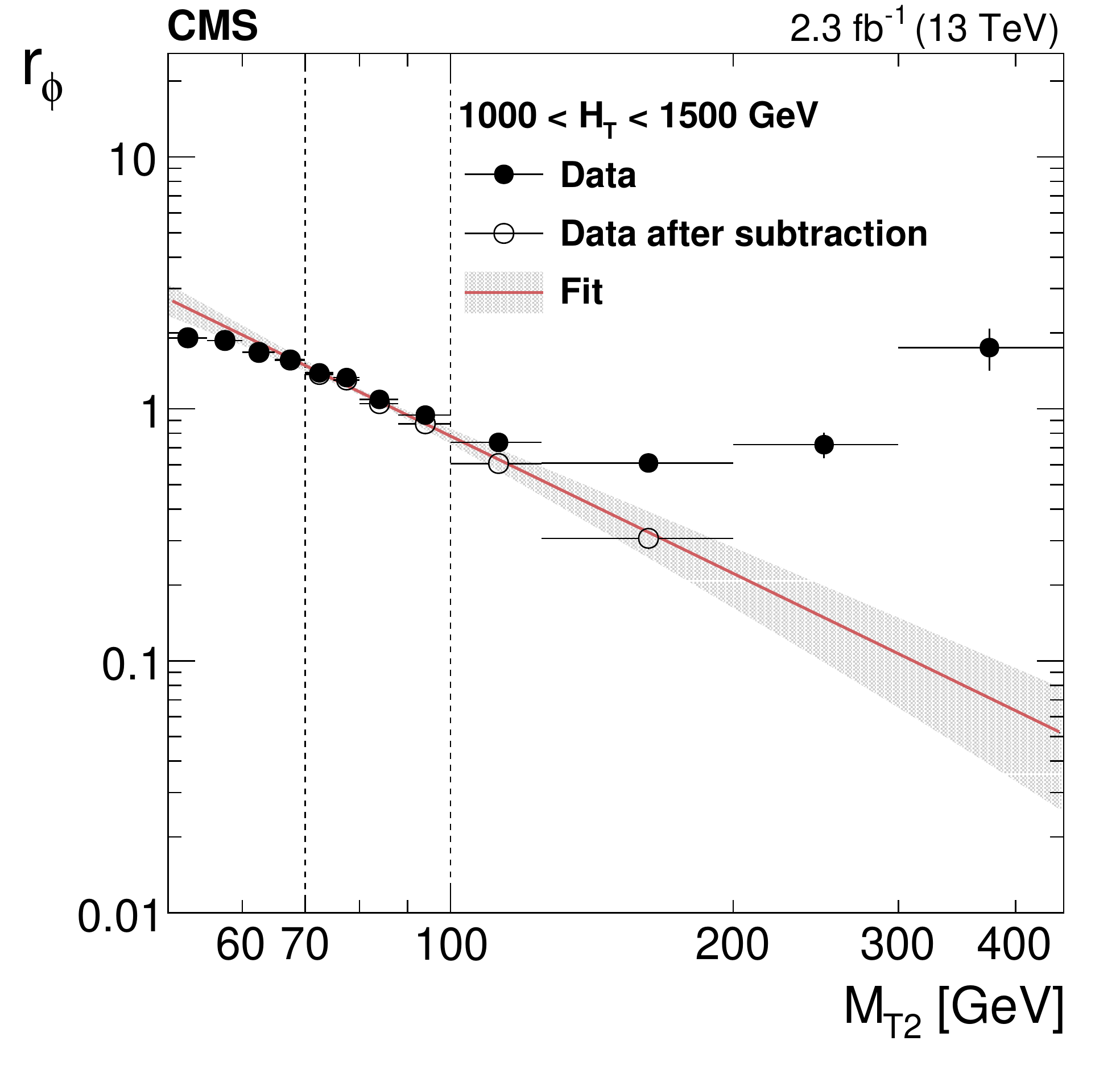}
    \caption{Distribution of the ratio $r_{\phi}$ as a function of \mttwo for the high-\HT region.
The fit is performed on the background-subtracted data points (open markers) in the
interval $70 < \mttwo< 100\GeV$ delimited by the two vertical dashed lines.
The solid points represent the data
before subtracting non-multijet backgrounds using simulation.  Data point uncertainties are statistical only.
The line and the band around it show the fit to a power-law function and the associated uncertainty.}
    \label{fig:qcd_rphi}
\end{figure}

\begin{figure}[htb]
  \centering
    \includegraphics[width=0.45\textwidth]{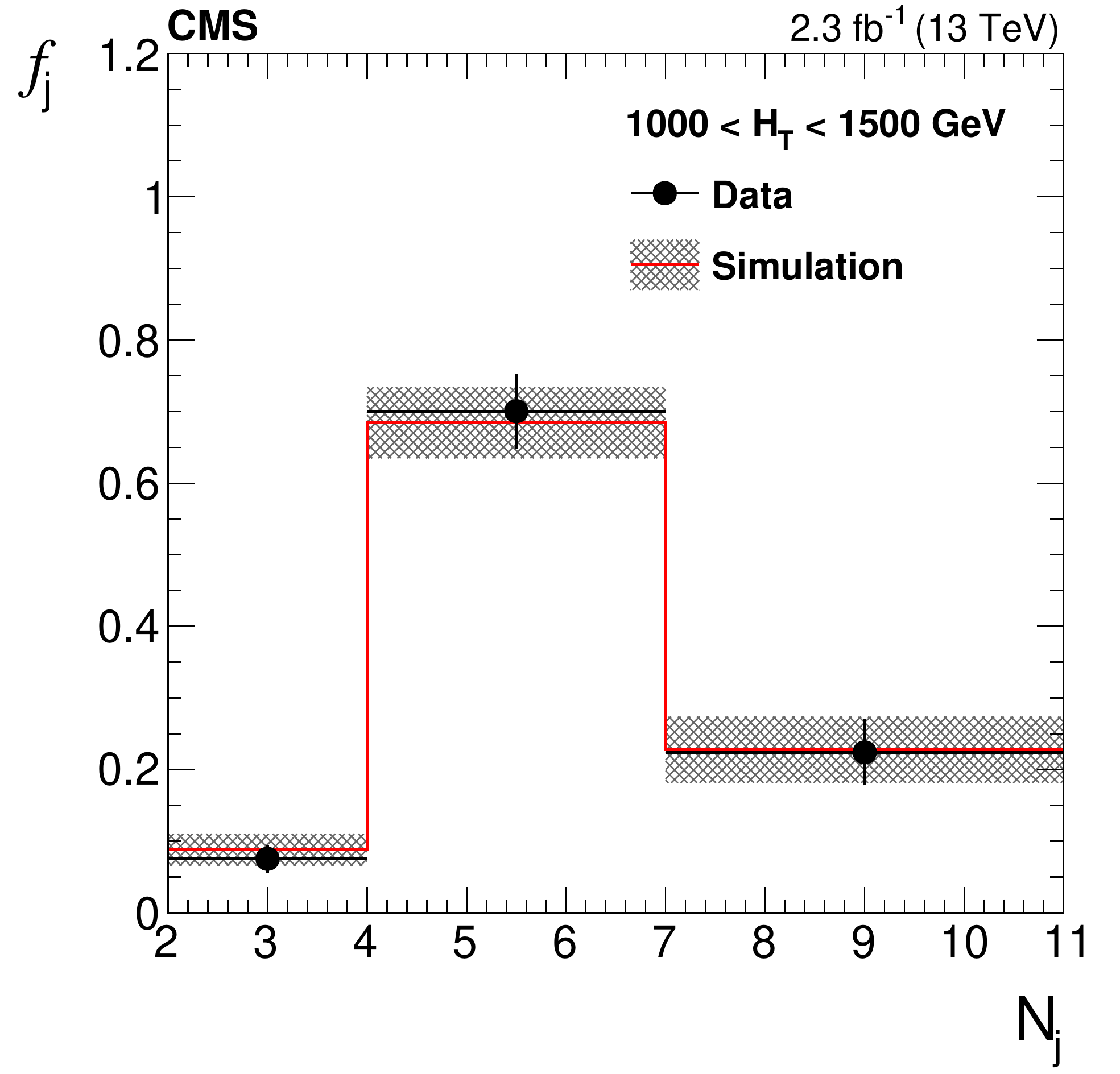}
    \includegraphics[width=0.45\textwidth]{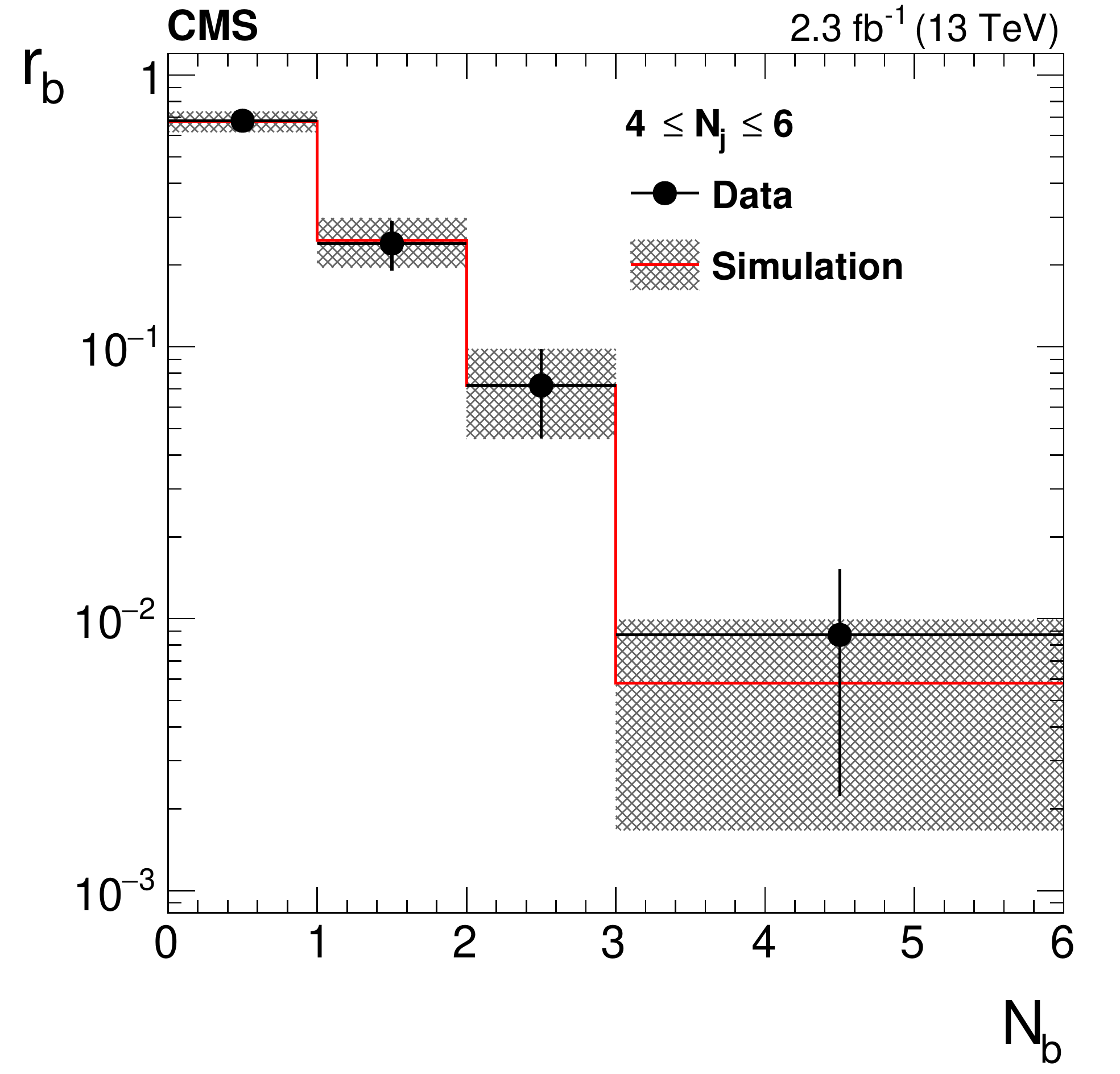}\\
    \caption{Fraction $f_{\mathrm{j}}$ of multijet events falling in bins of number of jets \njets (left) and fraction
      $r_{\PQb}$ of events falling in bins of number of b-tagged jets \nbtags (right). Values of $f_{\mathrm{j}}$ and $r_{\PQb}$
      are measured in data, after requiring $\dpmin< 0.3$ and $100<\mttwo<200\GeV$.
      The bands represent both statistical and systematic uncertainties of the estimate from simulation.}
    \label{fig:qcd_f_r}
\end{figure}

The extrapolation from low- to high-\dpmin is based on the following ratio:

\begin{equation}
\label{eq:qcd_ratio}
r_\phi(\mttwo) = N(\Delta\phi_{\min} > 0.3) / N(\Delta\phi_{\min} < 0.3).
\end{equation}
Studies in simulation show the ratio to be well described by a power
law function, $a \, (\mttwo)^{b}$.
The parameters $a, b$ are determined in each \HT bin by fitting the ratio $r_\phi(\mttwo)$ in a sideband in data, i.e. $60< \mttwo < 100\GeV$, after subtracting non-multijet contributions using simulation.
For the high- and extreme-\HT regions, the fit is performed in a
slightly narrower \mttwo window, with the lower edge increased to
70\GeV.  Data with lower values of \mttwo are not used, since in
these events the \Met no longer arises predominantly from underestimated
jet energies, but also receives important contributions from the measurement of
energy not clustered into jets. The high-\mttwo boundary of the fitting region is
chosen to minimize the effect of the non-multijet contributions
mentioned above.
An example in the high-\HT region is shown in Fig.~\ref{fig:qcd_rphi}. The inclusive multijet contribution in each \HT region, $N^{\mathrm{SR}}_{\mathrm{inc}}\left(\mttwo\right)$, is estimated using the fitted $r_\phi(\mttwo)$ and the number of events in the low-\dpmin control region, $N^{\mathrm{CR}}_{\mathrm{inc}}\left(\HT\right)$:

\begin{equation}
\label{eq:qcd_inc}
  N^{\mathrm{SR}}_{\mathrm{inc}}(\mttwo) = N^{\mathrm{CR}}_{\text{inc}}\left(\HT\right) \, r_\phi(\mttwo).
\end{equation}
From the inclusive multijet estimate in each \HT region, the predicted background in bins of \njets and \nbtags is obtained from the following equation

\begin{equation}
\label{eq:qcd_fjrb}
N^{\mathrm{SR}}_{\mathrm{j},\PQb}\left(\mttwo\right) = N^{\mathrm{SR}}_{\text{inc}}(\mttwo) \, f_{\mathrm{j}} \left(\HT\right) \, r_{\PQb} \left(\njets\right),
\end{equation}
where $f_{\mathrm{j}}$ is the fraction of multijet events falling in bin \njets, and $r_{\PQb}$ is the fraction of all events in bin \njets\ that fall in bin \nbtags. Simulation indicates that $f_{\mathrm{j}}$ and $r_{\PQb}$ attain similar values in low- and high-\dpmin regions, and that the values are independent of \mttwo.
We take advantage of this to measure the values of $f_{\mathrm{j}}$ and $r_{\PQb}$ using events with \mttwo between 100--200\GeV\ in the low-\dpmin sideband, where $f_{\mathrm{j}}$ is measured separately in each \HT bin, while $r_{\PQb}$ is measured in bins of \njets, integrated over \HT, as $r_{\PQb}$ is found to be independent of the latter.
Values of $f_{\mathrm{j}}$ and $r_{\PQb}$ measured in data are shown in Fig.~\ref{fig:qcd_f_r} compared to simulation.

An estimate based on $r_\phi(\mttwo)$ is not viable in the monojet search region so a different strategy must be employed.
Multijet events can pass the monojet event selections through rare fluctuations
in dijet events, as when the transverse momentum of one of the two jets is severely underestimated
because of detector response or
because of particularly energetic neutrinos from b and c quark decays. In these cases, the resulting
reconstructed jet can be assigned a transverse momentum below the jet-counting threshold ($\pt < 30$\GeV). In order to
estimate this background contribution, we define a control region by selecting dijet events in which the leading jet
has a transverse momentum $\pt > 200$\GeV (as in the monojet signal region), and the second jet has a transverse
momentum just above threshold, \ie $30 < \pt <60$\GeV. These events must further pass an inverted \dpmin requirement,
in order to ensure statistical independence
from the signal region. After subtracting non-multijet contributions, the data yield in the control region
is taken as an estimate of the background in the monojet search regions.
The rate of events with $30 <\pt<60$\GeV is expected to be larger than that of
events with $\pt<30$\GeV, as the latter would require even larger detector response fluctuations.
Closure tests on the simulation indicate a small overestimate.  Nevertheless, the multijet background
is not expected to exceed 8\% in any monojet search region.

Statistical uncertainties due to the event yields in the control regions, where the $r_\phi(\mttwo)$ fit
is performed and the $f_{\mathrm{j}}$ and $r_{\PQb}$ values are measured, are propagated to the final estimate.
The invariance of $f_{\mathrm{j}}$ with \mttwo and $r_{\PQb}$ with \mttwo and \HT is evaluated in simulation, and residual differences are taken as additional systematic uncertainties, which are shown in Fig.~\ref{fig:qcd_f_r}.
An additional uncertainty is assigned to cover the sensitivity of the $r_{\phi}$ value to variations in the fit window.
These variations result in an uncertainty that increases with \mttwo and ranges from 15 to 200\%.
The total uncertainty in the estimate covers the differences observed in closure tests based on simulation
and in data control regions. The latter is performed in the $100 < \mttwo<200$\GeV sideband.
For the monojet regions, the statistical uncertainty from the data yield
in the dijet sideband
is combined with a 50\% systematic uncertainty in all bins.
\subsection{Cross-check of multijet background estimation}
\label{sec:bkgds:rs}
\begin{figure}[!ht]
  \centering
    \includegraphics[width=0.99\textwidth]{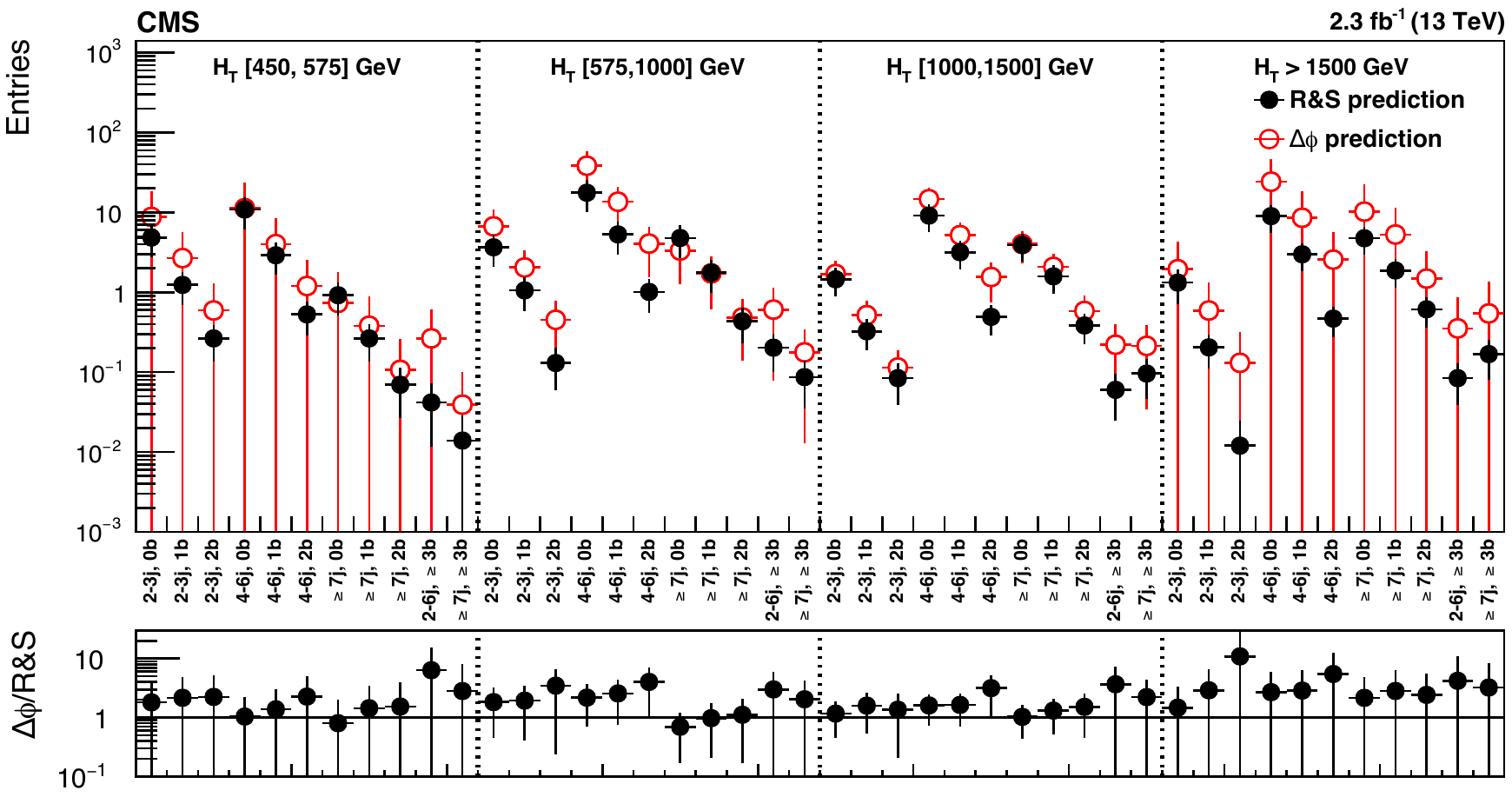}
    \caption{Comparison of the predictions of the multijet background
      in the topological regions ($\mttwo >200\GeV$) from the R\&S method
      and the $\dpmin$ ratio method. Both methods are described in Ref.~\cite{paperWithRS} and
      in the text, respectively. The uncertainties are combined statistical and systematic.
      Within each of the four \HT categories, the estimates from the \dpmin ratio method are
      correlated as they are derived from the same fit to the \dpmin ratio data.
      The lower plot shows the ratio of the estimates from the \dpmin and the R\&S methods.}
    \label{fig:rs_comparison}
\end{figure}
As a cross-check of the \dpmin ratio method described
in Section~\ref{sec:bkgds:qcd}, the multijet background is also
estimated using the ``rebalance and smear'' (R\&S) method
described in Ref.~\cite{paperWithRS}.
This method rebalances multijet events in data by adjusting the jet \pt values to minimize \ETmiss
and then smears them multiple times in order to build a large sample of multijet events with nonzero \ETmiss.
During both the rebalance and the smearing steps, the jet \pt values are varied according to a parameterization of
the jet energy response.
The performance of the method has been tested on
multijet simulation, as well as on data control regions
defined by inverting the \dpmin requirement or by
selecting a sideband of \mttwo (i.e. $100<\mttwo <200$\GeV).
Based on these studies, we assign total systematic
uncertainties of 50\% (low- and medium-\HT regions) and
40\% (high- and extreme-\HT regions) in the background
estimate based on R\&S for $\mttwo >200$\GeV.  These uncertainties also include a
small ($<$7\%) uncertainty due to contamination from \wjets and \zjets events
of the multijet data sample used in the R\&S procedure.

In Fig.~\ref{fig:rs_comparison}, we compare the multijet predictions from the
R\&S method with those from the \dpmin ratio method, \ie the estimation method used in our analysis
for multijet signal regions.
This comparison is done separately for each topological region, integrating over \mttwo bins.
The level of agreement between the two methods serves to further increase our confidence in
the multijet background estimation used for the final results of the analysis.

The R\&S method cannot be applied to the very-low-\HT region as
not enough data are available in the relevant multijet control sample
because of the small fraction of events accepted by the prescaled triggers
with very low thresholds in \HT.
\section{Results and interpretation}
\label{sec:results}
\begin{figure}[!htb]
  \centering
    \includegraphics[width=0.99\textwidth]{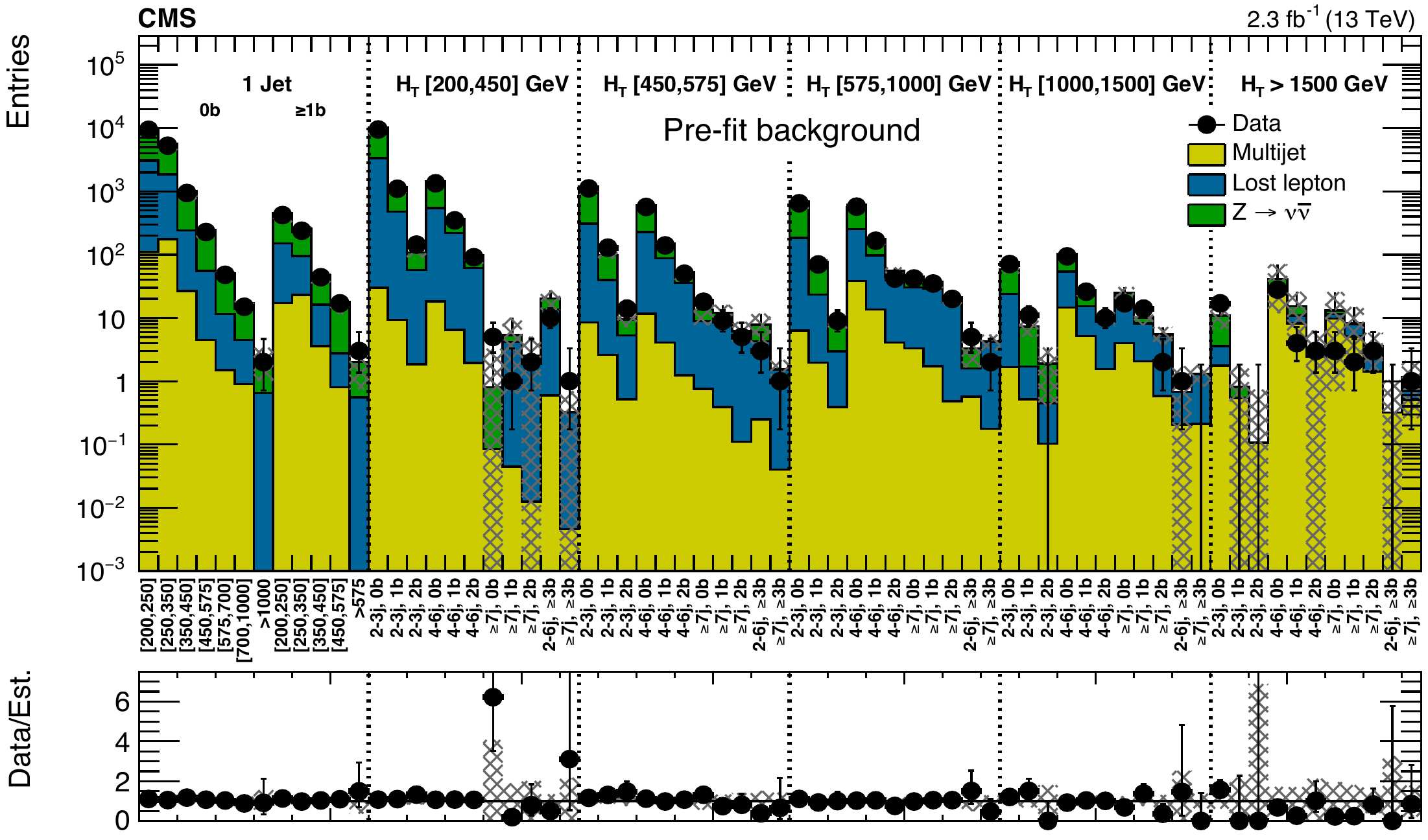}\\
    \includegraphics[width=0.99\textwidth]{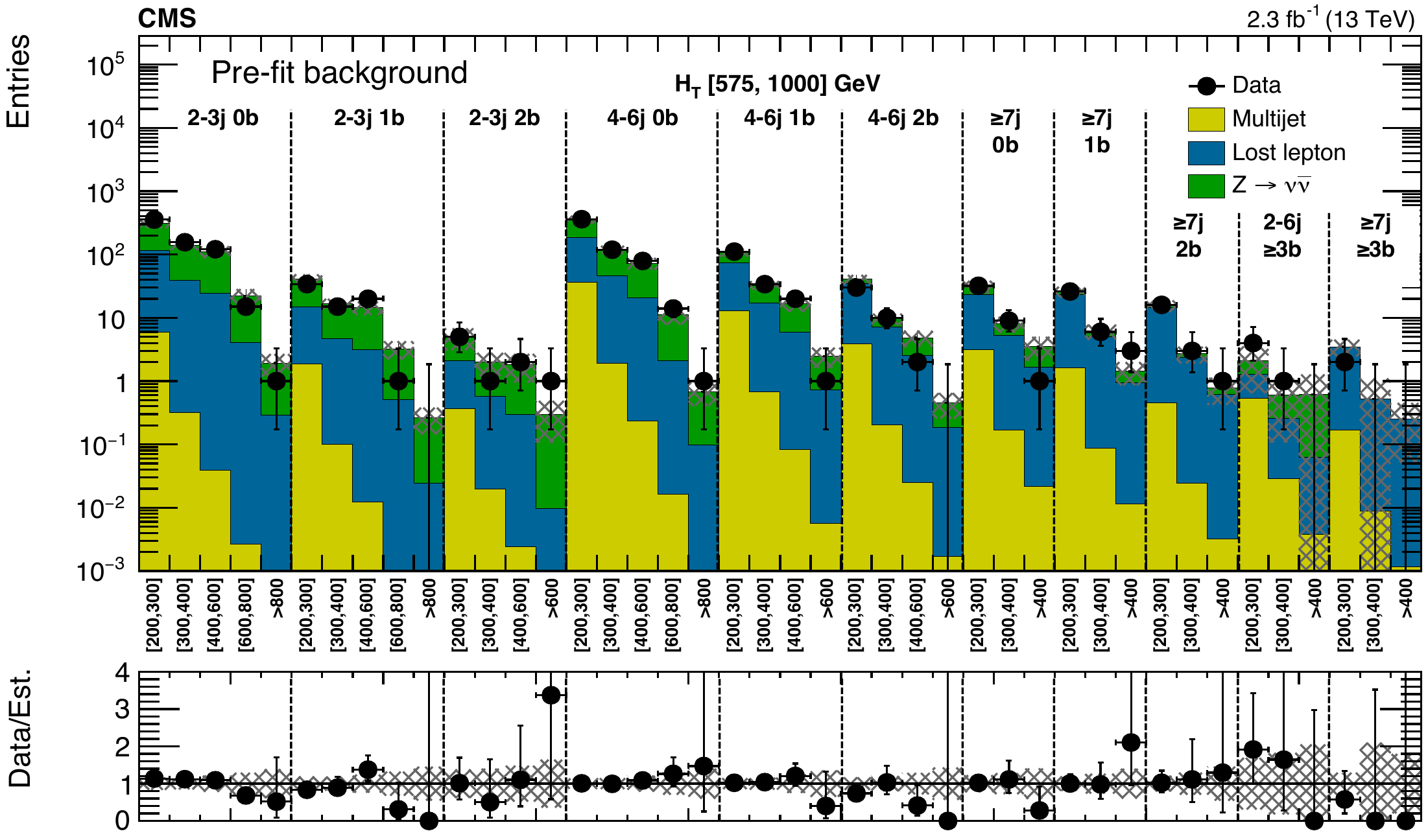}\\
    \caption{(Above) Comparison of estimated background (pre-fit) and observed data events in each topological region.
      The results shown for $\njets = 1$ correspond to the monojet search regions binned in jet \pt in \GeVns.
      For the multijet data, the notations j and b
      indicate the \njets and \nbtags multiplicity.
      Hatched bands represent the full uncertainty in the background estimate.
      (Below) Comparison for individual \mttwo signal bins in the medium \HT region.
      On the $x$-axis, the \mttwo range of each signal region is shown in \GeVns.
      Bins with no entry for data have an observed count of 0 events.}
    \label{fig:results}
\end{figure}

Figure~\ref{fig:results} shows a summary of the observed event yields in data, together with the predicted total SM background.
Each bin in the upper plot corresponds to a single (\HT, \njets, \nbtags) search region integrated over \mttwo.
The lower plot further breaks down the background estimates and observed data yields into all
\mttwo bins for the medium \HT region. The data are statistically compatible with the expected background
contributions, providing no evidence for new physics:
analyzing the 87 signal regions with a non-zero excess in the observed data, we see that three bins correspond
to a p-value~\cite{Agashe:2014kda} approximately equal to 2$\sigma$, zero have a p-value larger than 3$\sigma$,
and in general all p-values are compatible with a standard normal distribution.
The background estimates and corresponding uncertainties shown in these plots rely exclusively
on the inputs from control samples and simulation as described in Section~\ref{sec:bkgds} and are indicated
in the rest of the text as ``pre-fit background'' results.

We also estimate the backgrounds
in the signal regions performing a maximum-likelihood fit to the data in the signal regions themselves.
These fits are carried out under either the background-only or background+signal hypotheses.
The estimates from these fits, which still depend on the modeling of the backgrounds from the pre-fit procedure,
are indicated as ``post-fit'' results and are utilized to constrain models
of new physics as described below.
Similar comparisons between data and background predictions, for both pre- and post-fit estimates, are shown
for all the remaining \HT regions in Appendix~\ref{app:results}.

\begin{figure}
  \centering
    \includegraphics[width=0.3\textwidth]{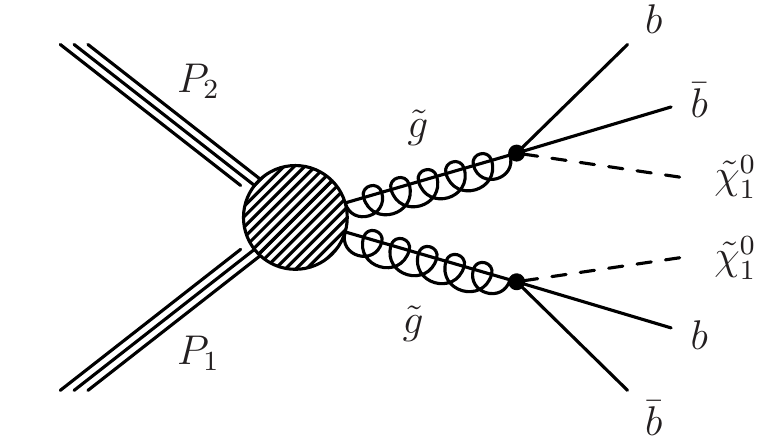}
    \includegraphics[width=0.3\textwidth]{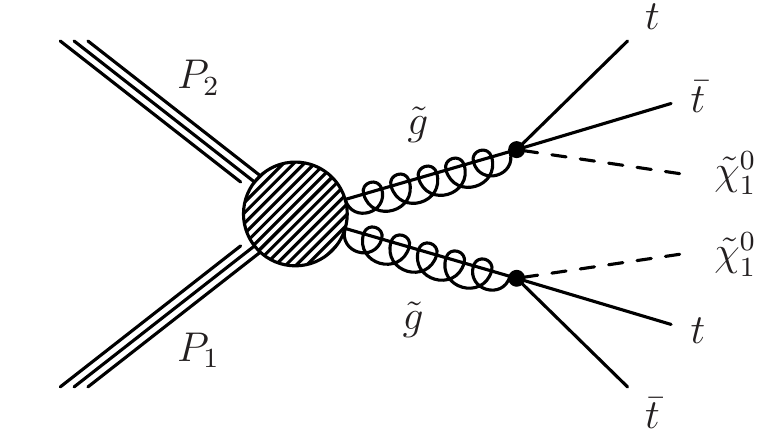}
    \includegraphics[width=0.3\textwidth]{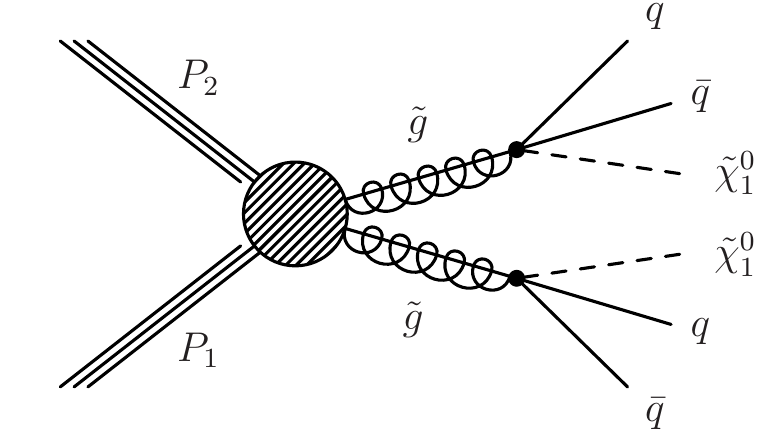}\\
    \includegraphics[width=0.3\textwidth]{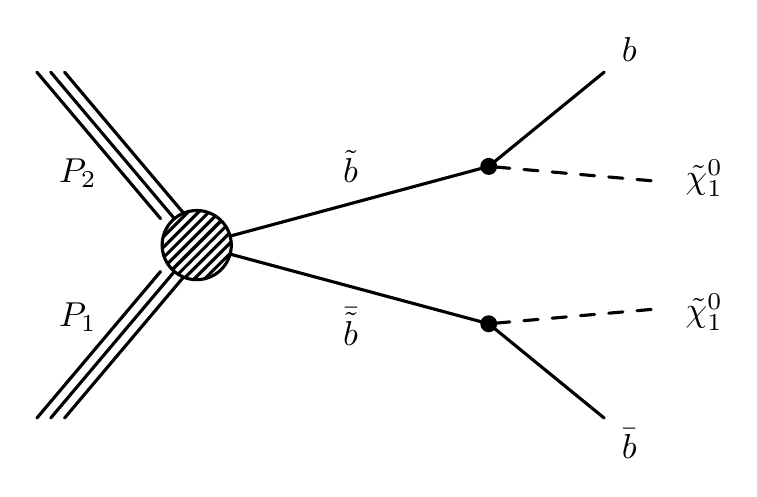}
    \includegraphics[width=0.3\textwidth]{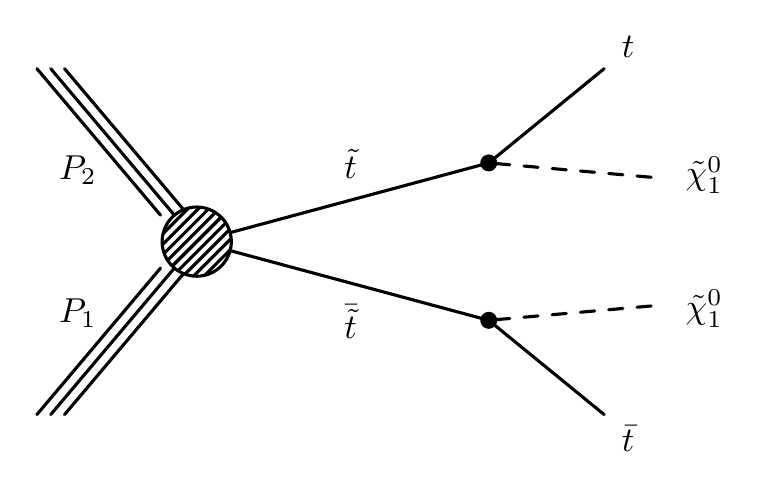}
    \includegraphics[width=0.3\textwidth]{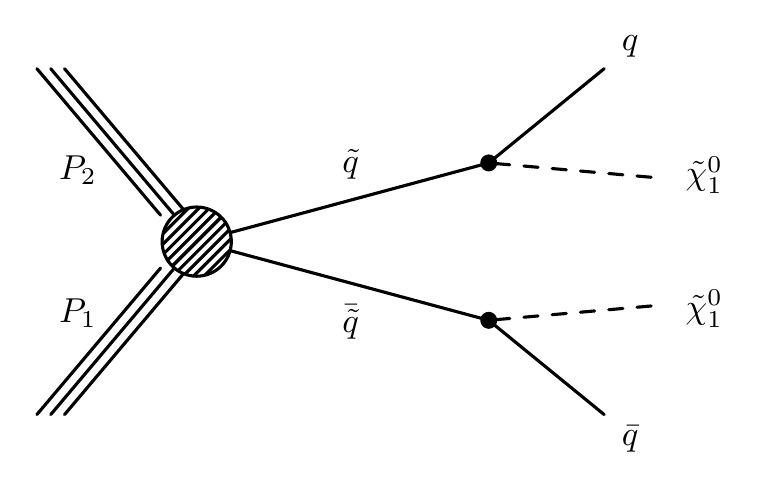}
    \caption{(Above) Diagrams for the three considered scenarios of gluino-mediated bottom squark, top squark, and light
    flavor squark production.  The depicted three-body decays are assumed to proceed through off-shell squarks.
    (Below) Diagrams for the three considered simplified models of direct pair production of bottom squarks,
    top squarks, and light flavor squarks. The top quarks in these processes are assumed to be produced unpolarized.}
    \label{fig:SMS_feynDiagrams}
\end{figure}

The results of the search are used to constrain specific models~\cite{bib-sms-1,bib-sms-2,bib-sms-3,bib-sms-4,Chatrchyan:2013sza} of new physics such as those identified
by the diagrams in Fig.~\ref{fig:SMS_feynDiagrams}.
For each scenario of gluino (squark) pair production, our simplified models assume that
all supersymmetric particles other than the gluino (squark)
and the lightest neutralino are too heavy to be produced directly, and that the gluino (squark) decays promptly.
For gluino pair production, the models assume that each gluino decays with a 100\% branching fraction into the
lightest supersymmetric particle (LSP) and either b quark pairs ($\PSg\to\bbbar\PSGczDo$),
top quark pairs ($\PSg\to\ttbar\PSGczDo$), or light-flavor quarks ($\PSg\to\qqbar\PSGczDo$),
proceeding respectively through an off-shell bottom, top, or light-flavor squark.

For the scenario of top squark pair production, the polarization of the top quark is model dependent and is a
function of the top-squark and neutralino mixing matrices. To remain agnostic to a particular model realization,
events are generated without polarization.  Also, for the region where $m_{\PSQt} - m_{\mathrm{LSP}} < m_{\PQt}$,
a uniform phase-space decay is assumed.

For a given signal scenario, limits are derived by combining all search regions using a modified
frequentist approach, employing the CL$_\mathrm{s}$ criterion and an asymptotic
formulation~\cite{Read:2002hq,Junk:1999kv,Cowan:2010js,ATLAS:2011tau}.

Typical values of the uncertainties
considered in the signal yield for one of the models are listed in Table~\ref{tab:sig_systs}.
The largest uncertainties come from the limited size of the MC samples for a small number of model points with low acceptance, and the uncertainty in the b tagging efficiency.
The uncertainty in the modeling of initial-state radiation (ISR)
can also be significant for model points with small mass splittings,
where some boost from ISR is necessary to observe the decay products of the initially produced sparticles.
The uncertainty is determined by comparing the simulated and measured \pt spectra
of the system recoiling against the ISR jets in \ttbar events,
using the technique described in Ref.~\cite{stop8tev}.
The two spectra are observed to agree below 400\GeV, and
the statistical precision of the comparison is used to define
an uncertainty of 15\% (30\%) for $400<\pt<600\GeV$ ($\pt>600\GeV$).
The uncertainty in the acceptance due to
the renormalization and factorization scales is found to be relatively small, and a constant value of 5\%
is used in the analysis.

The uncertainty due to the jet energy scale is found to be compatible with
statistical fluctuations for bins populated by few MC events, so a constant value of 5\% is taken, motivated by more populated search bins.
Uncertainties in the integrated luminosity, ISR, b tagging, and lepton efficiencies are treated as correlated across search bins.
No additional uncertainty due to variations of the PDF set is taken since the main effect on signal
acceptance is through modeling of the recoil \pt spectrum and the ISR uncertainty already accounts for this.

\begin{figure}[htb]
  \centering
    \includegraphics[width=0.49\textwidth]{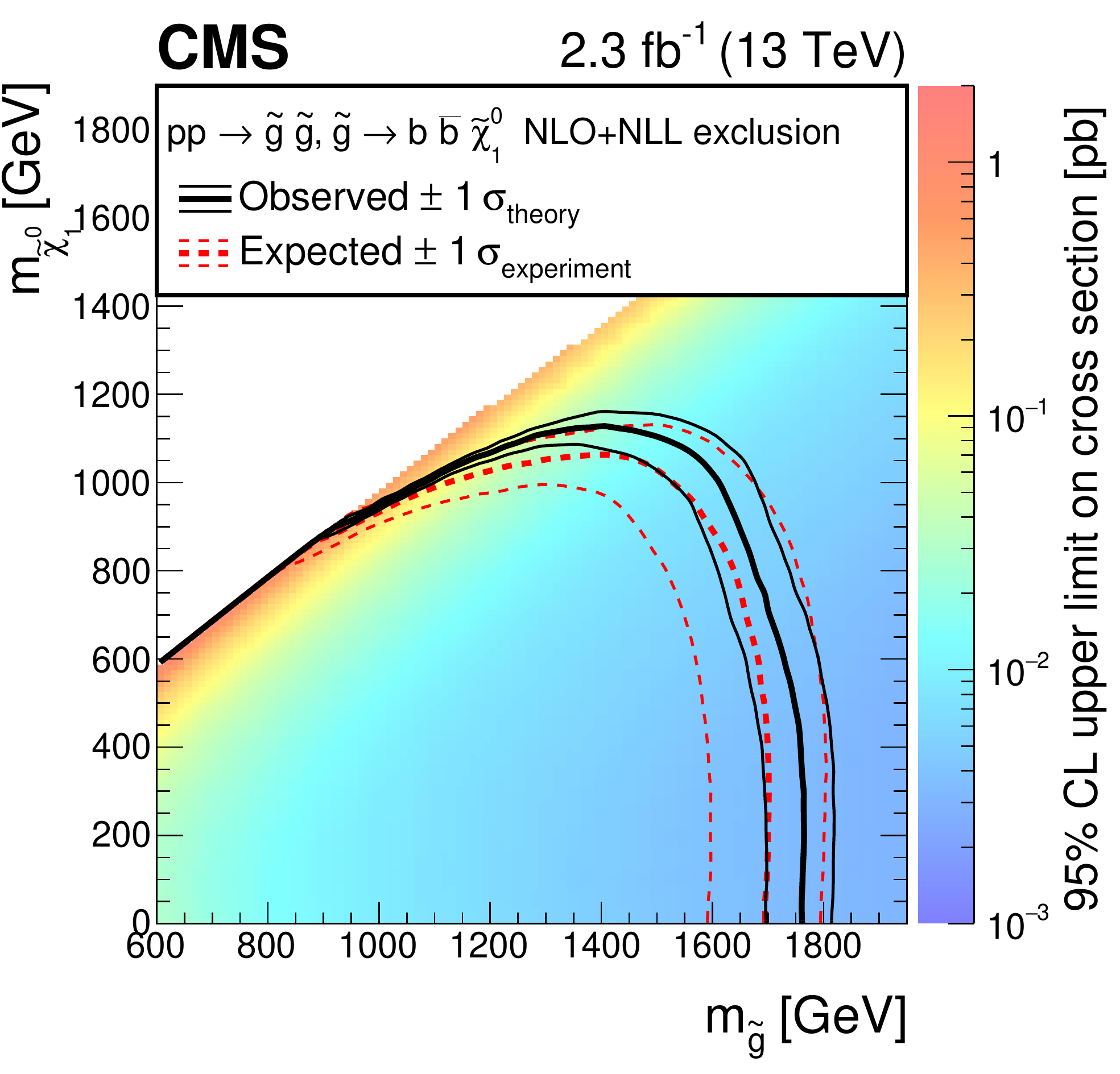}
    \includegraphics[width=0.49\textwidth]{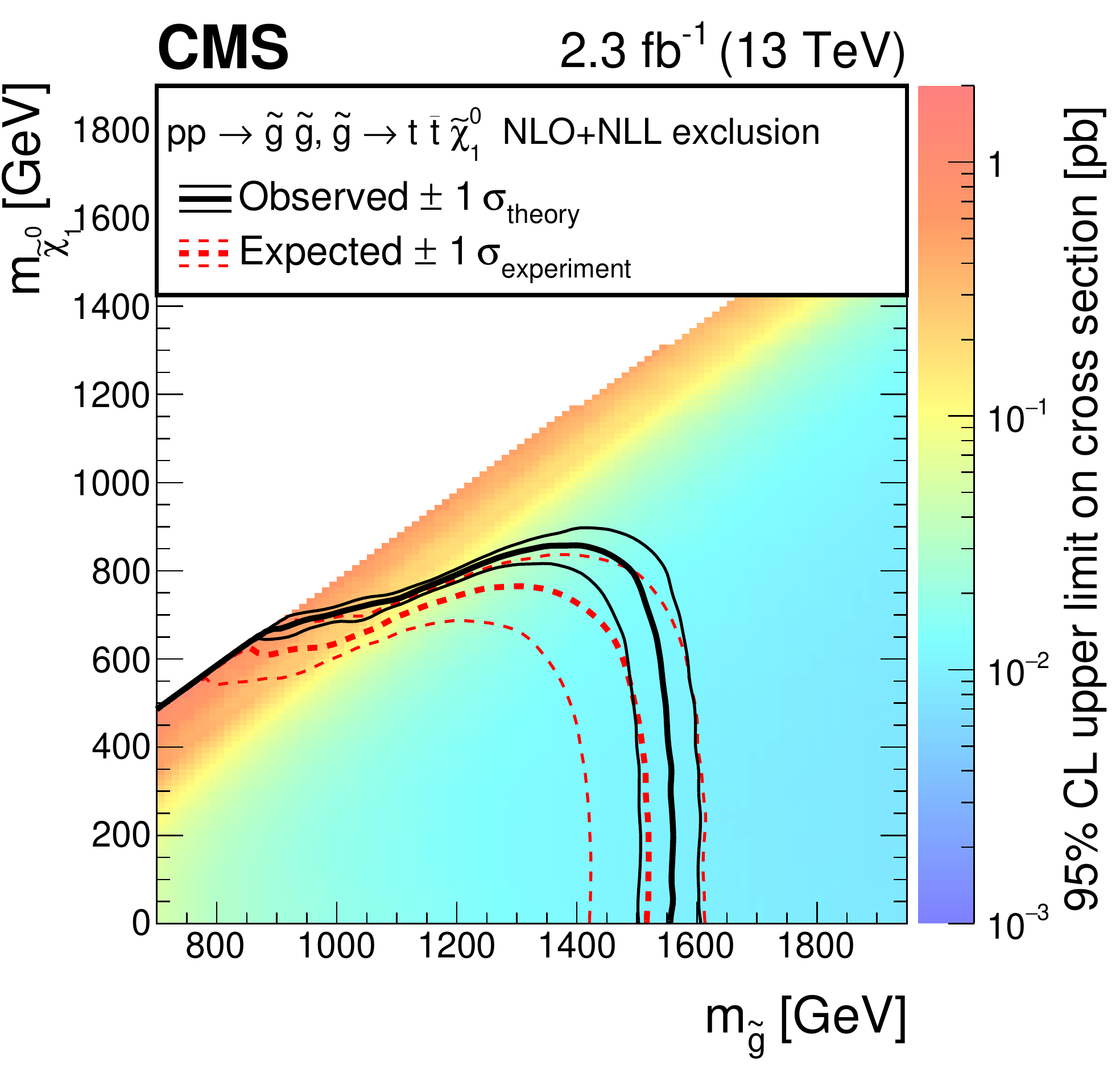}
    \includegraphics[width=0.49\textwidth]{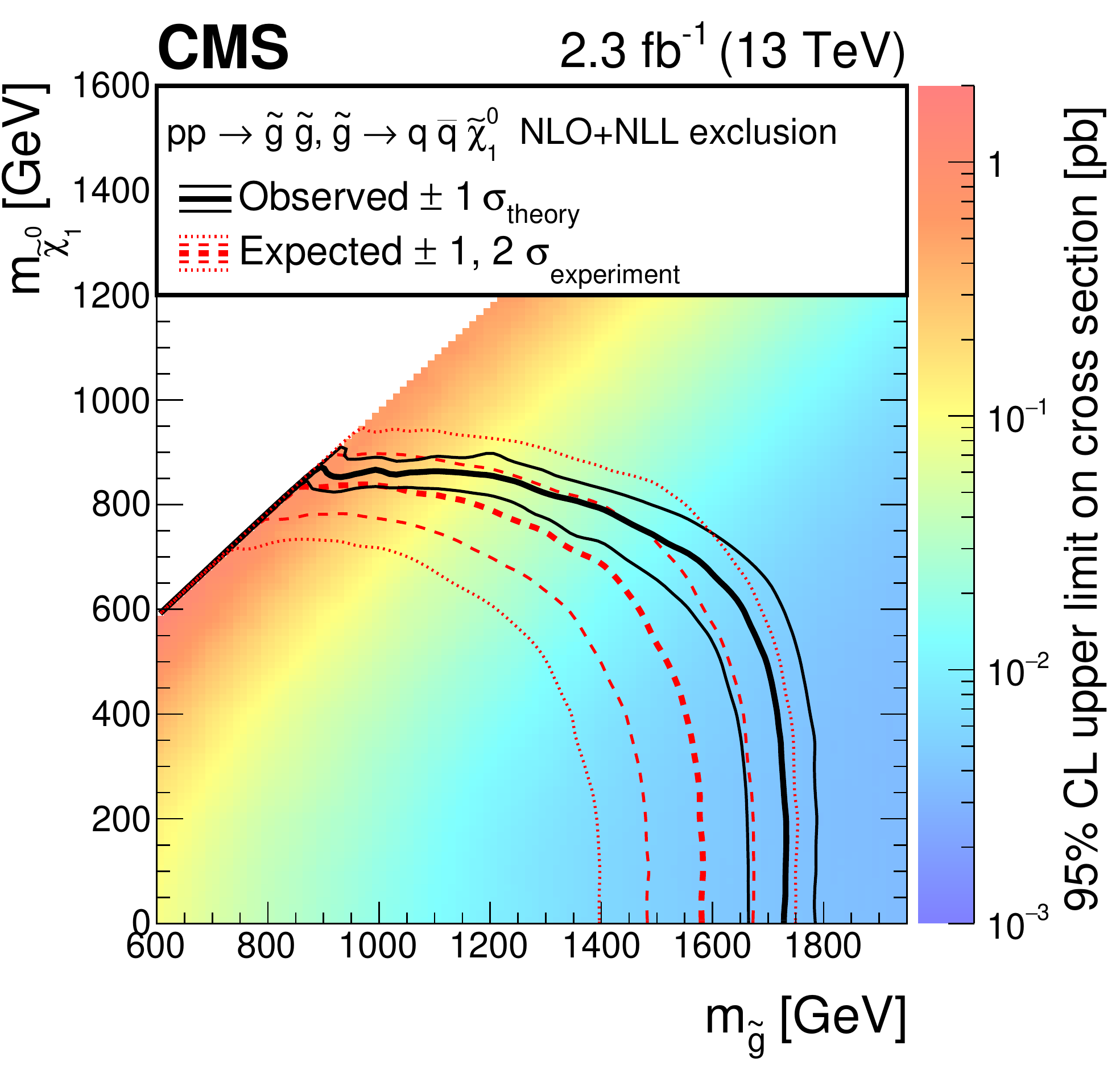}
    \caption{Exclusion limits at  95\% CL on the cross sections for gluino-mediated bottom squark production (above left),
    gluino-mediated top squark production (above right), and gluino-mediated light-flavor squark production (below).
      The area to the left of and below the thick black curve represents the observed exclusion region,
      while the dashed red lines indicate the expected limits and
      their $\pm$1 $\sigma_{\text{experiment}}$ standard deviation uncertainties.
      For the gluino-mediated light-flavor squark production plot, the $\pm$2 standard deviation
      uncertainties are also shown.
      The thin black lines show the effect of the theoretical
      uncertainties $\sigma_{\text{theory}}$ on the signal cross section.}
    \label{fig:t1x}
\end{figure}

\begin{figure}[htb]
  \centering
    \includegraphics[width=0.49\textwidth]{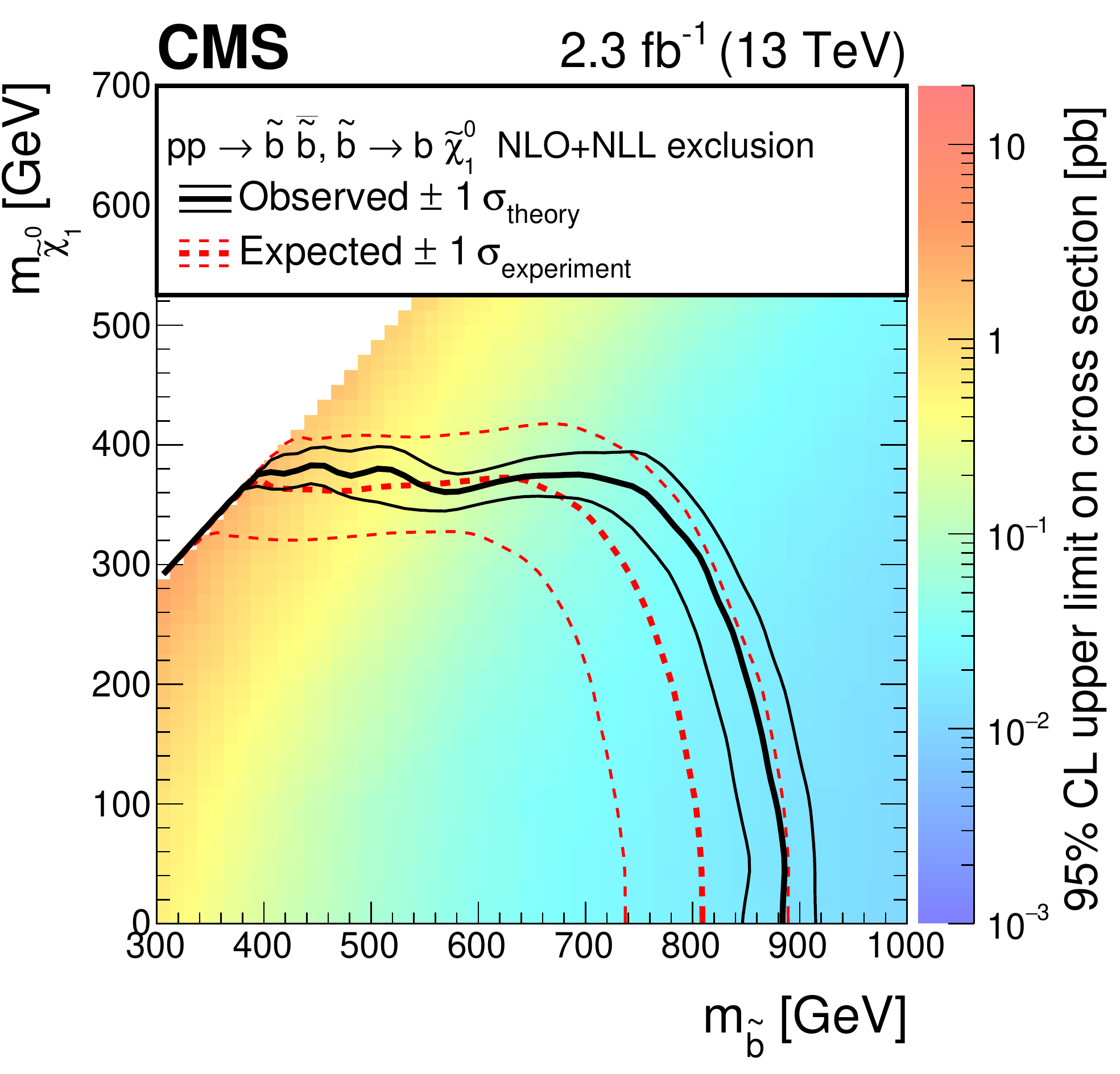}
    \includegraphics[width=0.49\textwidth]{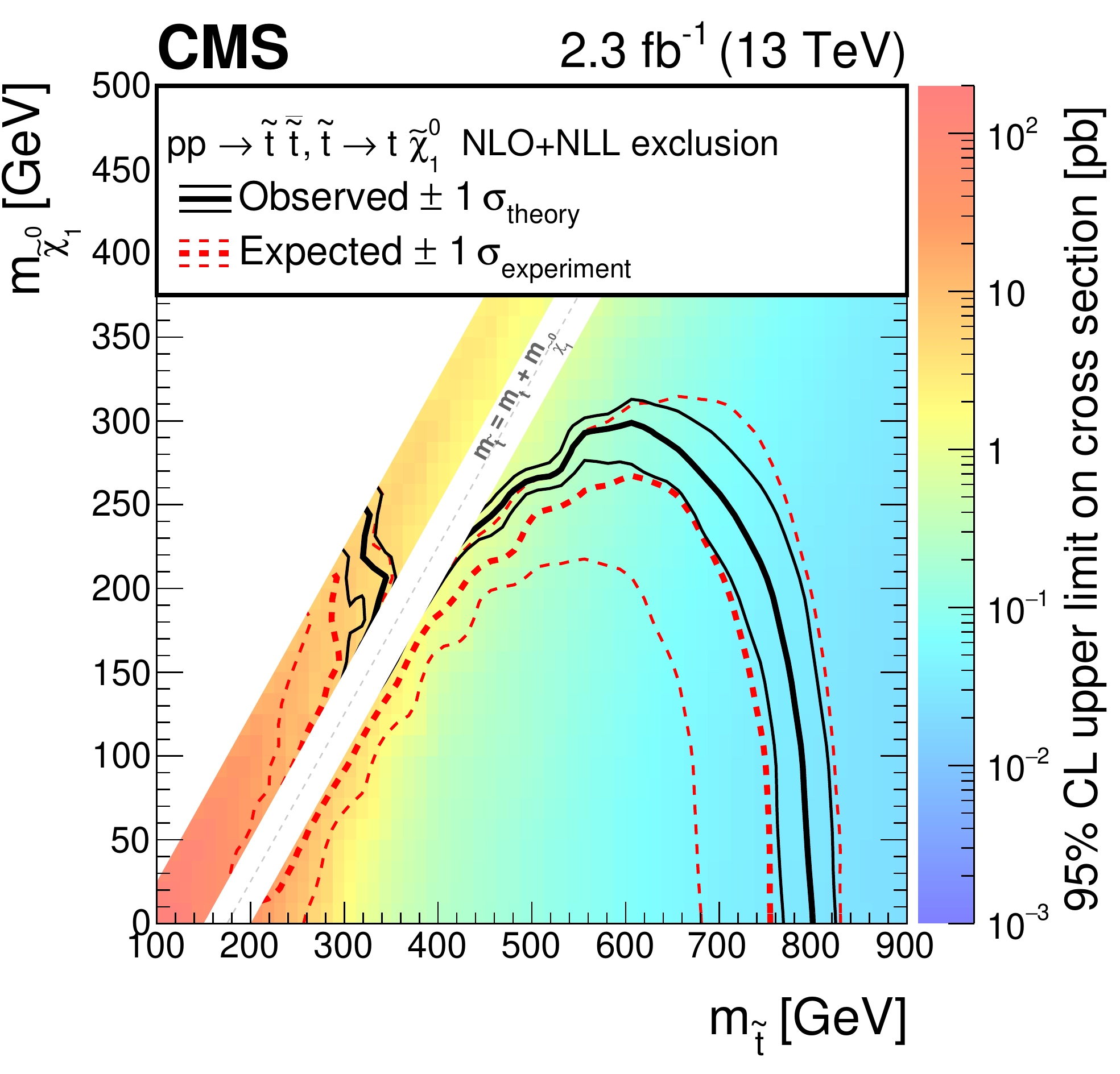}
    \includegraphics[width=0.49\textwidth]{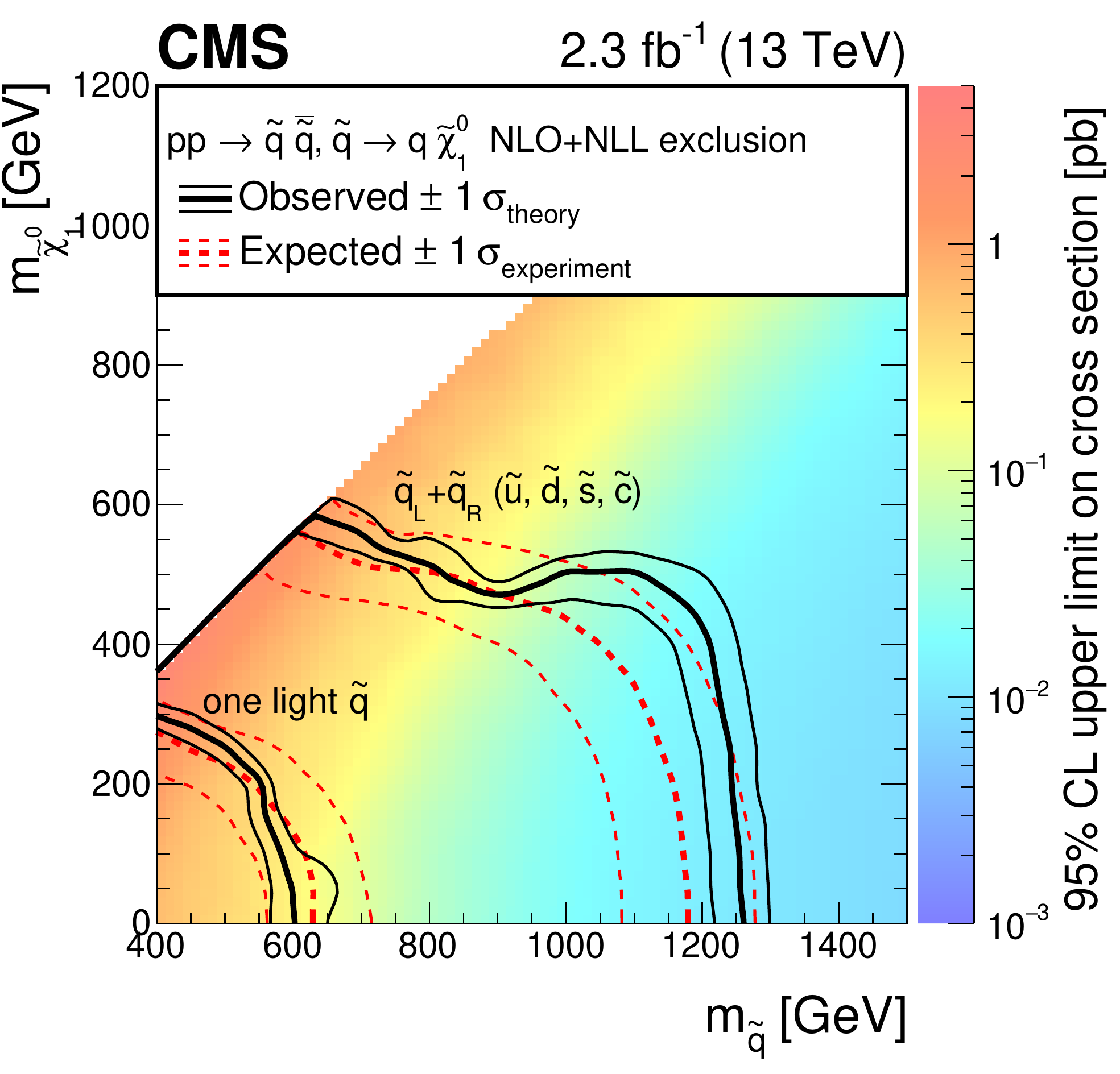}
    \caption{Exclusion limit at 95\% CL on the cross sections for bottom squark pair production (above left), top squark pair production (above right),
      and light-flavor squark pair production (below).
      The area to the left of and below the thick black curve represents the observed exclusion region,
      while the dashed red lines indicate the expected limits and
      their $\pm$1 $\sigma_{\text{experiment}}$ standard deviation uncertainties.
      The thin black lines show the effect of the theoretical
      uncertainties $\sigma_{\text{theory}}$ on the signal cross section.
      The white diagonal band in the upper right plot corresponds to the region
      $\abs{m_{\PSQt}-m_{\PQt}-m_{\mathrm{LSP}}}< 25$\GeV. Here the efficiency of the selection
      is a strong function of $m_{\PSQt}-m_{\mathrm{LSP}}$, and as a result the precise
      determination of the cross section upper limit is uncertain
      because of the finite granularity of the available
      MC samples in this region of the ($m_{\PSQt}$, $m_{\mathrm{LSP}}$)  plane.}
    \label{fig:t2x}
\end{figure}

Figure~\ref{fig:t1x} shows exclusion limits at 95\% confidence level (CL) for gluino-mediated bottom squark,
top squark, and light-flavor squark production.
Exclusion limits for the pair production of bottom, top and light-flavor squarks are shown in Fig.~\ref{fig:t2x}.
In the upper right plot of this figure, the white diagonal band corresponds to the region
$\abs{m_{\PSQt}-m_{\PQt}-m_{\mathrm{LSP}}}<25$\GeV, where the selection efficiency of top squark events
is a strong function of $m_{\PSQt}-m_{\mathrm{LSP}}$. As a result, the precise
determination of the cross section upper limit is uncertain
because of the finite granularity of the available
MC samples in this region of the ($m_{\PSQt}$, $m_{\mathrm{LSP}}$)  plane.

All mass limits shown are obtained using signal cross sections calculated at NLO+NLL
order in $\alpha_{\mathrm{s}}$~\cite{bib-nlo-nll-01,bib-nlo-nll-02,bib-nlo-nll-03,bib-nlo-nll-04,bib-nlo-nll-05}.
Table~\ref{tab:lim} summarizes the limits of the supersymmetric particles excluded in the simplified model
scenarios considered.
\begin{table}[!htb]
\centering
\topcaption{Ranges of typical values of the signal systematic uncertainties as evaluated for the $\PSg\to\bbbar\PSGczDo$
  signal model.
    Uncertainties evaluated on other signal models are consistent with these ranges of values.
    A large uncertainty from the limited size of the simulated sample only occurs for a small number of model
    points for which a small subset of search regions have very low efficiency.
    \label{tab:sig_systs}}
\begin{tabular}{lc}
\hline
Source & Typical values [\%] \\
\hline
Integrated luminosity                     & 5     \\
Limited size of MC samples                & 1--100  \\
Renormalization and factorization scales  & 5       \\
ISR                                       & 0--30   \\
b tagging efficiency, heavy flavor        & 0--40   \\
b tagging efficiency, light flavor        & 0--20   \\
Lepton efficiency                         & 0--20   \\
Jet energy scale                          & 5       \\
\hline
\end{tabular}
\end{table}
\begin{table}[htbp]
\centering
  \topcaption{Summary of 95\% CL observed exclusion limits for different SUSY simplified model scenarios.
    The limit on the mass of the produced sparticle is quoted for a massless LSP, while for the lightest
    neutralino the best limit on its mass is quoted.
    \label{tab:lim}}
\begin{tabular}{lcc}
\hline
Simplified & Limit on produced sparticle & Best limit on \\
model & mass [\GeVns{}] for $m_{\PSGczDo}=0$\GeV & LSP mass [\GeVns{}] \\
\hline
Direct squark production & & \\[\cmsTabSkip]
Bottom squark & 880 & 380 \\
Top squark & 800 & 300 \\
Single light squark & 600 & 300 \\
8 degenerate light squarks & 1260 & 580 \\[\cmsTabSkip]
Gluino mediated production & & \\[\cmsTabSkip]
$\PSg\to \bbbar\PSGczDo$ & 1750 & 1125 \\
$\PSg\to \ttbar\PSGczDo$ & 1550 & 825 \\
$\PSg\to \qqbar\PSGczDo$ & 1725 & 850 \\
\hline
\end{tabular}
\end{table}

To facilitate reinterpretation of our results in the context of other models,
we have also provided predictions and results in ``aggregated regions,'' made from summing up
our individual signal bins in topologically similar regions.  These results are presented in Appendix~\ref{app:aggregateregions}.
\section{Summary}
\label{sec:conclusions}
A search for new physics using events containing hadronic jets with transverse momentum
imbalance as measured by the \mttwo variable has been presented.
Results are based on a data sample of proton-proton collisions at $\sqrt{s} = 13\TeV$ collected with the
CMS detector and corresponding to an integrated luminosity of 2.3\fbinv.
No significant deviations from the standard model expectations are observed.

In the limit of a massless LSP, gluino masses of up to 1750\GeV are excluded, extending the reach of Run 1 searches
by more than 300\GeV.
For lighter gluinos, LSP masses up to 1125\GeV in the most favorable models are excluded, also increasing
previous limits by more than 300\GeV. Among the three
gluino decays considered, the strongest limits on gluino pair production are generally achieved for
the $\PSg\to\bbbar\PSGczDo$ channel.
Improved sensitivity is obtained in this scenario as selections requiring at least two b-tagged jets in the
final state retain a significant fraction of gluino-mediated bottom squark events, while strongly suppressing
the background from \wjets, \zjets, and multijet processes. Also, unlike for models
with $\PSg\to\ttbar\PSGczDo$ decays, which include leptonic decays, gluino-mediated bottom squark events
do not suffer from an efficiency loss due to the lepton veto.

For direct pair production of first- and second-generation squarks, each assumed to decay exclusively to a quark of the same flavor
and the lightest neutralino,
squark masses of up to about 1260\GeV and LSP masses up to 580\GeV are excluded.  If only a single squark is assumed to be light,
the limit on the squark and LSP masses is relaxed to 600 and 300\GeV, respectively.
For the pair prouction of third-generation squarks, each assumed to decay with 100\% branching fraction to a quark of the same flavor and
the lightest neutralino, a bottom (top) squark mass up to 880 (800)\GeV is excluded.

For gluino-induced and direct squark production models, the observed exclusion limits on the masses of the
sparticles are from 200 to about 300\GeV higher than those obtained by a similar analysis performed on
8\TeV data~\cite{MT2at8TeV}, which is therefore superseded by the current search.
In relative terms, the largest difference is in the limit on the mass of the top squark, which moves from about
500\GeV to 800\GeV for a massless LSP. This is mostly due to a fluctuation in the 8\TeV data that is not present in
the 13\TeV data.
\clearpage
\begin{acknowledgments}
We congratulate our colleagues in the CERN accelerator departments for the excellent performance of the LHC and thank the technical
and administrative staffs at CERN and at other CMS institutes for their contributions to the success of
 the CMS effort. In addition, we gratefully acknowledge the computing centers and personnel of the
Worldwide LHC Computing Grid for delivering so effectively the computing infrastructure essential to our analyses. Finally, we acknowledge
the enduring support for the construction and operation of the LHC and the CMS detector provided by the following funding agencies:
BMWFW and FWF (Austria); FNRS and FWO (Belgium); CNPq, CAPES, FAPERJ, and FAPESP (Brazil); MES (Bulgaria);
CERN; CAS, MoST, and NSFC (China); COLCIENCIAS (Colombia); MSES and CSF (Croatia); RPF (Cyprus); MoER, ERC IUT and ERDF (Estonia);
Academy of Finland, MEC, and HIP (Finland); CEA and CNRS/IN2P3 (France); BMBF, DFG, and HGF (Germany); GSRT (Greece); OTKA and
NIH (Hungary); DAE and DST (India); IPM (Iran); SFI (Ireland); INFN (Italy); MSIP and NRF (Republic of Korea); LAS (Lithuania);
MOE and UM (Malaysia); CINVESTAV, CONACYT, SEP, and UASLP-FAI (Mexico); MBIE
(New Zealand); PAEC (Pakistan); MSHE and NSC (Poland); FCT (Portugal); JINR (Dubna); MON, RosAtom, RAS and RFBR (Russia); MESTD (Serbia);
SEIDI and CPAN (Spain); Swiss Funding Agencies (Switzerland); MST (Taipei); ThEPCenter, IPST, STAR and NSTDA (Thailand);
TUBITAK and TAEK (Turkey); NASU and SFFR (Ukraine); STFC (United Kingdom); DOE and NSF (USA).
Individuals have received support from the Marie-Curie program and the European Research Council and EPLANET (European Union);
the Leventis Foundation; the A. P. Sloan Foundation; the Alexander von Humboldt Foundation; the Belgian Federal Science
Policy Office; the Fonds pour la Formation \`a la Recherche dans l'Industrie et dans l'Agriculture (FRIA-Belgium); the Agentschap
voor Innovatie door Wetenschap en Technologie (IWT-Belgium); the Ministry of Education, Youth and Sports (MEYS) of the Czech Republic;
the Council of Science and Industrial Research, India; the HOMING PLUS program of the Foundation for Polish Science, cofinanced from
European Union, Regional Development Fund; the OPUS program of the National Science Center (Poland); the Compagnia di San Paolo (Torino);
MIUR project 20108T4XTM (Italy); the Thalis and Aristeia programs cofinanced by EU-ESF and the Greek NSRF; the National Priorities
Research Program by Qatar National Research Fund; the Rachadapisek Sompot Fund for Postdoctoral Fellowship, Chulalongkorn
University (Thailand); the Chulalongkorn Academic into Its 2nd Century Project Advancement Project (Thailand); and the Welch Foundation,
contract C-1845.
\end{acknowledgments}
\bibliography{auto_generated}
\numberwithin{figure}{section}
\numberwithin{table}{section}
\appendix
\clearpage
\section{Detailed results}
\label{app:results}
\begin{figure}[h!t]
  \centering
    \includegraphics[width=0.99\textwidth]{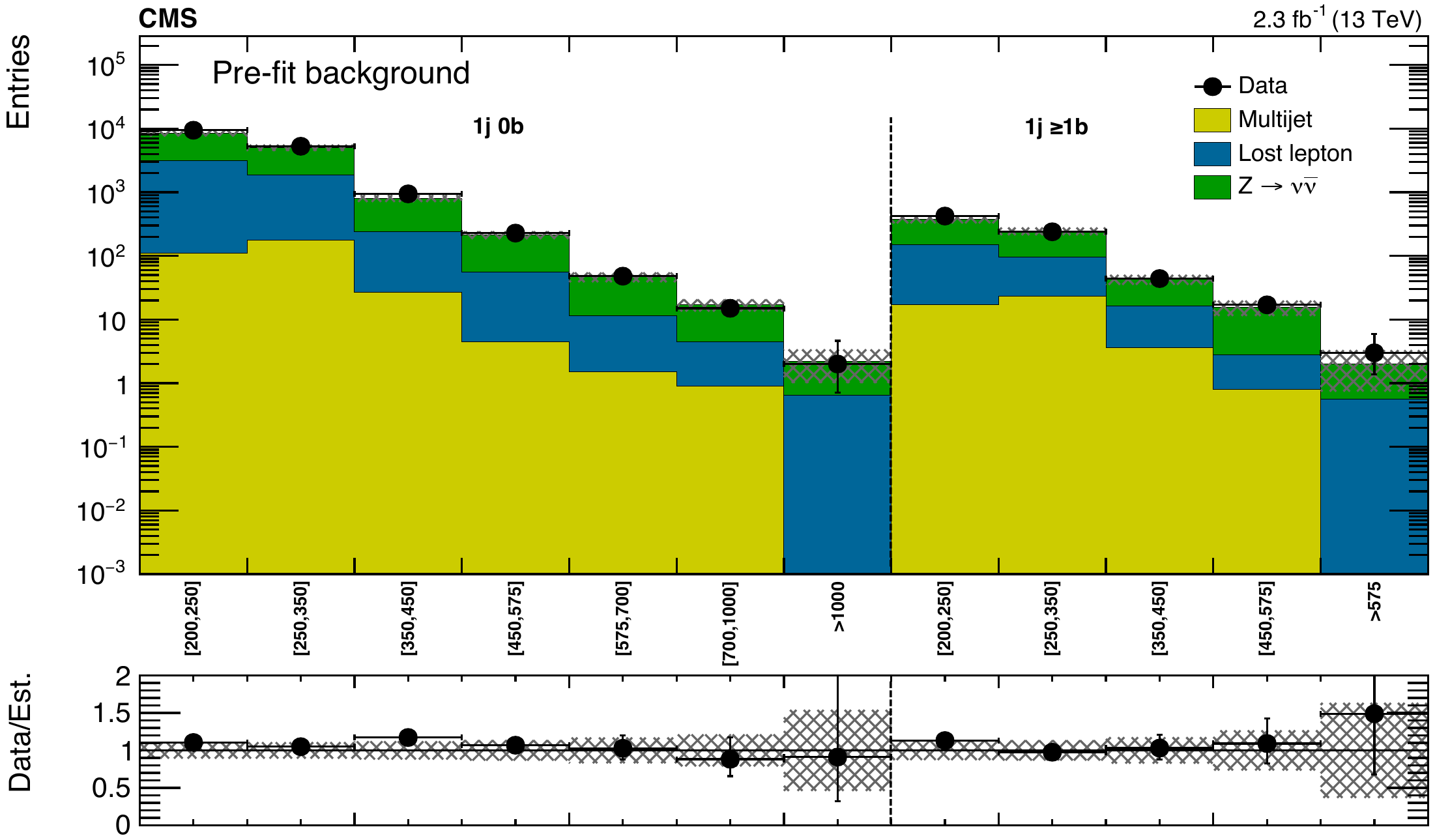}\\
    \includegraphics[width=0.99\textwidth]{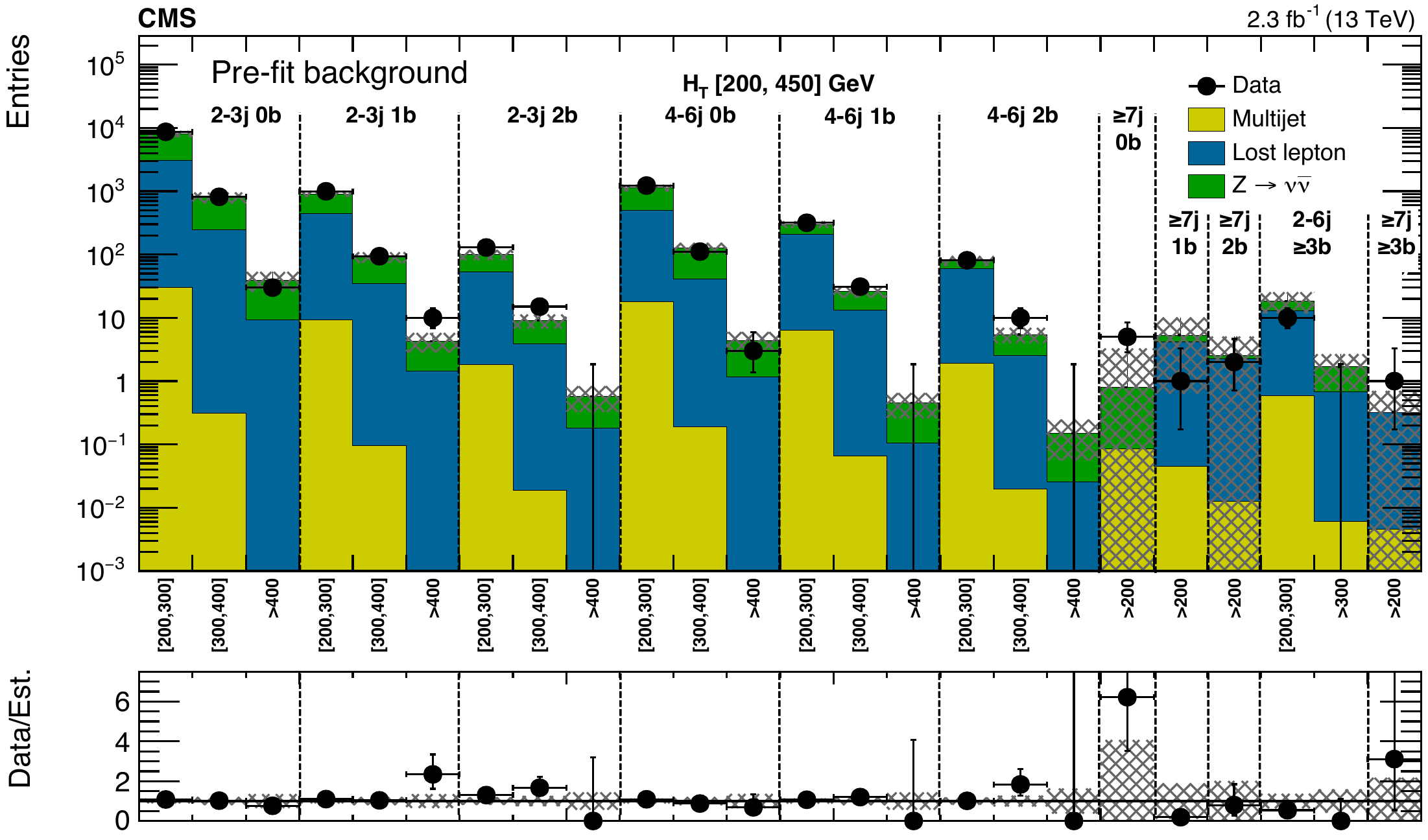}\\
    \caption{(Above) Comparison of the estimated background (pre-fit) and observed data events in each signal bin in the monojet region. On the $x$-axis, the jet \pt binning is shown (in\GeV). Hatched bands represent the full uncertainty in the background estimate.
    (Below) Same for the very-low-\HT region.  On the $x$-axis, the \mttwo binning is shown (in\GeV).
    Bins with no entry for data have an observed count of 0 events.}
    \label{fig:otherResults1}
\end{figure}
\begin{figure}
  \centering
    \includegraphics[width=0.99\textwidth]{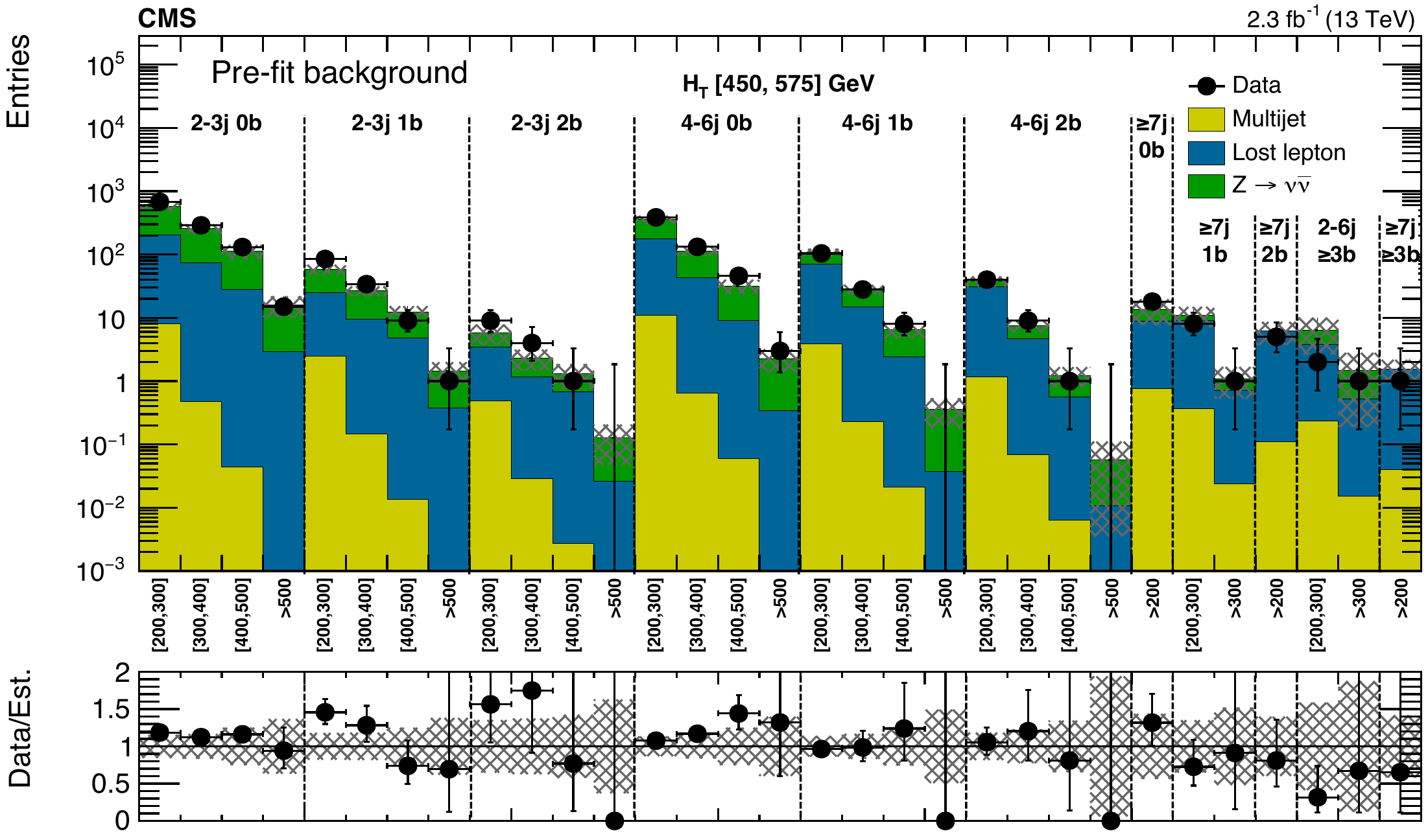}\\
    \includegraphics[width=0.99\textwidth]{figures/results/mt2_mediumHT_fullEstimate_fullRange.pdf}\\
    \caption{(Above) Comparison of the estimated background (pre-fit) and observed data events in each signal bin in the low-\HT region.  Hatched bands represent the full uncertainty in the background estimate.
    (Below) Same for the medium-\HT region. On the $x$-axis, the \mttwo binning is shown (in\GeV).
    Bins with no entry for data have an observed count of 0 events.}
    \label{fig:otherResults2}
\end{figure}
\begin{figure}
  \centering
    \includegraphics[width=0.99\textwidth]{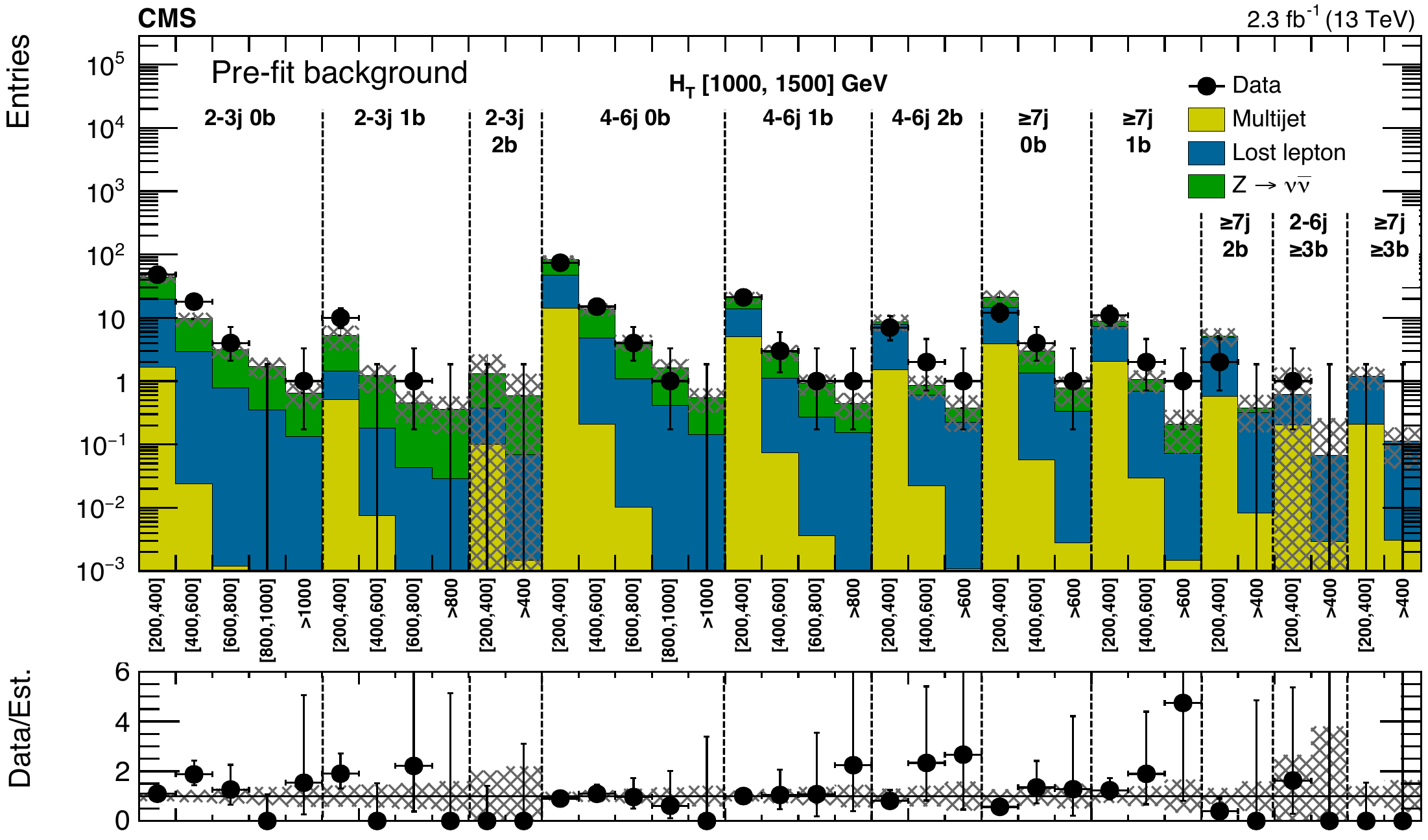}\\
    \includegraphics[width=0.99\textwidth]{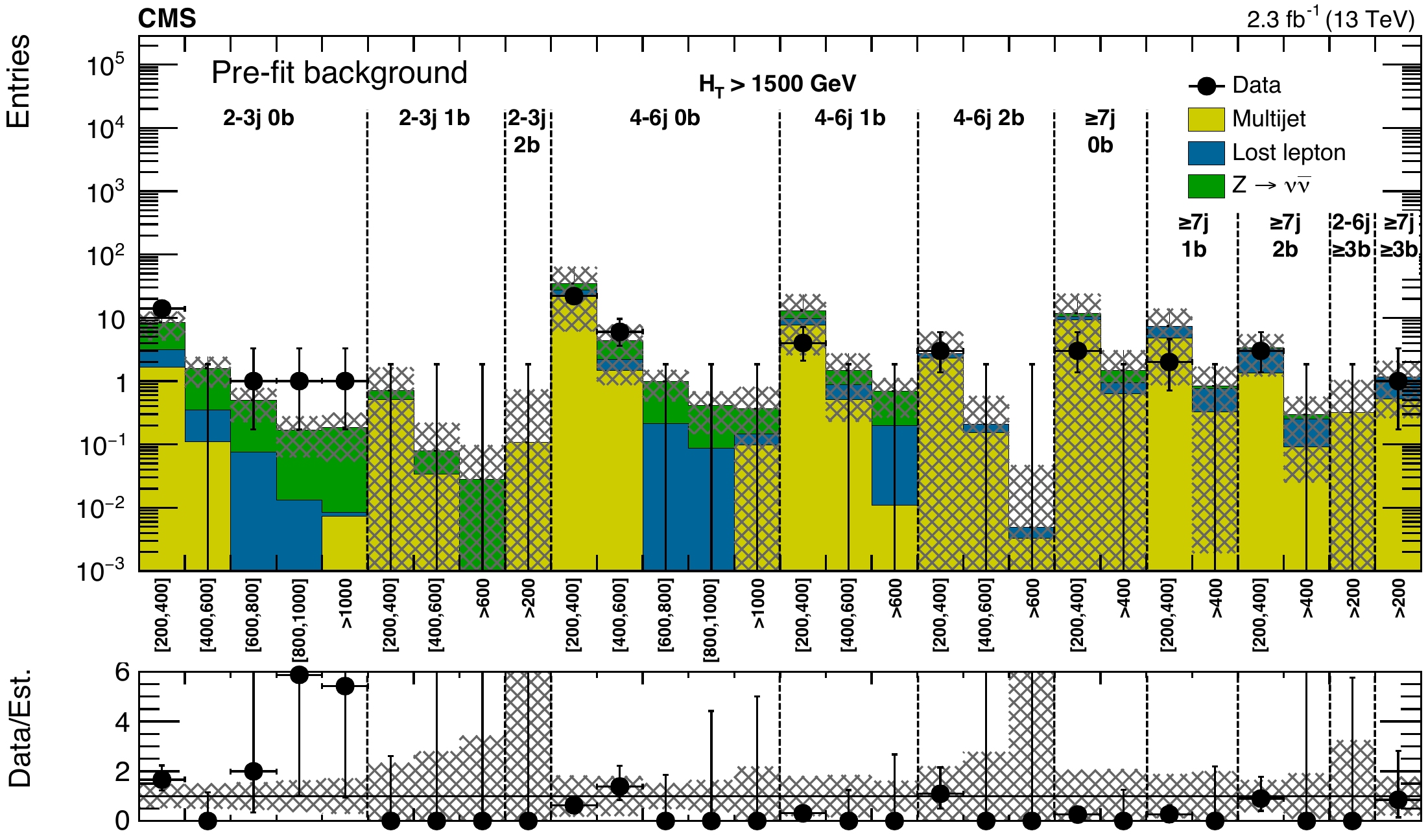}\\
    \caption{(Above) Comparison of the estimated background (pre-fit) and observed data events in each signal bin in the high-\HT region.  Hatched bands represent the full uncertainty in the background estimate.
    (Below) Same for the extreme-\HT region. On the $x$-axis, the \mttwo binning is shown (in\GeV).
    Bins with no entry for data have an observed count of 0 events.}
    \label{fig:otherResults2_split}
\end{figure}
\begin{figure}
  \centering
    \includegraphics[width=0.99\textwidth]{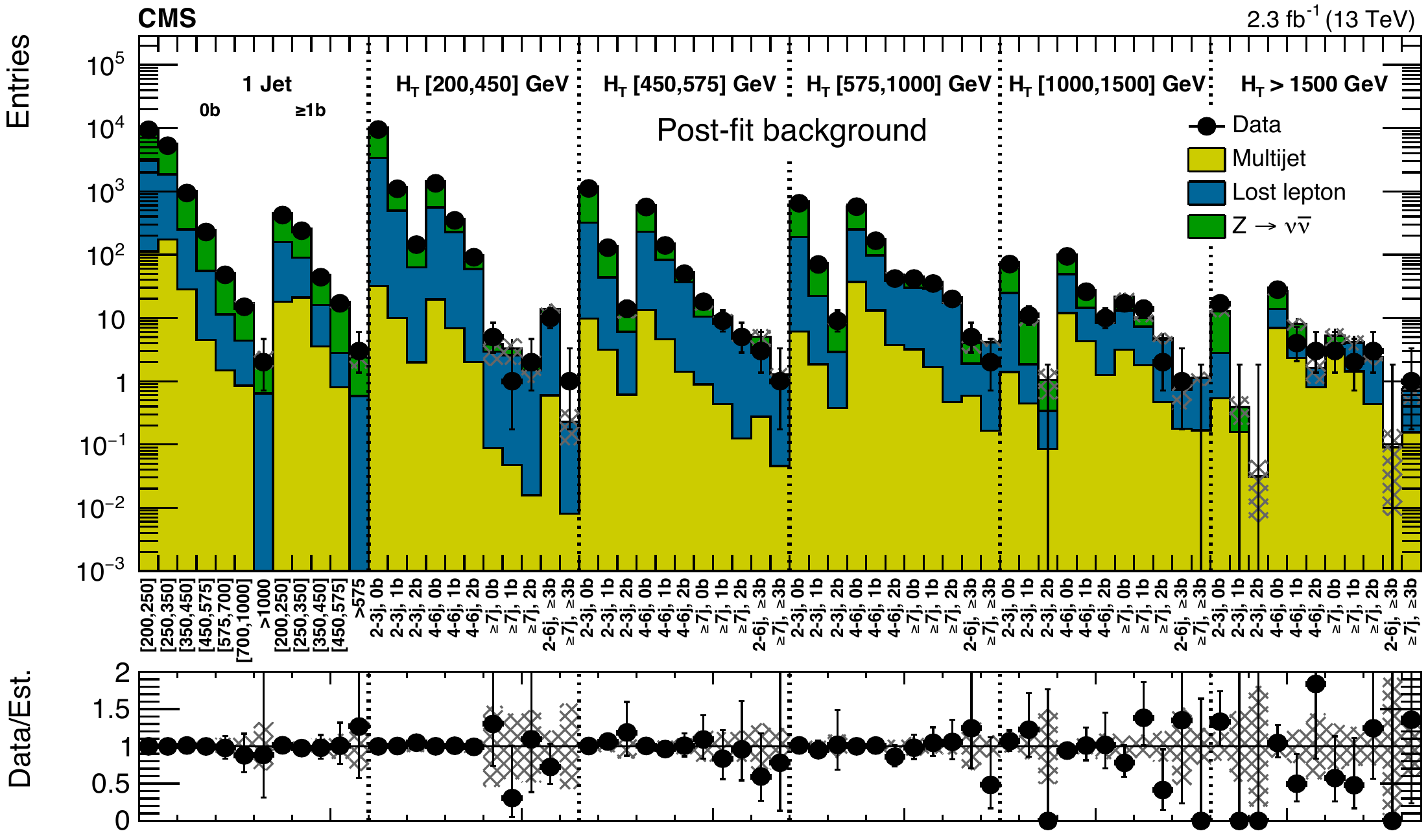}
    \caption{Comparison of post-fit background prediction and observed data events in each topological region.  Hatched bands represent the post-fit uncertainty in the background prediction.
For the monojet region, on the $x$-axis, the jet \pt binning is shown in \GeV.}
    \label{fig:otherResults3}
\end{figure}
\begin{figure}
  \centering
    \includegraphics[width=0.96\textwidth]{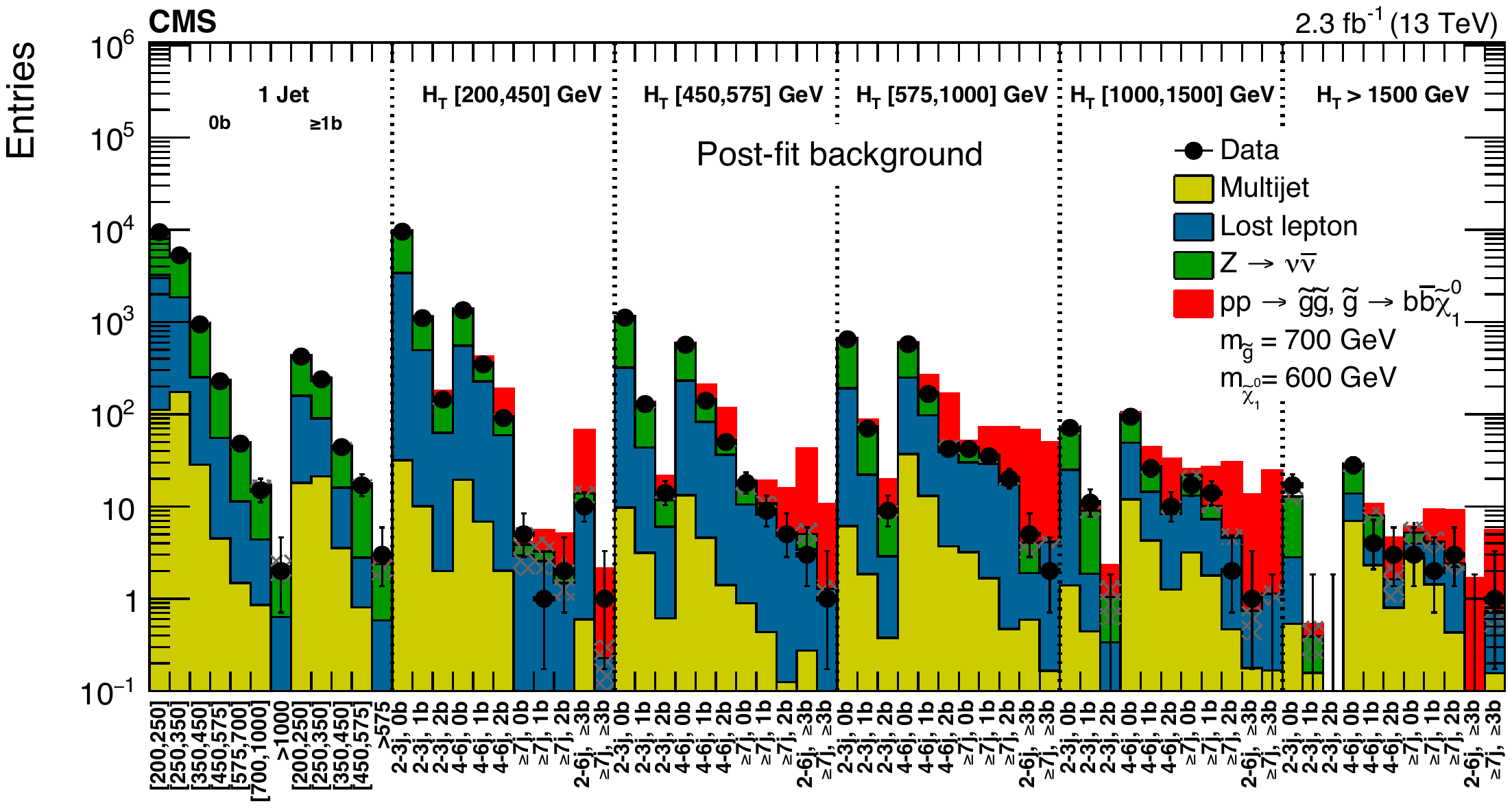}
    \caption{Same as Fig.~\ref{fig:otherResults3}  but also including the expected contribution from a compressed-spectrum signal model of gluino-mediated bottom squark production with the mass of the gluino and the LSP equal to 700 and 600 GeV, respectively. The signal model is described in Refs.~\cite{bib-sms-1,bib-sms-2,bib-sms-3,bib-sms-4,Chatrchyan:2013sza} and in the text.}
    \label{fig:otherResults4}
\end{figure}
\clearpage
\begin{figure}
  \centering
    \includegraphics[width=0.96\textwidth]{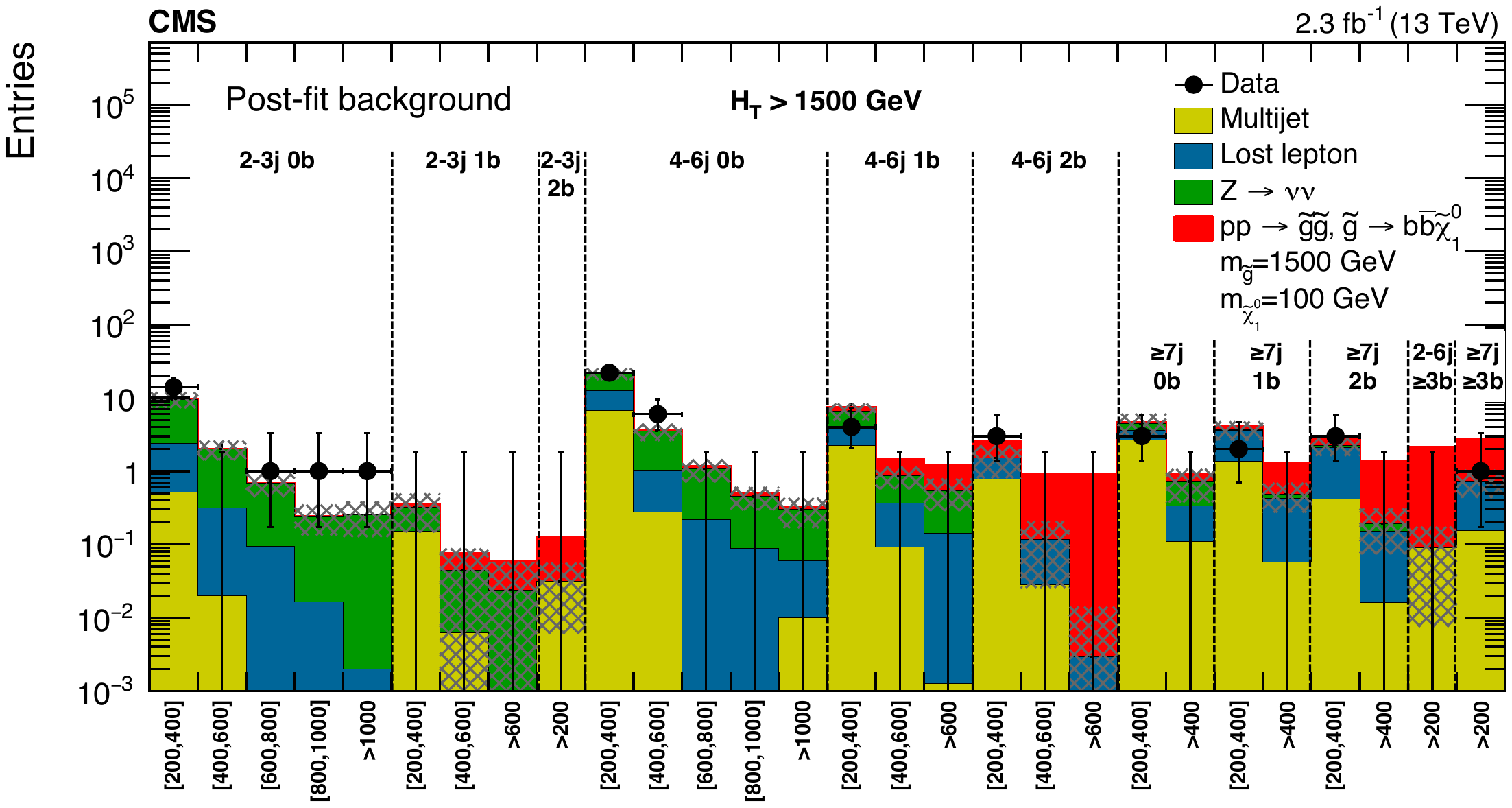}
    \caption{Post-fit background prediction, expected signal yields, and observed data events in each signal bin in the
extreme-\HT region. Hatched bands represent the post-fit uncertainty in the background prediction.
On the $x$-axis, the \mttwo binning is shown (in \GeV). The red histogram shows the expected contribution from an open-spectra signal model of gluino-mediated bottom squark production with the mass of the gluino and the LSP equal to 1500 and 100\GeV, respectively. The signal model is described in Refs.~\cite{bib-sms-1,bib-sms-2,bib-sms-3,bib-sms-4,Chatrchyan:2013sza} and in the text.}
    \label{fig:otherResults5}
\end{figure}
\begin{table}
\centering
\topcaption{Binning in \mttwo for each topological region of the multijet search regions with
very low, low, and medium \HT.} \label{tab:mt2bins1}
\begin{tabular}{lll}
\hline
\HT range [\GeVns{}] & \multicolumn{1}{c}{Jet multiplicities} & \multicolumn{1}{c}{Bin boundaries [\GeVns{}]} \\
\hline
\multirow{11}{*}{200--450} & 2--3j, $  0$b  &  200--300, 300--400, $>$400  \\
 & 2--3j, $  1$b  &  200--300, 300--400, $>$400  \\
 & 2--3j, $  2$b  &  200--300, 300--400, $>$400  \\
 & 4--6j, $  0$b  &  200--300, 300--400, $>$400  \\
 & 4--6j, $  1$b  &  200--300, 300--400, $>$400  \\
 & 4--6j, $  2$b  &  200--300, 300--400, $>$400  \\
 & $\geq$s7j, $  0$b  &  $>$200 \\
 & $\geq$7j, $  1$b  &  $>$200 \\
 & $\geq$7j, $  2$b  &  $>$200 \\
 & 2--6j, $\geq$3b  &  200--300, $>$300   \\
 & $\geq$7j, $\geq$3b  &  $>$200 \\[\cmsTabSkip]
\multirow{11}{*}{450--575} & 2-3j, $  0$b  &  200--300, 300--400, 400--500, $>$500   \\
 & 2--3j, $  1$b  &  200--300, 300--400, 400--500, $>$500   \\
 & 2--3j, $  2$b  &  200--300, 300--400, 400--500, $>$500   \\
 & 4--6j, $  0$b  &  200--300, 300--400, 400--500, $>$500   \\
 & 4--6j, $  1$b  &  200--300, 300--400, 400--500, $>$500   \\
 & 4--6j, $  2$b  &  200--300, 300--400, 400--500, $>$500   \\
 & $\geq$7j, $  0$b  &  $>$200 \\
 & $\geq$7j, $  1$b  &  200--300,  $>$300   \\
 & $\geq$7j, $  2$b  &  $>$200 \\
 & 2--6j, $\geq$3b  &  200--300,  $>$300   \\
 & $\geq$7j, $\geq$3b  &  $>$200 \\[\cmsTabSkip]
\multirow{11}{*}{575--1000} & 2-3j, $  0$b  &  200--300, 300--400, 400--600, 600--800,  $>$800   \\
 & 2--3j, $  1$b  &  200--300, 300--400, 400--600, 600--800,  $>$800   \\
 & 2--3j, $  2$b  &  200--300, 300--400, 400--600,  $>$600  \\
 & 4--6j, $  0$b  &  200--300, 300--400, 400--600, 600--800,  $>$800   \\
 & 4--6j, $  1$b  &  200--300, 300--400, 400--600,  $>$600  \\
 & 4--6j, $  2$b  &  200--300, 300--400, 400--600,  $>$600  \\
 & $\geq$7j, $  0$b  &  200--300, 300--400, $>$400   \\
 & $\geq$7j, $  1$b  &  200--300, 300--400, $>$400   \\
 & $\geq$7j, $  2$b  &  200--300, 300--400, $>$400   \\
 & 2--6j, $\geq$3b  &  200--300, 300--400, $>$400   \\
 & $\geq$7j, $\geq$3b  & 200--300, 300--400, $>$400   \\  \hline
 \end{tabular}
\end{table}
\begin{table}
\centering
\topcaption{Binning in \mttwo for each topological region of the multijet
search regions with high and extreme \HT.}\label{tab:mt2bins2}
\begin{tabular}{lll}
\hline
\HT range [\GeVns{}] & \multicolumn{1}{c}{Jet multiplicities} & \multicolumn{1}{c}{Bin boundaries [\GeVns{}]} \\
\hline
\multirow{11}{*}{1000--1500} & 2--3j, $  0$b  &  200--400, 400--600, 600--800, 800--1000, $>$1000   \\
 & 2--3j, $  1$b  &  200--400, 400--600, 600--800, $>$800   \\
 & 2--3j, $  2$b  &  200--400, $>$400   \\
 & 4--6j, $  0$b  &  200--400, 400--600, 600--800, 800--1000, $>$1000   \\
 & 4--6j, $  1$b  &  200--400, 400--600, 600--800, $>$800   \\
 & 4--6j, $  2$b  &  200--400, 400--600, $>$600   \\
 & $\geq$7j, $  0$b  &  200--400, 400--600, $>$600   \\
 & $\geq$7j, $  1$b  &  200--400, 400--600, $>$600   \\
 & $\geq$7j, $  2$b  &  200--400, $>$400   \\
 & 2--6j, $\geq$3b  &  200--400, $>$400   \\
 & $\geq$7j, $\geq$3b  &  200--400, $>$400   \\[\cmsTabSkip]
 \multirow{11}{*}{$>$1500} & 2-3j, $  0$b  &  200--400, 400--600, 600--800, 800--1000, $>$1000   \\
 & 2--3j, $  1$b  &  200--400, 400--600, $>$600   \\
 & 2--3j, $  2$b  &  $>$200 \\
 & 4--6j, $  0$b  &  200--400, 400--600, 600--800, 800--1000, $>$1000   \\
 & 4--6j, $  1$b  &  200--400, 400--600, $>$600   \\
 & 4--6j, $  2$b  &  200--400, 400--600, $>$600   \\
 & $\geq$7j, $  0$b  &  200--400, $>$400   \\
 & $\geq$7j, $  1$b  &  200--400, $>$400   \\
 & $\geq$7j, $  2$b  &  200--400, $>$400   \\
 & 2--6j, $\geq$3b  &  $>$200 \\
 & $\geq$7j, $\geq$3b  &  $>$200 \\
 \hline
\end{tabular}
\end{table}
\begin{table}
\centering
\topcaption{Binning in jet \pt for the monojet regions.}
\label{tab:mt2bins3}
\begin{tabular}{ll}
\hline
Jet multiplicities & Bin boundaries [\GeVns{}] \\
\hline
 1j, 0b  &  200--250, 250--350, 350--450 , 450--575, 575--700, 700--1000, $>$1000   \\
 1j, $\geq$1b  &  200--250, 250--350, 350--450 , 450--575, $>$575   \\
\hline
\end{tabular}
\end{table}
\clearpage
\section{Aggregated regions}
\label{app:aggregateregions}

To allow simpler reinterpretations, we also provide our results in ``aggregated regions,'' made from summing up
the event yields and the pre-fit background predictions for individual signal bins in topologically similar regions.
The uncertainty in the prediction in each aggregated region is
calculated taking into account the same correlation model used in the full analysis.
The definitions of these regions are given in Table~\ref{tab:macroregion_def}, while Table~\ref{tab:macroregion_yields} gives the predicted and observed number of events in each region
together with the 95\% CL upper limit on the number of signal events.

\begin{table}
\centering
\topcaption{Definitions of aggregated regions.  Each aggregated region is obtained by selecting all events that pass the logical OR of the listed selections. \label{tab:macroregion_def}}
\begin{tabular}{lcccc}
\hline
Region & \njets & \nbtags & \HT [\GeVns{}] & \mttwo [\GeVns{}] \\
\hline
\multirow{4}{*}{1j loose} & $=$1 & \NA & $>$450 & \NA \\
 & 2--3 & $\leq$2& 450--575 & $>$400 \\
 & 2--3 & $\leq$2& 575--1000 & $>$300 \\
 & 2--3 & $\leq$2& $>$1000 & $>$200 \\[\cmsTabSkip]
\multirow{3}{*}{1j medium} & $=$1 & \NA & $>$575 & \NA \\
 & 2--3 & $\leq$2& 575--1000 & $>$600 \\
 & 2--3 & $\leq$2& $>$1000 & $>$200 \\[\cmsTabSkip]
\multirow{8}{*}{1j tight} & $=$1 & $=$0 & $>$1000 & \NA \\
 & $=$1 & $\geq$1& $>$575 & \NA \\
 & 2--3 & $=$0 & 575--1000 & $>$800 \\
 & 2--3 & 1--2 & 575--1000 & $>$600 \\
 & 2--3 & 0--1 & 1000--1500 & $>$800 \\
 & 2--3 & $=$2 & 1000--1500 & $>$400 \\
 & 2--3 & 0--1 & $>$1500 & $>$400 \\
 & 2--3 & $=$2 & $>$1500 & $>$200 \\[\cmsTabSkip]
\multirow{4}{*}{2j tight} & 2--3 & \NA & $>$1000 & $>$600 \\
 & 2--3 & \NA & $>$1500 & $>$400 \\
 & 4--6 & \NA & $>$1000 & $>$800 \\
 & 4--6 & \NA & $>$1500 & $>$600 \\[\cmsTabSkip]
4j medium & $\geq$4 & \NA & $>$575 & $>$400 \\[\cmsTabSkip]
\multirow{2}{*}{4j tight} & $\geq$4 & \NA & $>$1000 & $>$600 \\
 & $\geq$7 & \NA & $>$1500 & $>$400 \\[\cmsTabSkip]
7j tight & $\geq$7 & \NA & $>$575 & $>$400 \\[\cmsTabSkip]
\multirow{3}{*}{7j very tight} & $\geq$7 & 0--1 & $>$1000 & $>$600 \\
 & $\geq$7 & $\geq$2 & $>$1000 & $>$400 \\
 & $\geq$7 & \NA & $>$1500 & $>$400 \\[\cmsTabSkip]
2b medium & $\geq$2 & $\geq$2 & $>$575 & $>$200 \\[\cmsTabSkip]
2b tight & $\geq$2 & $\geq$2 & $>$575 & $>$400 \\[\cmsTabSkip]
2b very tight & $\geq$2 & $\geq$2 & $>$1000 & $>$400 \\[\cmsTabSkip]
3b medium & $\geq$2 & $\geq$3 & $>$200 & $>$200 \\[\cmsTabSkip]
3b tight & $\geq$2 & $\geq$3 & $>$575 & $>$200 \\[\cmsTabSkip]
3b very tight & $\geq$2 & $\geq$3 & $>$1000 & $>$200 \\
\hline
\end{tabular}
\end{table}

\begin{table}
\centering
\topcaption{Predictions and observations for the aggregated regions defined in Table~\ref{tab:macroregion_def}, together with the observed 95\% CL limit on the number of signal events contributing to each region ($N_{95}^{\text{obs}}$).  An uncertainty of either 15 or 30\% in the signal efficiency is assumed for calculating the limits.\label{tab:macroregion_yields}}
\renewcommand*{\arraystretch}{1.2}
\begin{tabular}{lcccc}
\hline
Region & Prediction & Observation & $N_{95}^{\text{obs}}$, 15\% unc. & $N_{95}^{\text{obs}}$, 30\% unc. \\
\hline
1j loose & $833\pm95$ & 902 & 246 & 273 \\
1j medium & $175\pm22$ & 185 & 60 & 66 \\
1j tight & $15.9^{+3.2}_{-2.9}$ & 12 & 7.9 & 8.4 \\
2j tight & $15.7^{+4.0}_{-3.9}$ & 12 & 8.9 & 9.5 \\
4j medium & $159\pm25$ & 165 & 60 & 66 \\
4j tight & $16.2^{+5.0}_{-4.9}$ & 11 & 8.7 & 9.3 \\
7j tight & $15.3^{+4.6}_{-4.5}$ & 14 & 11 & 12 \\
7j very tight & $5.3^{+3.3}_{-3.2}$ & 3 & 5.7 & 6.1 \\
2b medium & $119\pm14$ & 98 & 21 & 23 \\
2b tight & $13.5^{+3.3}_{-3.1}$ & 10 & 7.7 & 8.2 \\
2b very tight & $4.5^{+2.3}_{-2.1}$ & 4 & 6.3 & 6.8 \\
3b medium & $40.9^{+9.9}_{-8.8}$ & 24 & 11 & 11 \\
3b tight & $11.0^{+3.2}_{-2.5}$ & 9 & 7.7 & 8.2 \\
3b very tight & $3.5^{+1.9}_{-1.4}$ & 2 & 4.3 & 4.5 \\
\hline
\end{tabular}
\end{table}

If these aggregated regions are used to derive cross section limits on the signals considered in this paper,
they typically yield results that are less stringent by a factor of about two compared to the full binned analysis.
This is shown in more detail for few signal models in Table~\ref{tab:macroRegions_plusSignal}.
The expected upper limit on the signal cross section as obtained from the full analysis is compared
to the one obtained from the aggregated region that has the best sensitivity to the signal model considered.
A 15\% uncertainty in the signal selection efficiency is assumed for calculating these limits.
The same table also provides the expected signal yields in the given aggregated regions.

\begin{table}
\topcaption{Expected upper limits on the cross section of several signal models, as determined from the full binned analysis,
are compared to the upper limits obtained using only the aggregated region that has the best sensitivity to each considered
signal model. A 15\% uncertainty in the signal selection efficiency is assumed for calculating these limits.
The signal yields expected for an integrated luminosity of 2.3\fbinv
are also shown.}\label{tab:macroRegions_plusSignal}
\centering
\resizebox{\textwidth}{!}{
\renewcommand{\arraystretch}{1.5}
\begin{tabular}{lcccc}
\hline
Signal &  Expected limit [fb] & Best aggregated & Signal yield (best & Expected limit [fb] (best \\%
       &  (full analysis)&      region                  & aggregated region)             & aggregated region)\\
\hline
$\Pp\Pp\to\PSg\PSg, \PSg\to \bbbar\PSGczDo$
& \multirow{2}{*}{4.80} & \multirow{2}{*}{2b very tight} & \multirow{2}{*}{3.19} & \multirow{2}{*}{9.83}\\
($m_{\PSg}=1700$\GeV, $m_{\PSGczDo}=0$\GeV) & & & \\[\cmsTabSkip]
$\Pp\Pp\to\PSg\PSg, \PSg\to \bbbar\PSGczDo$
& \multirow{2}{*}{393} & \multirow{2}{*}{2b tight} & \multirow{2}{*}{4.79} & \multirow{2}{*}{667}\\
($m_{\PSg}=1000$\GeV, $m_{\PSGczDo}=950$\GeV) & & & \\\hline
$\Pp\Pp\to\PSg\PSg, \PSg\to \qqbar\PSGczDo$
& \multirow{2}{*}{8.67} & \multirow{2}{*}{4j tight} & \multirow{2}{*}{5.31} & \multirow{2}{*}{17.2}\\
($m_{\PSg}=1600$\GeV, $m_{\PSGczDo}=0$\GeV) & & & \\[\cmsTabSkip]
$\Pp\Pp\to\PSg\PSg, \PSg\to \qqbar\PSGczDo$
& \multirow{2}{*}{357} & \multirow{2}{*}{7j tight} & \multirow{2}{*}{7.33} & \multirow{2}{*}{536}\\
($m_{\PSg}=1000$\GeV, $m_{\PSGczDo}=850$\GeV) & & & \\\hline
$\Pp\Pp\to\PSg\PSg, \PSg\to \ttbar\PSGczDo$
& \multirow{2}{*}{12.9} & \multirow{2}{*}{7j very tight} & \multirow{2}{*}{4.48} & \multirow{2}{*}{20.7}\\
($m_{\PSg}=1500$\GeV, $m_{\PSGczDo}=0$\GeV) & & & \\[\cmsTabSkip]
$\Pp\Pp\to\PSg\PSg, \PSg\to \ttbar\PSGczDo$
& \multirow{2}{*}{555} & \multirow{2}{*}{3b tight} & \multirow{2}{*}{5.55} & \multirow{2}{*}{1100}\\
($m_{\PSg}=900$\GeV, $m_{\PSGczDo}=600$\GeV) & & & \\\hline
$\Pp\Pp\to\PSQt\PASQt, \PSQt \to \PQt\PSGczDo$
& \multirow{2}{*}{41.8} & \multirow{2}{*}{2b tight} & \multirow{2}{*}{5.79} & \multirow{2}{*}{73.7}\\
($m_{\PSQt}=750$\GeV, $m_{\PSGczDo}=0$\GeV) & & & \\[\cmsTabSkip]
$\Pp\Pp\to\PSQt\PASQt, \PSQt\to \PQt\PSGczDo$
& \multirow{2}{*}{151} & \multirow{2}{*}{2b medium} & \multirow{2}{*}{17.5} & \multirow{2}{*}{321}\\
($m_{\PSQt}=600$\GeV, $m_{\PSGczDo}=250$\GeV) & & & \\[\cmsTabSkip]
$\Pp\Pp\to\PSQt\PASQt, \PSQt\to \PQt\PSGczDo$
& \multirow{2}{*}{18600} & \multirow{2}{*}{2b medium} & \multirow{2}{*}{9.37} & \multirow{2}{*}{73900}\\
($m_{\PSQt}=250$\GeV, $m_{\PSGczDo}=150$\GeV) & & & \\\hline
$\Pp\Pp\to\tilde{\PQb}\bar{\tilde{\PQb}}, \PSQb\to \PQb\PSGczDo$
& \multirow{2}{*}{26.9} & \multirow{2}{*}{2b tight} & \multirow{2}{*}{5.83} & \multirow{2}{*}{48.1}\\
($m_{\PSQb}=800$\GeV, $m_{\PSGczDo}=0$\GeV) & & & \\[\cmsTabSkip]
$\Pp\Pp\to\tilde{\PQb}\bar{\tilde{\PQb}}, \PSQb\to \PQb\PSGczDo$
& \multirow{2}{*}{451} & \multirow{2}{*}{2b medium} & \multirow{2}{*}{21.3} & \multirow{2}{*}{777}\\
($m_{\PSQb}=500$\GeV, $m_{\PSGczDo}=350$\GeV) & & & \\\hline
$\Pp\Pp\to\PSQ\PASQ, \PSQ\to \PQq\PSGczDo$, $\PSQ_{\mathrm{L}}+\PSQ_{\mathrm{R}} (\PSQu, \PSQd, \PSQs, \PSQc)$
& \multirow{2}{*}{14.0} & \multirow{2}{*}{2j tight} & \multirow{2}{*}{7.85} & \multirow{2}{*}{18.3}\\
($m_{\PSQ}=1200$\GeV, $m_{\PSGczDo}=0$\GeV) & & & \\[\cmsTabSkip]
$\Pp\Pp\to\PSQ\PASQ, \PSQ\to \PQq\PSGczDo$, $\PSQ_{\mathrm{L}}+\PSQ_{\mathrm{R}} (\PSQu, \PSQd, \PSQs, \PSQc)$
& \multirow{2}{*}{148} & \multirow{2}{*}{4j medium} & \multirow{2}{*}{300} & \multirow{2}{*}{267}\\
($m_{\PSQ}=600$\GeV, $m_{\PSGczDo}=0$\GeV) & & & \\[\cmsTabSkip]
$\Pp\Pp\to\PSQ\PASQ, \PSQ\to \PQq\PSGczDo$, $\PSQ_{\mathrm{L}}+\PSQ_{\mathrm{R}} (\PSQu, \PSQd, \PSQs, \PSQc)$
& \multirow{2}{*}{493} & \multirow{2}{*}{4j medium} & \multirow{2}{*}{34.0} & \multirow{2}{*}{902}\\
($m_{\PSQ}=700$\GeV, $m_{\PSGczDo}=500$\GeV) & & & \\
\hline
\end{tabular}
}
\end{table}
\clearpage
\section{Summary plots}
\label{app:macroplots}

The figures in this appendix summarize in fewer bins the results shown in
Figs~\ref{fig:results}, \ref{fig:otherResults1}, and \ref{fig:otherResults2}.
The observed data are compared to estimated backgrounds as a function of \mttwo in more inclusive regions.
The aggregated regions presented in these figures are different from those in Appendix~\ref{app:aggregateregions},
being instead formed by summing pre-fit values for all signal regions contained in the inclusive \HT,\njets,\nbtags selection displayed
in the upper left corner of each plot.

\begin{figure}[htb]
  \centering
    \includegraphics[width=0.49\textwidth]{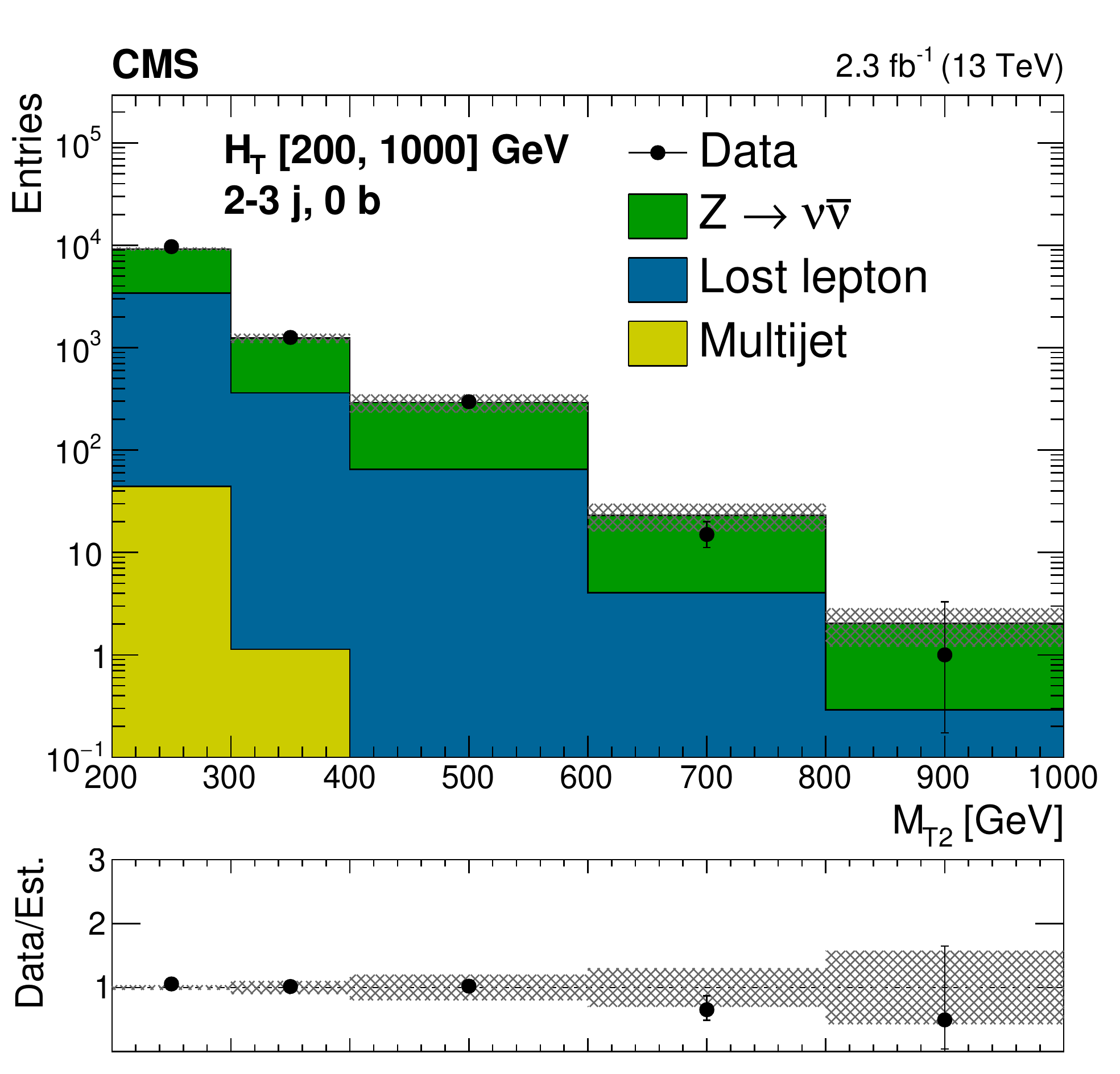}
    \includegraphics[width=0.49\textwidth]{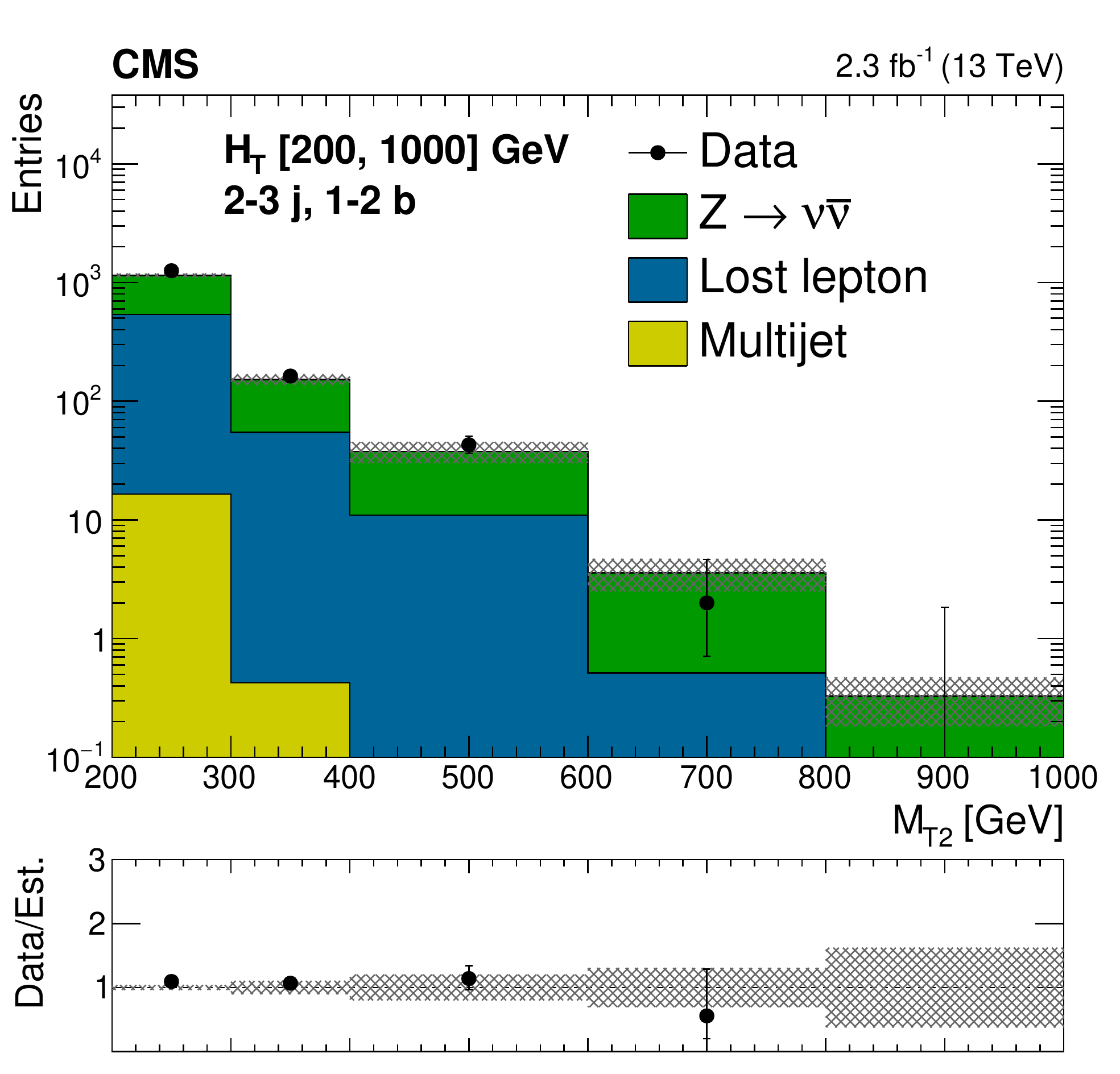}\\
    \includegraphics[width=0.49\textwidth]{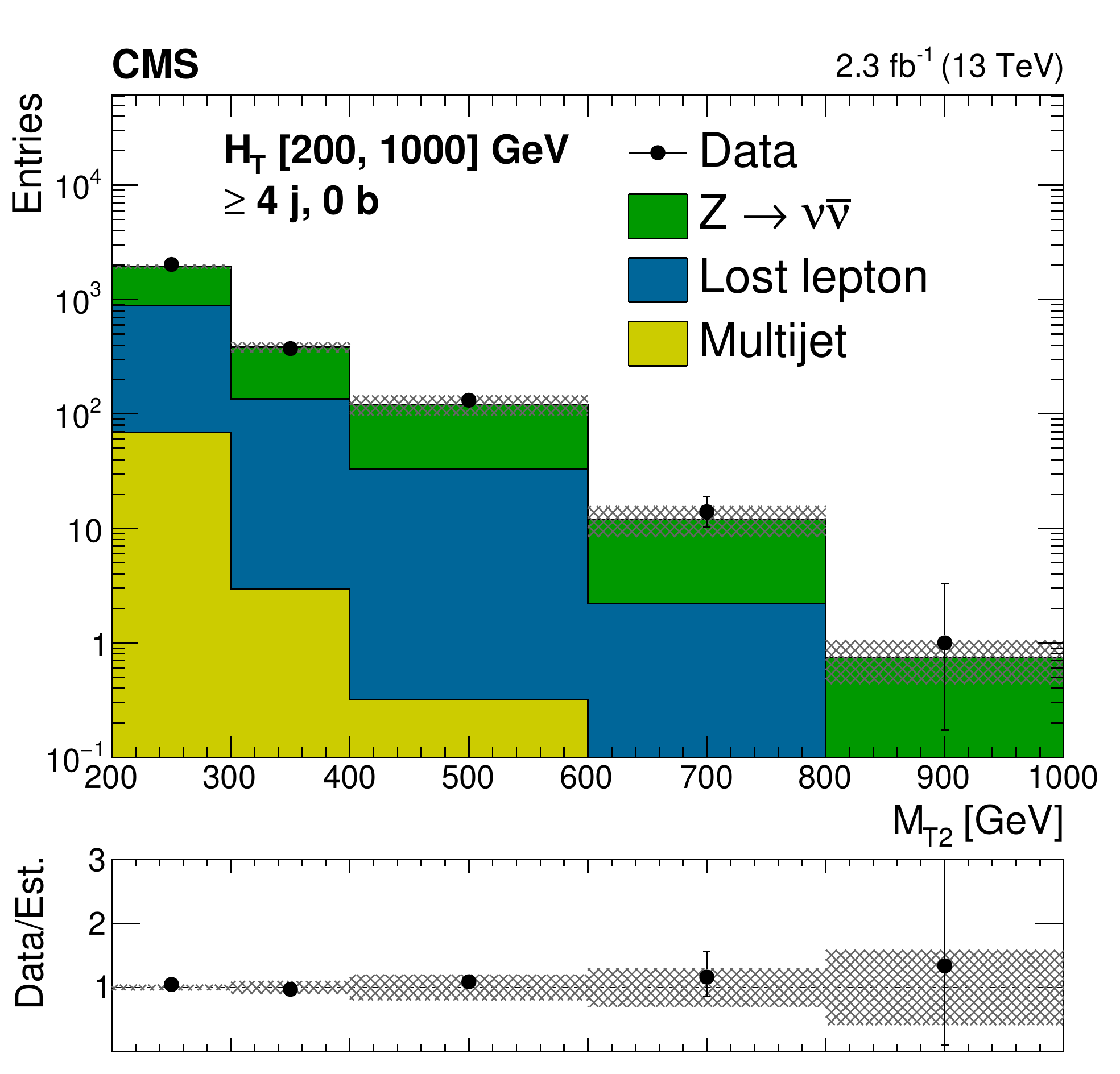}
    \includegraphics[width=0.49\textwidth]{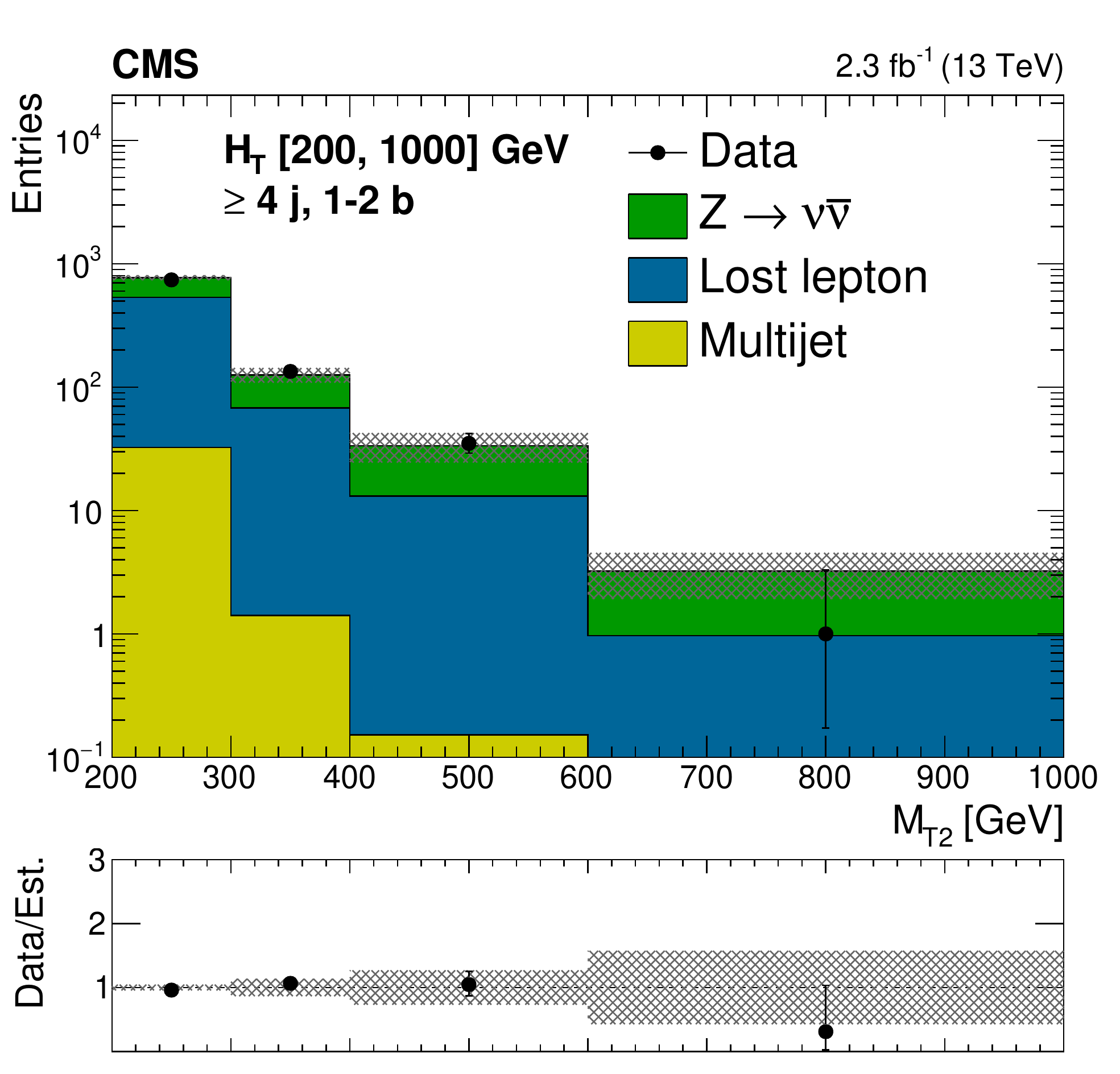}\\
    \caption{ Comparison of estimated background and observed data events in inclusive topological
    regions, as labeled in the legends, as a function of \mttwo,
    for events with $200 < \HT < 1000\GeV$. The background prediction is formed by summing pre-fit values for all signal regions
    included in each plot. Hatched bands represent the full uncertainty in the background estimate. }
    \label{fig:macroplots1}
\end{figure}

\begin{figure}[htb]
  \centering
    \includegraphics[width=0.49\textwidth]{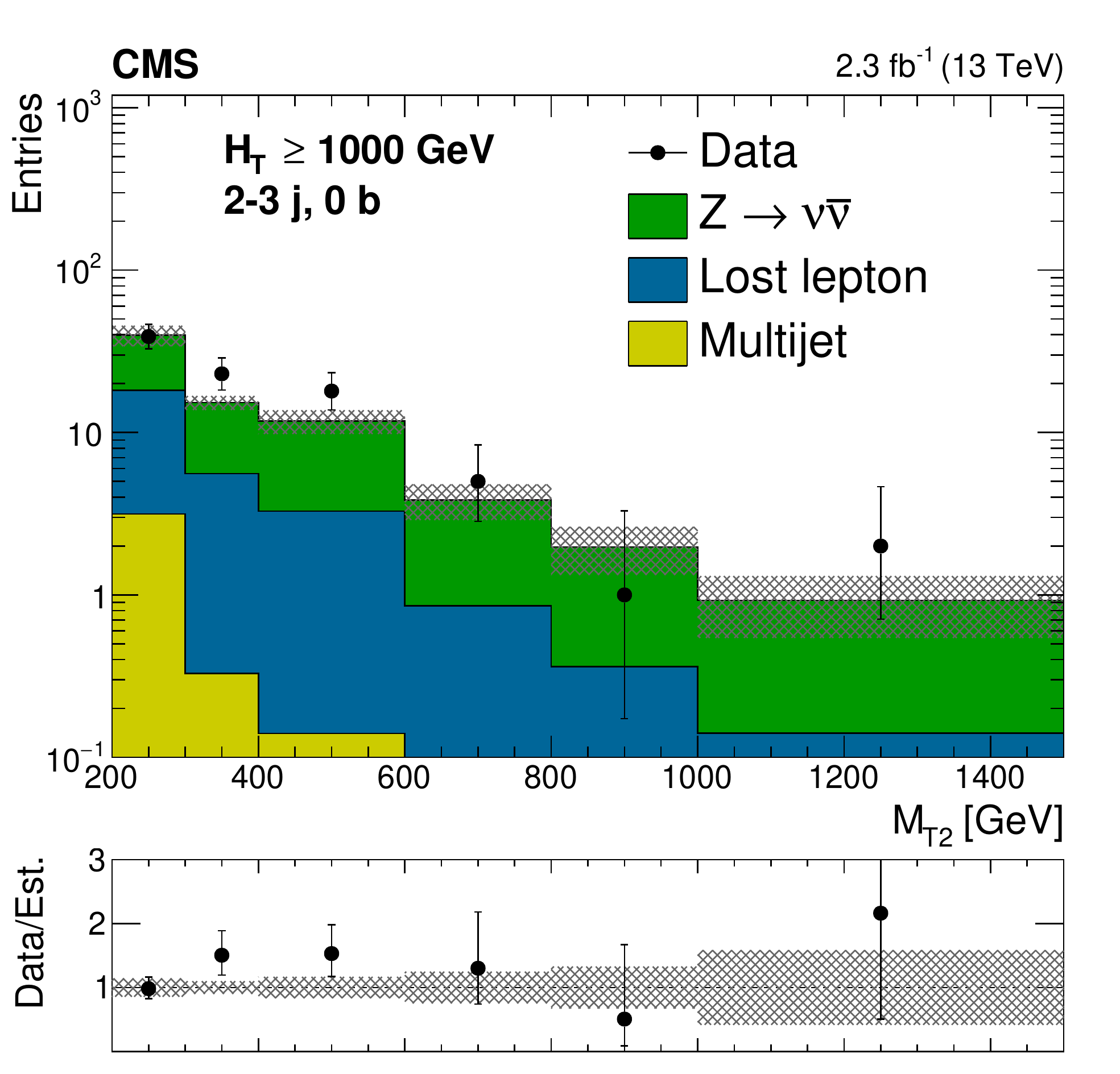}
    \includegraphics[width=0.49\textwidth]{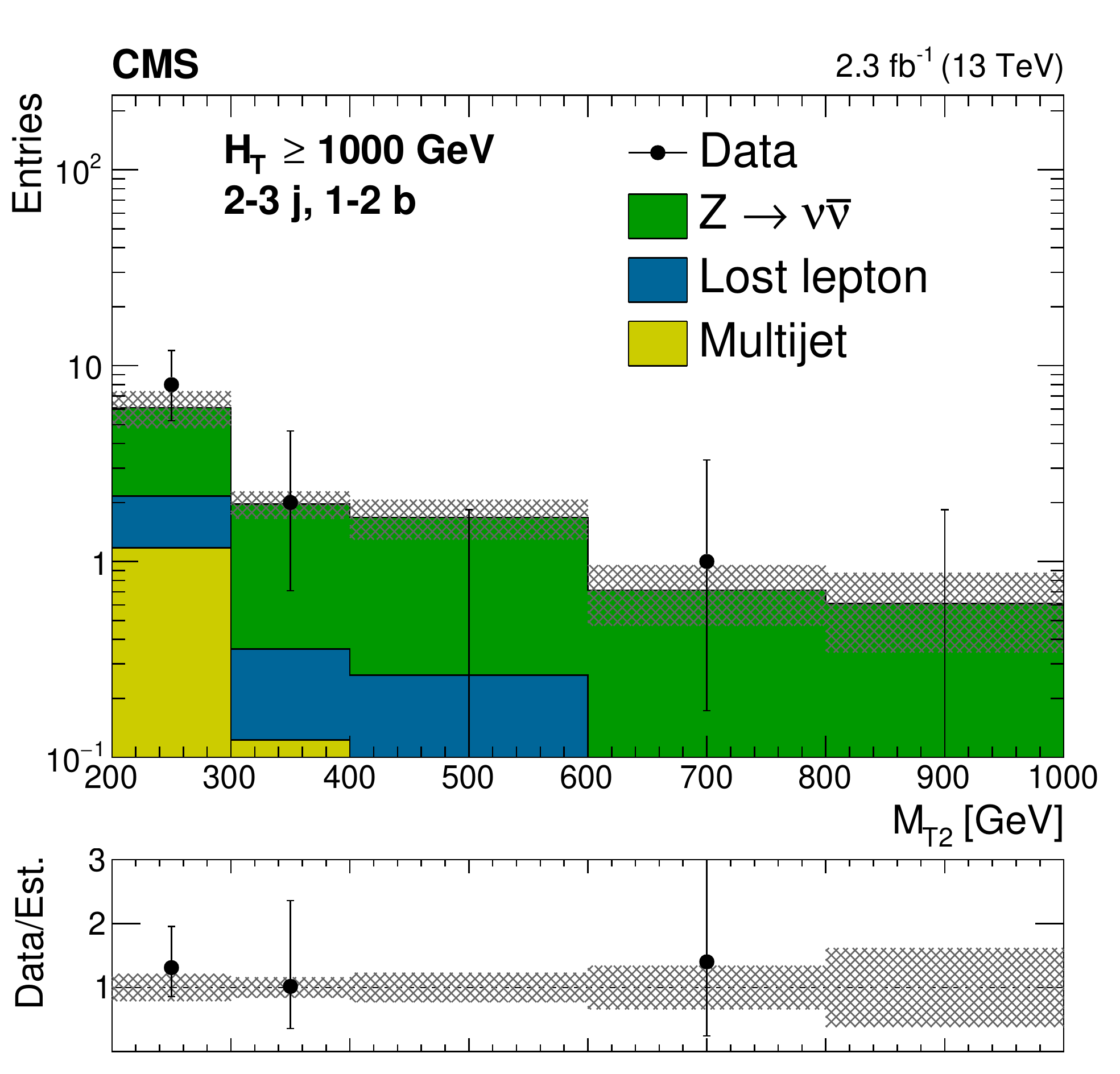}\\
    \includegraphics[width=0.49\textwidth]{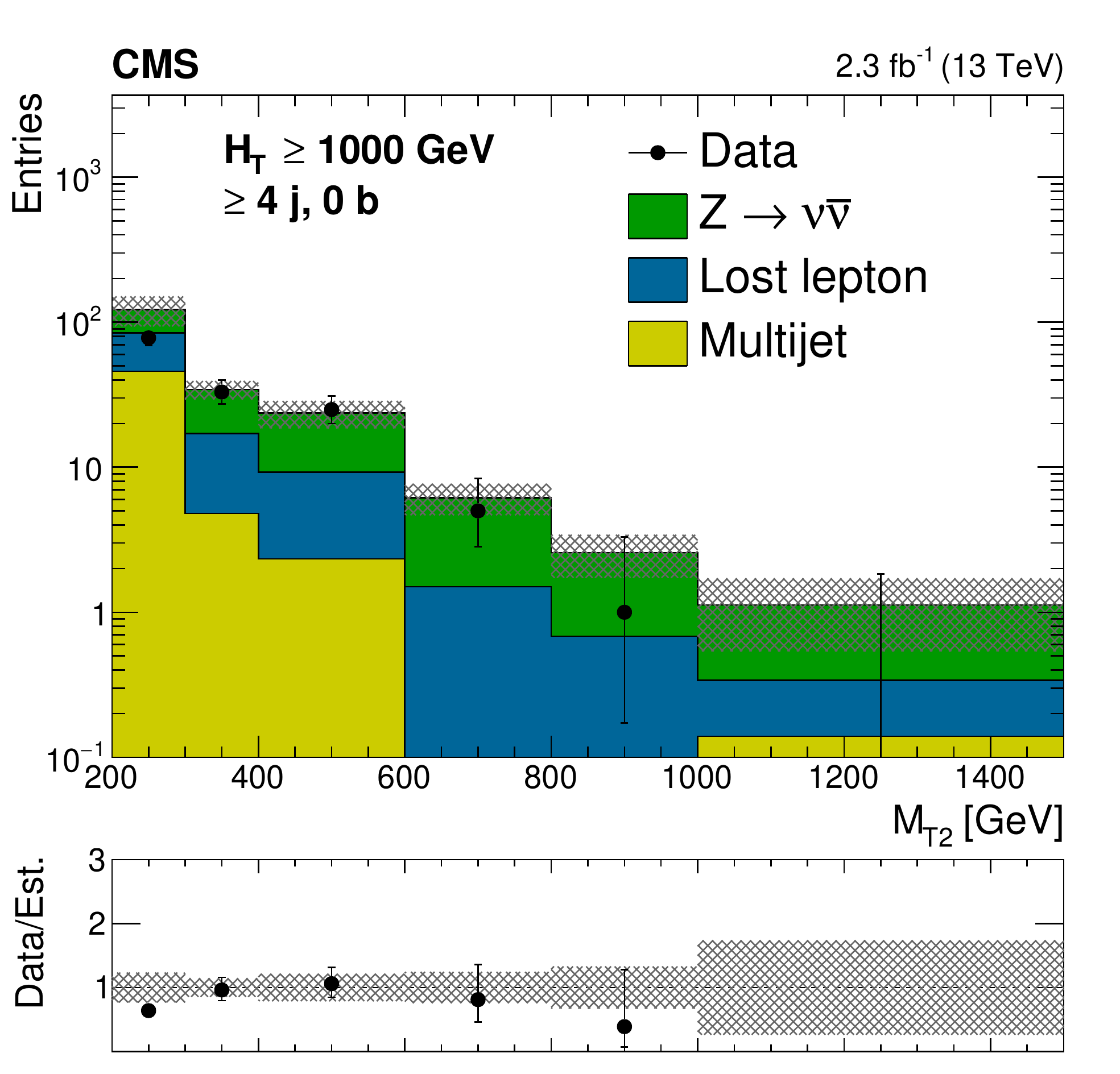}
    \includegraphics[width=0.49\textwidth]{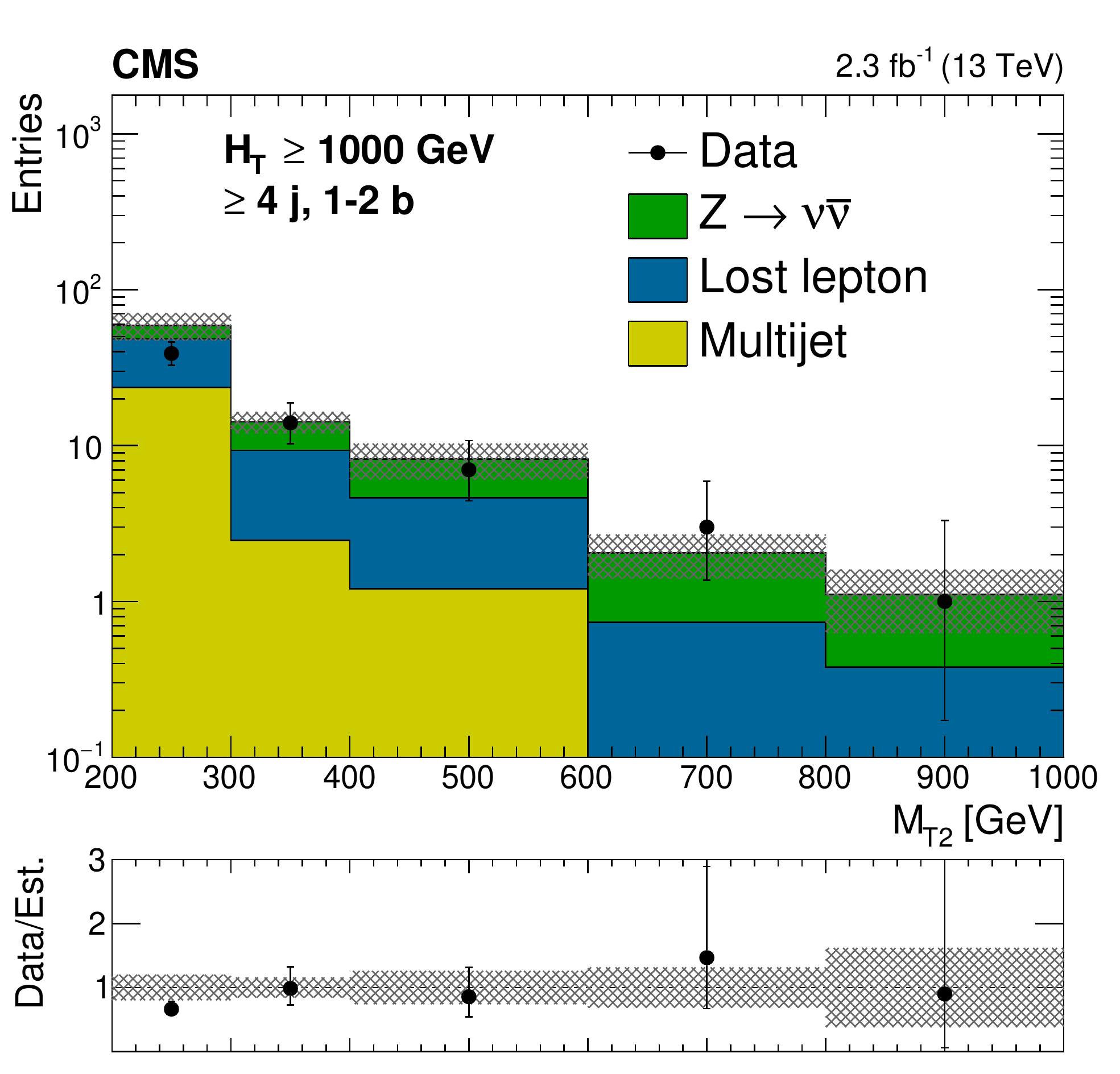}\\
    \caption{ Comparison of estimated background and observed data events in inclusive topological regions,
    as labeled in the legends, as a function of \mttwo,
    for events with $\HT > 1000\GeV$. The background prediction is formed by summing pre-fit values for all signal regions
    included in each plot. Hatched bands represent the full uncertainty in the background estimate. }
    \label{fig:macroplots2}
\end{figure}

\cleardoublepage \section{The CMS Collaboration \label{app:collab}}\begin{sloppypar}\hyphenpenalty=5000\widowpenalty=500\clubpenalty=5000\textbf{Yerevan Physics Institute,  Yerevan,  Armenia}\\*[0pt]
V.~Khachatryan, A.M.~Sirunyan, A.~Tumasyan
\vskip\cmsinstskip
\textbf{Institut f\"{u}r Hochenergiephysik der OeAW,  Wien,  Austria}\\*[0pt]
W.~Adam, E.~Asilar, T.~Bergauer, J.~Brandstetter, E.~Brondolin, M.~Dragicevic, J.~Er\"{o}, M.~Flechl, M.~Friedl, R.~Fr\"{u}hwirth\cmsAuthorMark{1}, V.M.~Ghete, C.~Hartl, N.~H\"{o}rmann, J.~Hrubec, M.~Jeitler\cmsAuthorMark{1}, A.~K\"{o}nig, I.~Kr\"{a}tschmer, D.~Liko, T.~Matsushita, I.~Mikulec, D.~Rabady, N.~Rad, B.~Rahbaran, H.~Rohringer, J.~Schieck\cmsAuthorMark{1}, J.~Strauss, W.~Treberer-Treberspurg, W.~Waltenberger, C.-E.~Wulz\cmsAuthorMark{1}
% ...
\vskip\cmsinstskip
% ...
1:~~Also at Vienna University of Technology, Vienna, Austria\\
% ...

\end{sloppypar}
\end{document}